\documentclass[conf]{new-aiaa}
\usepackage[utf8]{inputenc}

\usepackage{graphicx}
\usepackage{amsmath}
\usepackage[version=4]{mhchem}
\usepackage{siunitx}
\usepackage{longtable,tabularx}
\usepackage{placeins}
\newcommand{\FST}{free--stream turbulence}
\newcommand{\FS}{free stream}
\newcommand{\LE}{leading edge}

\newcommand{\SPF}{surface pressure fluctuations}
\usepackage{framed} 
\usepackage{multicol} 
\usepackage{nomencl} 
\makenomenclature
\renewcommand*\nompreamble{\begin{multicols}{2}}
\renewcommand*\nompostamble{\end{multicols}}

\title{The Increase of the Airfoil Trailing Edge Noise and Unsteady Surface Pressure due to High Inflow Turbulence}

\author{Laura Botero-Bolívar\footnote{PhD candidate, Engineering Fluid Dynamics, Department of Thermal Fluid Engineering, l.boterobolivar@utwente.nl.}, Fernanda L. dos Santos\footnote{PhD candidate, Engineering Fluid Dynamics, Department of Thermal Fluid Engineering, f.l.dossantos@utwente.nl}, Cornelis H. Venner\footnote{Chair Engineering Fluid Dynamics group, Department of Thermal Fluid Engineering, c.h.venner@utwente.nl} and Leandro de Santana\footnote{Assistant professor, Engineering Fluid Dynamics, Department of Thermal Fluid Engineering, leandro.desantana@utwente.nl}}
\affil{University of Twente, Enschede, Netherlands}

\begin{document}

\maketitle

\begin{abstract}
Human factors, specifically visual impact and noise production, are the current main limitations for broader urban wind energy exploitation. Trailing edge noise, caused by the turbulent boundary layer interacting with the airfoil surface, is the primary source of noise of modern horizontal and vertical axis wind turbines. Low inflow turbulence levels do not affect the trailing edge noise. However, an inflow of high turbulence intensity has flow structures that can penetrate the boundary layer, increasing the velocity fluctuations inside the boundary layer and, consequently, the surface pressure fluctuations and trailing edge noise. This research investigates the effect of high \FST, observed in the atmospheric boundary layer of urban zones, in the trailing edge noise generation. This was performed by measuring the increment of the turbulence inside of the boundary layer and surface pressure fluctuations near the trailing edge when an airfoil is submitted to high inflow turbulence. 
Experimental measurements were performed in the Aeroacoustic Wind Tunnel of the University of Twente on a NACA~0012 airfoil subjected to a turbulent inflow. The chord--based Reynolds number of the experiments ranged from 170 $\times$ 10\textsuperscript{3} to 500 $\times$ 10\textsuperscript{3}. Results showed that high inflow turbulence significantly increases the velocity fluctuations and the integral length scale along the entire boundary layer, resulting in an increment of the surface pressure spectrum more than 6~dB and 10~dB in the entire frequency range for 10\% and 20\% of \FST, respectively. The 10\% \FST~increases the velocity fluctuations just in the low--frequency range, whereas the 20\% inflow turbulence influences the velocity spectrum in the entire frequency range, increasing the size of the smallest structures of the turbulence. 
Amiet's theory is applied to predict the trailing edge far--field noise.
\end{abstract}

\section{Nomenclature}

{\renewcommand\arraystretch{1.0}
\noindent\begin{longtable*}{@{}l @{\quad=\quad} l@{}}
$\gamma^2$  & Coherence  [-] \\
$\delta$  & Boundary layer thickness  [m] \\
$\delta^*$  & Boundary layer displacement thickness  [m] \\
$\theta$  & Boundary layer momentum thickness  [m] \\
$\kappa$ & von Kármán constant (= 0.38) [-] \\
$\kappa_{x, y, z}$ & Wavenumber in the x--, y-- and z--direction [m] \\
$\kappa_e$ & Wavenumber of the main energy bearing eddies [m] \\
$\Lambda_{z|PP}$  & Spanwise correlation length  [m] \\
$\Lambda$  & Integral length scale in the x--direction [m] \\
$\Lambda_{y|vv}$  & Integral length scale in the y--direction [m] \\
$\nu$  & Kinematic viscosity [m\textsuperscript{2}/s]\\
$\Pi_\omega$ & Surface pressure frequency spectrum [Pa\textsuperscript{2}/Hz] \\
$\Pi_w$ & Cole's wake factor [-] \\
$\rho$ & Density [kg/m\textsuperscript{3}] \\
$\sigma$ & Flow corrected radial distance of microphone position [m] \\
$\phi_{ii}$  & i--direction velocity spectrum  [1/m\textsuperscript{2}] \\
$\phi_{\mathrm{FFM,FFM}}$  & Auto--spectrum of far--field microphone measurement  [1/Pa\textsuperscript{2}] \\
$\phi_{\mathrm{SM,SM}}$  & Auto--spectrum of surface microphone measurements  [1/Pa\textsuperscript{2}] \\
$\phi_{\mathrm{FFM,SM}}$  & Cross--spectrum of surface microphone and far--field microphone measurements  [1/Pa\textsuperscript{2}] \\
$\phi_{i,j}$  & Phase between microphones i and j [rad] \\
$\omega$  & Angular frequency [rad/seg] \\
$b$  & Span [m] \\
$b_c$  & Corcos' constant [m] \\
$c$  & Chord [m] \\
$c_o$ & Speed of sound [m/s] \\
$c_f$ & Friction coefficient [-] \\
$C_p$ & Pressure coefficient [-] \\
$H$ & Boundary layer shape factor (= $\delta^*$/$\theta$) [-] \\
$k$ & Trip heigth [m] \\
$\mathscr{L}$ & Airfoil response function [-]\\
$M$     & Mesh size [m]\\
$n_{x, z}$     & Distance between the microphones in the x-- and z--direction [m]\\
$P_\omega$ & Wall pressure wavenumber--frequency spectral density [Pa\textsuperscript{2}/m]\\
$P_{\mathrm{ref}}$ & Reference pressure to calculate the spectrum (= 20 \textmu Pa)  [Pa]\\
$q$ & Dynamic pressure [Pa]\\
$S_{\mathrm{pp}}$  & Far--field power spectral density [Pa\textsuperscript{2}/Hz]\\
$Tu$  & Turbulence in the x--direction [\%] \\
$U, V, W$  & Free-stream velocity in the x--, y-- and z--direction [m/s] \\
$u, v, w$  & Velocity fluctuations in the x--, y-- and z--direction [m/s] \\
$U_c$ & Convection velocity [m/s] \\
$U_e$  & Edge velocity  [m/s] \\
$u_\tau$  & Friction velocity  [m/s] \\
$x, y, z$  & Coordinate reference: chordwise, normal to the wall and spanwise [m] \\
$x_o, y_o, z_o$  & Microphone position [m]\\
$y^+$  & Non--dimensional distance from the wall (= y$\nu$/u\textsubscript{$\tau$}) [-] \\
$\mathrm{FFM}$  & Far--field microphone \\
$\mathrm{OTS}$  & Open test section \\
$\mathrm{SM}$  & Surface microphone \\
$\mathrm{TE}$  & Trailing edge \\
$\mathrm{TF_{\mathrm{FFM,SM}}}$  & Transfer function between surface microphone and far--field microphone \\

\end{longtable*}}


\section{Introduction}
\lettrine{I}{n} recent years, the global concerns about reducing CO\textsubscript{2} emissions have driven research efforts towards further extension of the exploitation of wind resources to produce energy. Harvesting the available energy of the natural movement of the air is among the best current options to produce clean energy due to the high--efficiency, low--cost, and abundant wind resources around the world. Urban wind energy is currently developing as an interesting alternative due to the reduction in the cost of energy transmission and the increase in the efficiency of wind turbines. The optimal exploitation of urban wind resources highly depends on societal acceptance, which is a current scientific and industrial challenge that urgently requires innovative breakthroughs. Community noise is the most perceived societal impact of wind turbines, and its abatement is crucial for the growth of wind power participation in a countries' energy matrix.

Vertical axis wind turbines installed in urban areas and horizontal axis wind parks placed at the vicinity of urban zones are an essential component of the power matrix of future environmentally--friendly smart cities. The urban canopy develops pressure build--up and canyon effects that are unique favorable conditions capable of boosting the wind turbines' aerodynamic performance~\cite{abohela2013effect}. Nevertheless, the atmospheric--scale inhomogeneous roughness, caused by buildings of different heights, and horseshoe vortices formed around and over tall buildings, produce extremely high atmospheric turbulence intensity levels, up to 50\%~\cite{ricciardelli2006some}~\cite{carpman2011turbulence} that affect the aerodynamic performance, structural loads, and the wind turbine noise production. Accurate prediction of wind turbine performance is needed to control the community noise and guarantee the quality of life of the population living near wind turbines. Therefore, real atmospheric flow conditions must be incorporated in the current methods to predict the noise generated by wind turbines. 

Mechanical and aerodynamic noise are the primary noise sources of a typical modern wind turbine. The mechanical noise is generated by the wind turbine power generator, the gearbox, and the mechanical components and is significantly reduced by acoustic treatments in the nacelle and specific mechanical components and couplings. The aerodynamic noise is caused by the interaction between the blades and the air. Therefore, it is generated at all operating conditions~\cite{hansen2017wind} and its abatement has consequences to the wind turbine aerodynamic performance and power production. Five aerodynamic wind turbine noise sources have been identified, i.e., inflow--turbulence, trailing edge, vortex shedding, tip, and blunt trailing edge noise~\cite{hansen2017wind}. Inflow--turbulence interacting with the wind turbine blade leading edge produces the leading--edge noise~\cite{amiet1975acoustic}. The other four sources are considered airfoil self--noise since they exist even in uniform turbulence--free incoming flow conditions~\cite{brooks1989airfoil}. Tip noise originates from the blade tip vortex interacting with the structure. Blunt trailing edge and vortex shedding noise are mechanisms less critical and depend on the boundary layer's specific conditions and airfoil geometry~\cite{brooks1989airfoil}. Trailing edge noise is recognized as the primary noise source of modern vertical and horizontal axis wind turbines~\cite{oerlemans2011wind}. The trailing edge imposes a sudden change of impedance, scattering the unsteady pressure of the surface caused by the turbulent boundary layer. An airfoil immersed in a uniform flow with a turbulent and attached boundary layer has the trailing edge as a unique noise source~\cite{brooks1989airfoil}. A slight increase of the incoming turbulence, i.e., 1$\%$ causes the leading edge to become the most important noise source~\cite{amiet1975acoustic}, particularly at low frequencies. However, the dominant airfoil noise source for very high inflow turbulence is still an open question. The central hypothesis and main conclusion of this research is that for high inflow turbulence levels, the penetration of the \FST~significantly increases the unsteady surface pressure fluctuations of the airfoil, consequently increasing the trailing edge noise. 

The most current trailing edge noise prediction methods are based on Amiet's theory~\cite{amiet1976noise}. 
Amiet postulated that the surface pressure spectrum upstream of the trailing edge is the appropriate input for the prediction of the trailing edge noise. The pressure fluctuations are induced by turbulent structures convected inside the boundary layer. Several semi--empirical and semi--analytical methods have been developed to calculate the pressure fluctuation in a turbulent boundary layer~\cite{goody2004empirical, rozenberg2012wall, kamruzzaman2012semi, catlett2016empirical, hu2016characteristics}. Particularly, Parchen and Blake~\cite{parchen1998prediction} developed a model in the wavenumber frequency domain by solving the Poisson equation. The model calculates the surface pressure spectrum by predicting the velocity fluctuations, length scale, and velocity spectrum inside the boundary layer. 

Previous researches have demonstrated that high \FST~levels can penetrate the boundary layer and change the steady and unsteady characteristics of the boundary layer~\cite{dogan2016interactions, thole1996high,blair1983influence,castro1984effects,hancock1983effect,stefes2004skin,hutchins2007large}. The influence of the \FST~in the boundary layer depends on the turbulence intensity and the characteristic length scale~\cite{dogan2016interactions, thole1996high, hancock1983effect}. The length scale of the \FST~should be significantly larger than the turbulent length scale inside of the boundary layer to occur penetration of the \FST. Hutchins and Maurisc ~\cite{hutchins2007large}, Sharp et al.~\cite{sharp2009effects} and Dogan et al.~\cite{dogan2016interactions} demonstrated that large coherent structures present in the turbulent inflow change the nature of the turbulence and interaction between the turbulent scales inside the boundary layer. These researches showed that the \FST~could modulate the amplitude of the small--scale fluctuations close to the wall. The amplitude of the modulation depends on the level of \FST. Dogan et al. (2017)~\cite{dogan2016interactions}, performed hot--wire measurements close to the wall of a flat--plate, i.e., y\textsuperscript{+} $\approx$ 5, and decomposed the velocity fluctuations caused by the smaller and larger scales inside the boundary layer using a cut--off wavelength filter, as did by Hutchins and Marusic ~\cite{hutchins2007evidence} and Mathis et al.~\cite{mathis2009large}. The authors concluded that the modulation effect of larger scales on the near--wall region increases as the length scale of the inflow turbulence increases. Although the previous analyses demonstrated the penetration of the \FST~in the boundary layer and the increment of the velocity fluctuations, the effect on the surface pressure fluctuations and trailing edge noise had not been studied yet. Furthermore, modeling the turbulence inside the boundary layer in the presence of inflow turbulence is still a challenging and urgent task.

Understanding and modeling the influence of the \FST~on the trailing edge noise generation is essential for developing more accurate noise prediction methods used for the design of more socially acceptable wind turbines. Wind turbines located inside or in the vicinity of urban environments are subjected to conditions where atmospheric turbulence plays an essential role in power and noise production. This study provides new insights into the influence of high \FST~on the turbulence inside the boundary layer and on the surface pressure fluctuations near the trailing edge through experiments conducted in an open--jet wind tunnel on a NACA~0012 airfoil. Two passive grids were used to generate uniform turbulence of 10\% and 20\%. The experiments involved measurements of the boundary layer and of the surface pressure fluctuations using hot--wire anemometry and flushed--mounted microphones, respectively. Amiet's theory was used to predict the increase of the far--field noise caused by the presence of high \FST. The input for such a model is the measured surface pressure spectrum.

The remaining part of the paper proceeds as follows: Section III. shows the methodology , the wind tunnel facility, the airfoil instrumentation, the calibration, and the calculation procedures; section IV. presents the results and discussion divided into free--stream turbulence, boundary layer, and surface pressure fluctuations; this subsection also addresses the predicted increment of the trailing edge far--field noise due to inflow turbulence. Finally, section V. addresses the main conclusions of this work.

\section{Methodology}
\subsection{Experimental set--up}
Wind tunnel experiments were conducted in the closed--circuit Aeroacoustic Wind Tunnel of the University of Twente~\cite{deSantana2018}. The facility allows measurements in a closed test section and open--jet configuration of 0.9~m x 0.7~m width and height, respectively. An anechoic chamber of  6~m  $\times$ 6~m  $\times$ 4~m and 160~Hz cut--off frequency encloses the test region. The wind tunnel can reach velocities up 60~m/s with a turbulence intensity lower than 0.08\%~\cite{deSantana2018}. 

A 200~mm chord NACA~0012 airfoil was located in the open test section (OTS). The velocity at the OTS center ranged from 10~m/s to 30~m/s, which corresponds to a Reynolds number of 170$\cdot$ 10\textsuperscript{3} to 500$\cdot$ 10\textsuperscript{3} based on the chord of the airfoil. A zig--zag trip of 60\textsuperscript{o} angle and 12~mm width was positioned at x/c = -0.8 (x = 0 corresponds to the trailing edge) on both the upper and the lower side. The height of the trip (k) varied from 0.7~mm at 10~m/s to 0.30~mm at 20~m/s and 30~m/s, i.e., $k/\delta$ = 0.7, 0.5, and 0.6,  respectively, to guarantee the transition of the boundary layer without increasing the turbulence of the boundary layer~\cite{dossantos2021}. The angle of attack was 0\textsuperscript{o} in all cases presented in this paper.  The coordinate reference system considers the chordwise direction axis \textit{x}, with x = 0 located at the trailing edge, the axis normal to the wall \textit{y}, with y = 0 the airfoil wall, and the axis aligned with the span \textit{z}, with z = 0 at midspan.

\begin{figure}[hbt!]
\centering
\includegraphics[width=.7\textwidth]{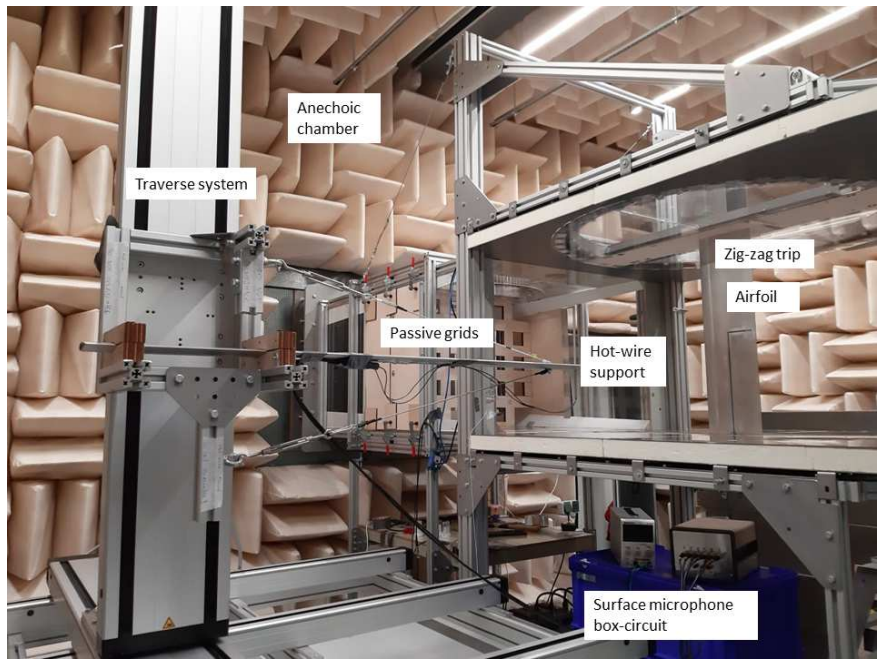}
\caption{ \label{Fig: setup} Picture of the experimental set--up.}
\end{figure}

The experimental campaign is organized in three parts, i.e., characterization of the passive grids by an X hot--wire sensor, measurements of the boundary layer by a single--wire hot--wire sensor, and measurements of the surface pressure fluctuations near the airfoil trailing edge by flush--mounted microphones. Figure~\ref{Fig: setup} depicts the experimental setup composed of the wind tunnel anechoic chamber, the grid to generate turbulence, the airfoil with a zig--zag trip, the traverse system, and the hot--wire support. Table~\ref{tab:exp_matrix} presents the test matrix.

\begin{table}[htbp!]
\caption{\label{tab:exp_matrix} Experimental test matrix.}
\centering
\begin{tabular}{lcccccccccccccc}
\hline
        & \multicolumn{1}{c}{Flow mapping}          && \multicolumn{1}{c}{Boundary layer } & & \multicolumn{1}{c}{Surface pressure fluctuations}   \\ \cline{2-6} 
       
Uniform flow & 10, 20, 30 [m/s] & & 10, 20, 30 [m/s] & &  10, 20, 30 [m/s] \\
Grid 1 (Tu = 20\%)      & 10 [m/s] & & 10 [m/s] & &  10 [m/s]  \\
Grid 2 (Tu = 10\%)  & 10, 20, 30 [m/s] & & 10, 20, 30 [m/s] & &  10, 20 [m/s] \\ 
\hline
\end{tabular}
\end{table}

\subsection{Passive grids}
\begin{figure}[hbt!]
\centering
 \begin{tabular}{cc}
{\footnotesize{}\includegraphics[width=.45\textwidth]{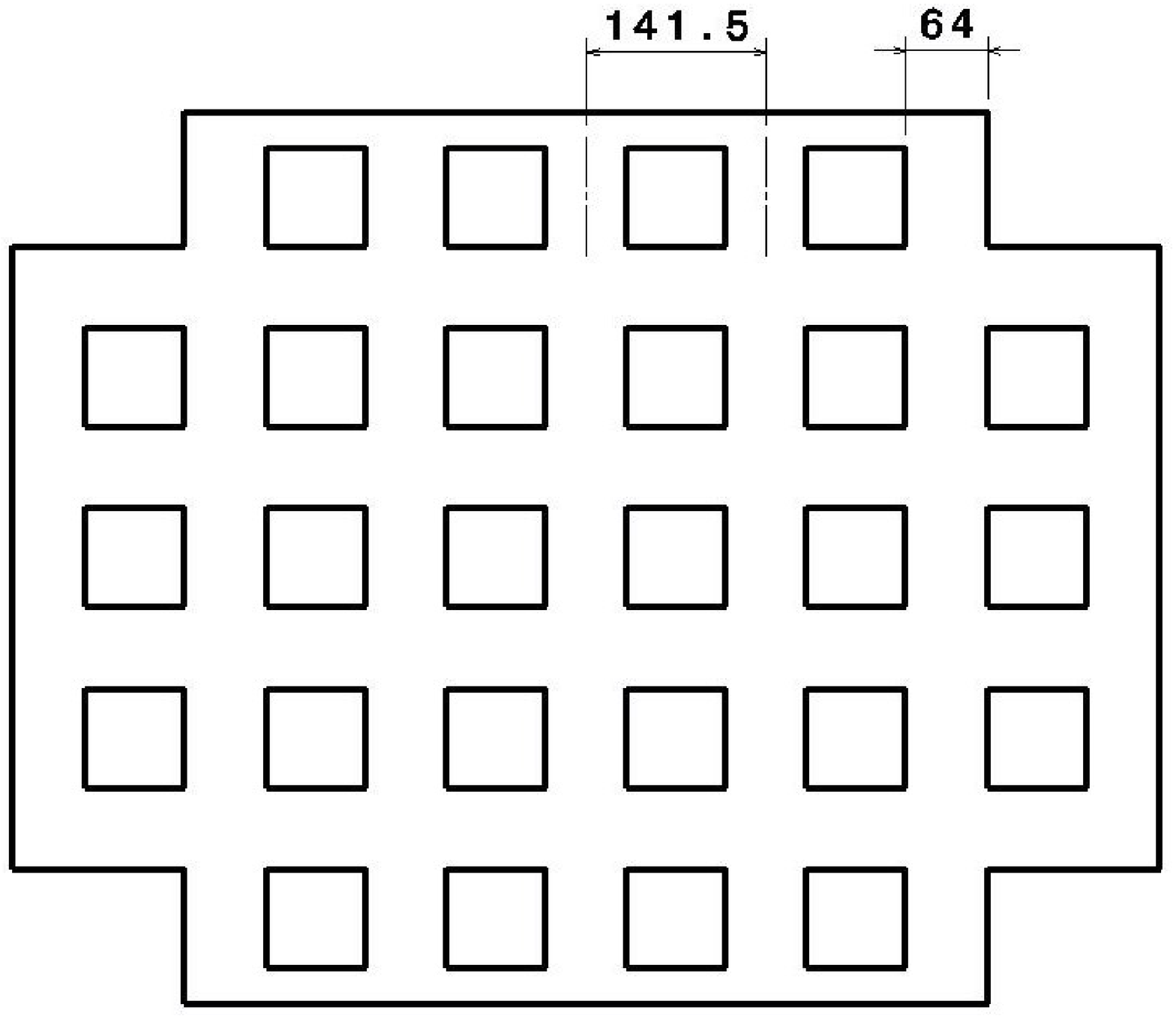}} &
{\footnotesize{}\includegraphics[width=.45\textwidth]{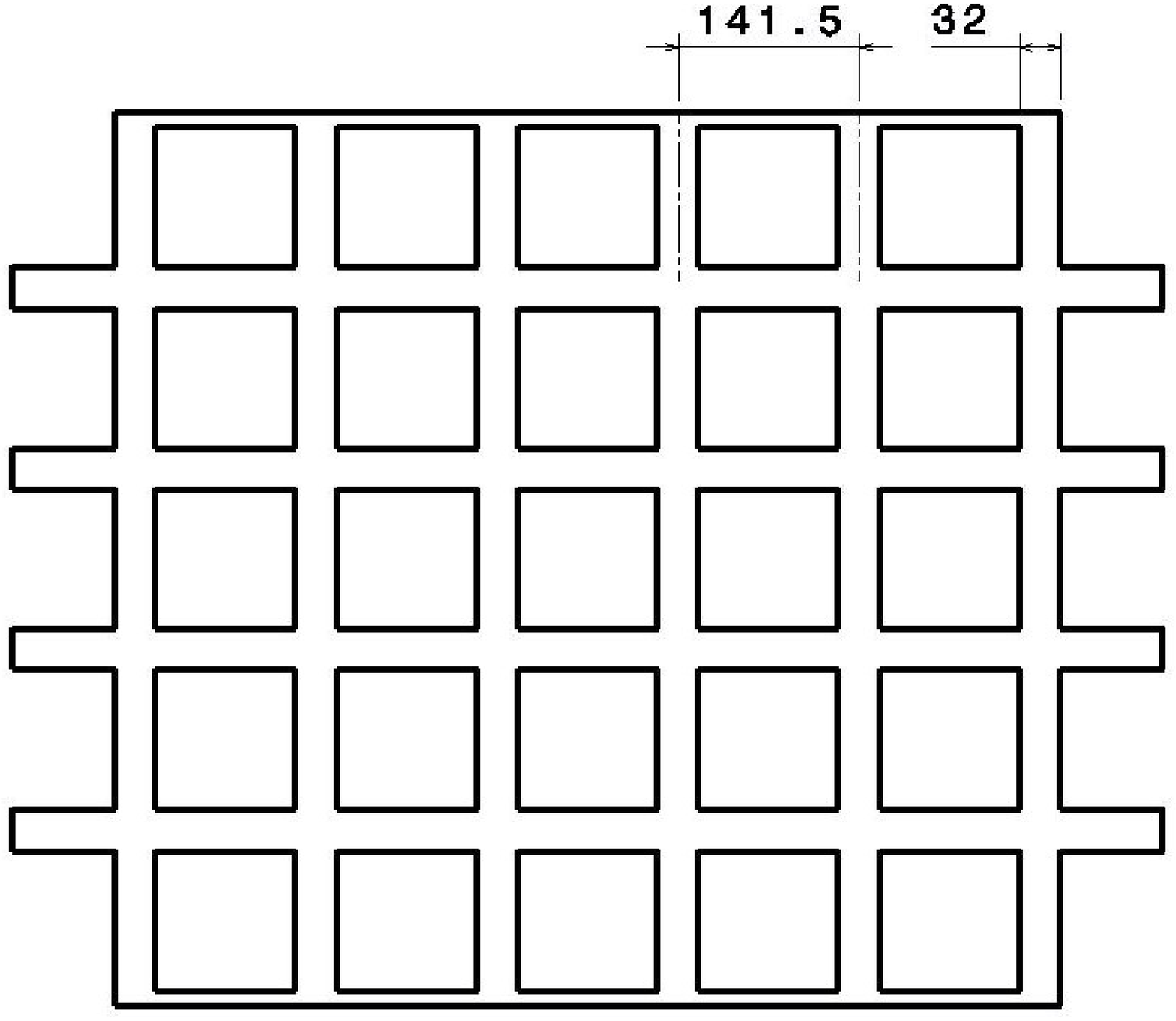}}
\tabularnewline
{\footnotesize{} (a) Grid 1. $\mathbf{\beta}$ = 0.3. Tu = 20\%} & {\footnotesize{} (b) Grid 2. $\mathbf{\beta}$ = 0.6. Tu = 10\%}
\end{tabular}
\caption{ \label{Fig: Grids} Passive grids to generate turbulence. All dimensions are in mm.}
\end{figure}

Square mesh passive grids produced out of wood were used to generate turbulence in the \FS. Figure~\ref{Fig: Grids} shows the dimensions of both grids used. The porosity is 0.3 and 0.6 for grid 1 and 2, respectively. The grids are located at x/M = 10 upstream of the \LE~plane of the airfoil.

\subsection{Hot--wire measurements}
Hot--wire measurements were performed to verify the uniformity of the flow velocity, named flow mapping in Tab. \ref{tab:exp_matrix} and the generated turbulence in the plane of the airfoil leading edge and to measure the statistics of the boundary layer in the vicinity of the airfoil trailing edge. The flow--mapping measurements adopted an X--wire probe ref. 55P51 from Dantec Dynamics. The second campaign used a single--wire boundary layer type ref. 55P15 from Dantec Dynamics. The X--wire and single--wire probes were mounted in 55H26 and 55H20 supports from Dantec Dynamics, respectively. A 3D  traverse system from Dantec Dynamics positioned the probes at each  specific location with a precision of 6.2~\textmu m. The Dantec Dynamics StreamLine Pro CTA system maintained the hot wires at constant temperature. The Dantec StreamWare software acquires the probe's measurements digitized by a 9215 A/D converter from National Instrument. The sampling frequency for all hot--wire measurements was 20~kHz. The overheat ratio was 0.8. Low-- and high--pass cut--off filters of 10~kHz and 10~Hz were applied in the acquired data to eliminate alias and effects related to flow buffeting instability naturally present in the open test section wind tunnels. 

A Prandtl tube served as the reference for the in--situ calibration of both hot--wire sensors. A fourth--order polynomial curve fitted 20 calibration velocities between 1.5~m/s and 60~m/s  distributed logarithmically, besides the no--velocity point. For the directional calibration of the X--wire probe, a K10CR1/M  Motorized Rotation Mount from ThorLabs precisely varied the angle from -45\textsuperscript{o} to +45\textsuperscript{o}, with an increment in the angle of 5\textsuperscript{o}. 

The X--wire probe of 5~\textmu m diameter and 3~mm length measured the x and y components of the velocity of one--quarter of the wind tunnel open test section at the plane in which the leading edge of the airfoil would be located. For the flow mapping, the velocity is sampled during 5~s. The distance between the points was 16~mm in each direction, which implied 609 measurements in the mapped region. The single wire of 4~\textmu m diameter and 3~mm length measured the x velocity at different boundary layer positions. 

\subsubsection{Turbulence spectrum}
The velocity time--history measured by the hot--wire probes at the center of the test section and inside of the boundary layer is adopted for calculating the velocity spectrum. To analyze the turbulence generated by the grids, the acquisition time was 30~s and for the boundary layer measurements, the velocity was acquired during 40~s. The spectral estimation is performed by Welch's method. For 50\% overlapping blocks of 2\textsuperscript{11} samples, Hanning's window is applied, yielding to a bin size of 9.7~Hz.

\begin{figure}[hbt!]
\centering
\includegraphics[width=0.45\textwidth]{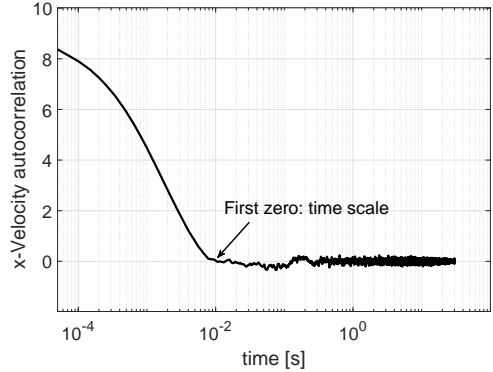}
\caption{\label{Fig: LS} X--velocity autocorrelation for Grid 2 (Tu = 10\%). U$_\infty$ = 30~m/s.}
\end{figure}

The streamwise von Kármán spectrum is calculated as shown in Eq.~\ref{Eq: vonkarman}. The streamwise integral length scale ($\Lambda$) is calculated according to the methodology proposed by Hinze~\cite{hinze19720}. The turbulent time--scale corresponds to the first zero of the autocorrelation of the velocity time--history, as shown in Fig.~\ref{Fig: LS}. Assuming Taylor's frozen turbulence hypothesis~\cite{taylor1938spectrum}, the integral turbulent length scale is determined considering the convection velocity and time scale. 
\begin{equation}\label{Eq: vonkarman}
\begin{split}
   \phi_{uu}(\kappa_x) &= \frac{2}{\sqrt{\pi}}\frac{ \Gamma \left(5/6 \right)}{\Gamma \left(1/3 \right)}\frac{u_{rms}^2}{\kappa_e}(1+\kappa_x^2)^{--5/6}  \\
    \kappa_e& = \frac{\sqrt{\pi}}{\Lambda}\frac{ \Gamma \left( 5/6 \right)}{\Gamma \left(1/3 \right)} \\
    \kappa_x &= \frac{2\pi f}{U_c}
\end{split}
\end{equation}

\subsubsection{Boundary layer}
The measurements inside of the boundary layer consisted of 45 points differently spaced in the y--direction, i.e., perpendicular to the airfoil's wall. Measurements were conducted at x/c = -0.9517 and z/c = 0, i.e., the same location of the central surface microphone closest to the trailing edge, ref. Fig.~\ref{Fig: airfoil}. The measurement closest to the airfoil's wall is at y = 1.1~mm. An additional measurement outside the boundary layer, i.e., at the middle between the airfoil and the shear--layer of the OTS, was performed and is considered the condition of the \FS~ for the following analyses.  

The boundary layer thickness is calculated using the criterion  $U/U_\infty = 0.98$. The boundary layer displacement thickness ($\delta$\textsubscript{*}) and the boundary layer momentum thickness ($\theta$) are calculated according to Eq.~\ref{eq: bldt}, where $\infty$ is a position outside of the boundary layer. The shape factor (H) is the ratio between $\delta$\textsuperscript{*} and $\theta$. The integration was done using the trapezium rule.
\begin{equation}\label{eq: bldt}
\begin{split}
      \delta^* &= \int_{0}^{\infty} \left( 1-\frac{U}{U_e}\right) dy \\
       \theta &=\int_{0}^{\infty}  \frac{U}{U_e} \cdot \left( 1-\frac{U}{U_e} \right) dy
\end{split}
\end{equation}

The friction velocity is calculated by fitting the mean velocity profile with the Prandtl -- von Kármán log--law coupled with Cole's wake law shown in Eq.~\ref{eq: mean_velocity_profile}. The fitting also calculated the factor $\Pi_w$, due to the lack of experimental information for the $dC_p/dx$. $\kappa$ is the von Kármán constant, equal to 0.38, and B is a level-constant equal to 5.
\begin{equation}\label{eq: mean_velocity_profile}
\begin{split}
\frac{U}{u_\tau}& = \frac{1}{\kappa}\log \left( y^+ \right) + B + \frac{2\Pi_w}{\kappa}\sin^2\left( \frac{\pi y}{2\delta} \right) \\
\Pi_w &= 0.8 \left( \frac{\delta^* }{\tau_w}\frac{\mathrm{d} C_p}{\mathrm{d} x} + 0.5 \right)^{3/4} \\
u_\tau &= \sqrt{\frac{1}{2}c_fU_\infty ^2}
 \end{split}   
\end{equation}

\subsection{Unsteady surface pressure}
\begin{figure}[hbt!]
\centering
\includegraphics[width=.5\textwidth]{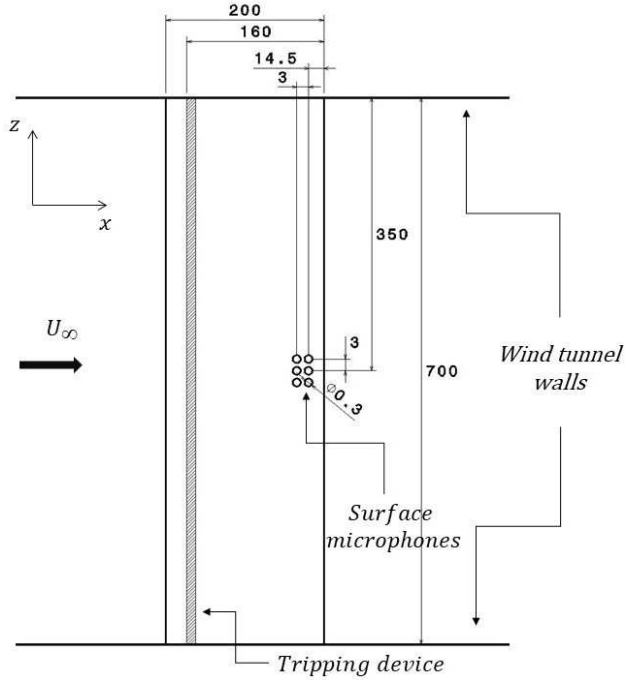}
\caption{ \label{Fig: airfoil} Scheme of airfoil instrumentation. All dimensions are in mm.}
\end{figure}

The airfoil is instrumented with 6 Knowles electret condenser analog microphones FG-23329-P07, which are placed close to the trailing edge under a pinhole of 0.3~mm diameter. Figure~\ref{Fig: airfoil} shows a scheme of the airfoil instrumentation. 
%

 The surface microphones were calibrated in--situ by an in--house calibrator equipped with a GRAS 40PH free-field microphone, used as the reference, and a Visaton FR8 loudspeaker, used as the noise source. The sensitivity of the reference microphone was calibrated by a GRAS 40 AG Sound Calibration, which uses an excitation of a 94~dB SPL tone of 1~kHz frequency. The calibration procedure of the surface microphones follows a two steps approach where the sensitivity is obtained in the first step and the frequency response in the second. The sensitivity of the microphones is determined by 1~kHz tone excitation. The ratio of the standard deviation of the signal of the reference microphone and the surface microphone corresponds to the conversion factor of the surface pressure sensors in Pa/V. The frequency response is calibrated adopting white noise as a noise source. A transfer function between the free-field microphone (FFM) and the surface microphone (SM) is calculated according to Eq.~\ref{eq: transfer_function}, where $\phi_{\mathrm{FFM,FFM}}$ and $\phi_\mathrm{{FFM,SM}}$ are the auto--and cross--spectrum of microphone signals. The equivalent spectrum of the surface pressure is calculated according to Eq.~\ref{eq: phi_eq}. The transfer function is determined as reported by Berntsen~\cite{berntsen2014remote}.
 \begin{equation}\label{eq: transfer_function}
     TF_{\mathrm{FFM,SM}} = \frac{\phi_{\mathrm{FFM,FFM}}}{\phi_{\mathrm{FFM,SM}}}
 \end{equation}
 \begin{equation}\label{eq: phi_eq}
     \phi = \frac{\phi_{\mathrm{SM,SM}}}{\lvert TF_{\mathrm{FFM,SM}} \rvert ^2}
 \end{equation}
 
A single PXIe--4499 Sound and Vibration module installed on a NI PXIe--1073 chassis was used to acquire the surface microphone data. The acquisition time was 30~s and the sampling frequency 102.4~kHz. Welch's method estimates the experimental spectrum. A Hanning window is applied to blocks of 2\textsuperscript{14} samples, and the average is computed considering 50\% overlap between subsequent blocks, yielding to a bin size of 6.2~Hz. 

\subsubsection{Spanwise correlation length}
The intensity of the trailing edge noise is directly related to the spanwise correlation length. This quantity is defined as the distance in which the surface pressure fluctuations in the spanwise direction have a significant level of coherence and is calculated as shown in Eq.~\ref{eq: corr_length}, where $n_z$ is the distance between the microphones in the spanwise direction. Measuring the spanwise correlation length is problematic since it requires having multiple pressure sensors located in the spanwise direction. Hence, Corcos~\cite{corcos1964structure} proposed a model to calculate the coherence decay between two spanwise points in a turbulent boundary layer as a function of the frequency. The Corcos model is presented in Eq.~\ref{eq: corcos_model}, where  n\textsubscript{z} is the distance between the two points, $U_c$ is the convection velocity, and b\textsubscript{c} is the Corcos' constant, which is commonly assumed to be 1.4~\cite{stalnov2016towards}.
\begin{equation}\label{eq: corr_length}
\Lambda_{z\mid PP} = \int_{0}^{\infty}\sqrt{\gamma ^2(f,n_z)}dn_z
\end{equation}
\begin{equation}\label{eq: corcos_model}
    \gamma ^2(f, n_z)= \exp{\left(\frac{-4\pi f}{b_c U_c} n_z\right)}
\end{equation}

Equation~\ref{eq: corr_lengt_t} is obtained after Eq.~\ref{eq: corcos_model} is substituted in~Eq.~\ref{eq: corr_length} yielding the theoretical value of the spanwise correlation length. In this work, the Corcos' constant is obtained by fitting Eq.~\ref{eq: corcos_model} to the coherence from a pair of microphones located in the spanwise direction. The calculation of the convection velocity is explained in the following subsection.
\begin{equation}\label{eq: corr_lengt_t}
    \Lambda_{z\mid PP} = b_c \frac{U_c}{2\pi f}
\end{equation}

\subsubsection{Convection velocity}
The convection velocity is defined as the velocity at which the vortices convect. The average of the convection velocity is calculated using the gradient of the phase spectrum of the signal of a pair of microphones in the chordwise direction, as presented in Eq.~\ref{eq: conv_velocity}. The convection velocity is usually in the range of 0.6--0.8 U$_\infty$ \cite{stalnov2016towards}.
\begin{equation}\label{eq: conv_velocity}
    U_c = \frac{2\pi n_x}{\partial \phi_{i,j}/\partial f}
\end{equation}

\subsection{Trailing edge airfoil noise prediction}
Amiet's theory~\cite{amiet1976noise} is adopted for the trailing edge noise prediction. The theory calculates the far--field acoustic pressure spectrum using the wall pressure wavenumber--frequency spectral density in the vicinity of the trailing edge. The theory assumes a large span, a stationary observer and airfoil, a uniform flow, and the boundary layer turbulence is convecting over the trailing edge as a frozen pattern, i.e., the turbulence is not affected by the discontinuity of the trailing edge. Equation~\ref{Eq:Amiet} presents the far--field power spectral density of an airfoil of chord \textit{c} and span \textit{b} for an observer perpendicular to the trailing edge at midspan (z\textsubscript{0} = 0)~\cite{stalnov2016towards}. For the variables nomenclature, the reader is referred to section I.
\begin{equation}\label{Eq:Amiet}
    S_{pp} (x_o, y_o, z_o = 0, \omega) = \left(\frac{\omega c y_o}{4\pi c_o \sigma^2} \right)\frac{\pi b}{2 \overline{U_c}} \left|\mathscr{L}(\kappa_x,\kappa_z = 0, x,y,U_\infty,\overline{U_c})\right|^2 P_\omega (\kappa_x,\kappa_z=0,\omega)
\end{equation}

This paper follows the modified TNO--Blake model~\cite{parchen1998prediction} proposed by Stalnov et al.~\cite{stalnov2016towards} to predict the surface pressure fluctuations. The surface pressure fluctuations beneath a turbulent boundary layer can be modeled by the Poisson equation \cite{stalnov2016towards}. The model solves the Poisson equation to calculate the wall pressure fluctuations in the wavenumber--frequency domain. The model adopts the steady and unsteady velocities in the streamwise and perpendicular--to--the--wall directions inside the boundary layer to calculate the single point wall pressure frequency spectrum ($\Pi(\omega)$) in the wavenumber--frequency domain, as shown in Eq.~\ref{Eq:TNO}. For the meaning of each variable, the reader is referred to section I. The calculated $\Pi(\omega)$ is compared with a single pressure transducer located on the airfoil's surface. The integration of Eq. \ref{Eq:TNO} is done by the trapezium rule.
\begin{equation}\label{Eq:TNO}
\Pi(\omega) = \frac{4\pi \rho^2}{\Lambda_{z\mid PP}(\omega)}\int_{0}^{\delta} \Lambda_{y\mid vv}(y) U_c(y)\left[ \frac{\partial U(y)}{\partial y}\right]^2 \frac{\bar{u_y^2}(y)}{U_c^2(y)}\phi_{vv}(\kappa_x,\kappa_z=0)e^{-2\mid \kappa \mid y}dy
\end{equation}
\begin{equation}\label{eq: Pw}
    P_{\omega} = \frac{U_c\Lambda_{z\mid PP}\Pi(\omega)}{\pi}
\end{equation}

\section{Results and discussion}
This section is organized as follows: section A presents the characterization of the turbulence generated by the passive grids and sections B and C discuss the influence of the \FST~on the boundary layer characteristics and on the surface pressure fluctuations and predicted far--field noise, respectively.  

\subsection{Free stream Turbulence}
Both passive grids generated uniform turbulence in the airfoil \FS~plane. The drag force generated by the grids and the consequent pressure drop in the wind tunnel circuit limited the maximum velocity of the wind tunnel to 10~m/s for grid~1 and 30~m/s for grid~2. Therefore, comparisons between both grids are shown at 10~m/s inflow velocity, and analyses at higher velocities are only presented for grid~2. 

\begin{figure}[hbt!]
 \begin{centering}
    \begin{tabular}{cc}
{\footnotesize{}\includegraphics[width=.45\textwidth]{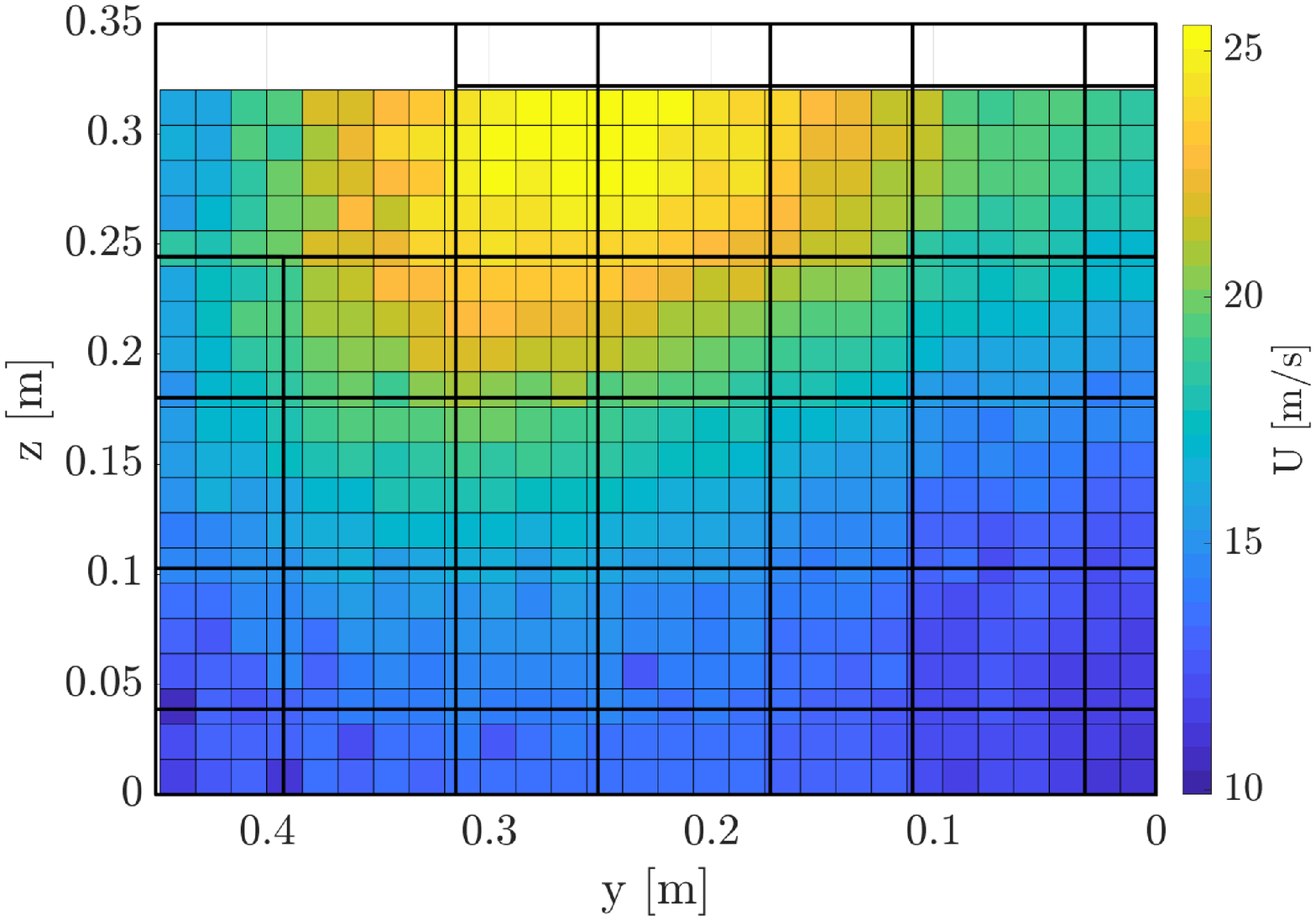}} &
{\footnotesize{}\includegraphics[width=.45\textwidth]{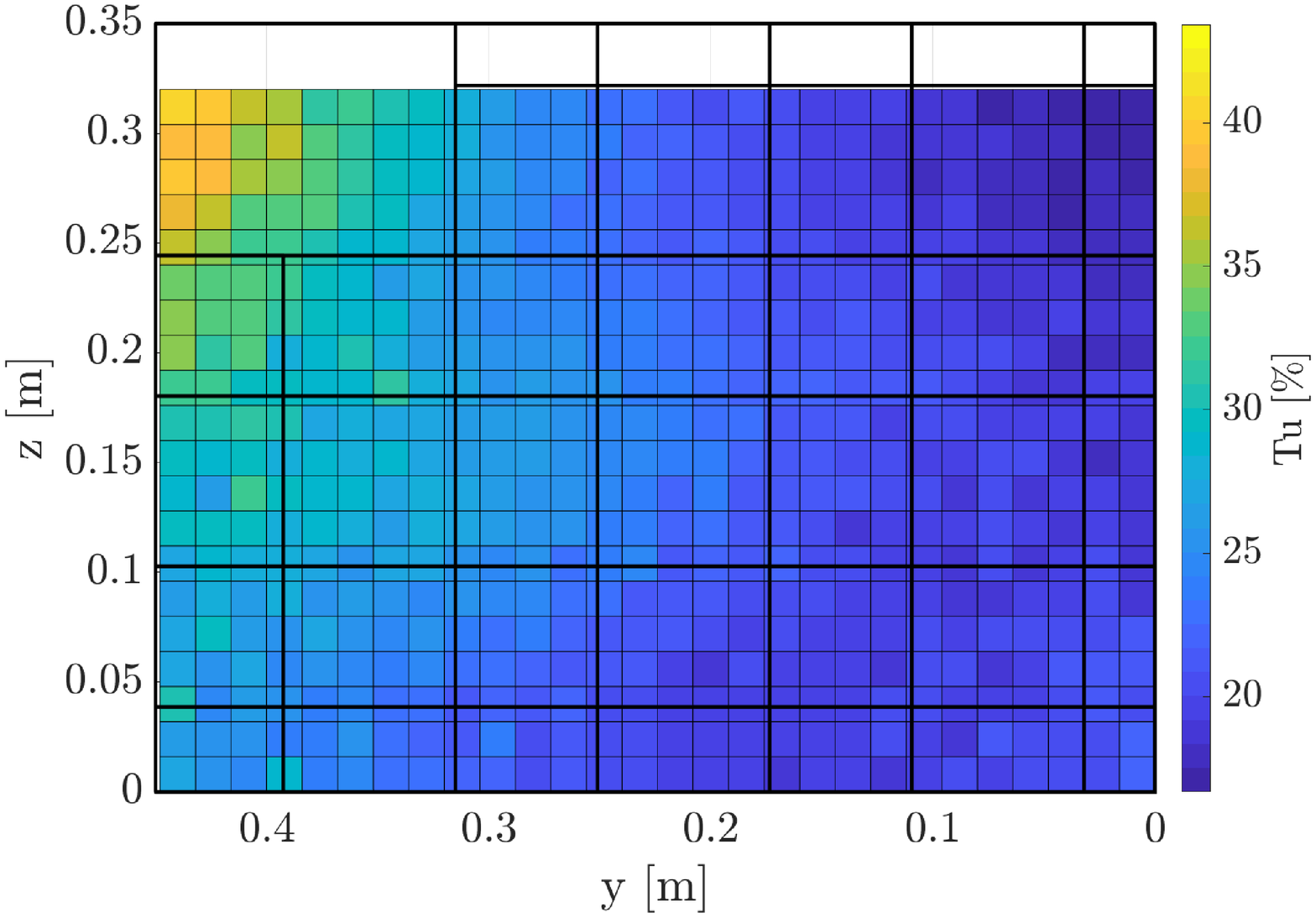}}
\tabularnewline
{\footnotesize{} (a) Grid 1. X--velocity} & {\footnotesize{} (b) Grid 1. Turbulence intensity}
\tabularnewline
{\footnotesize{}\includegraphics[width=.45\textwidth]{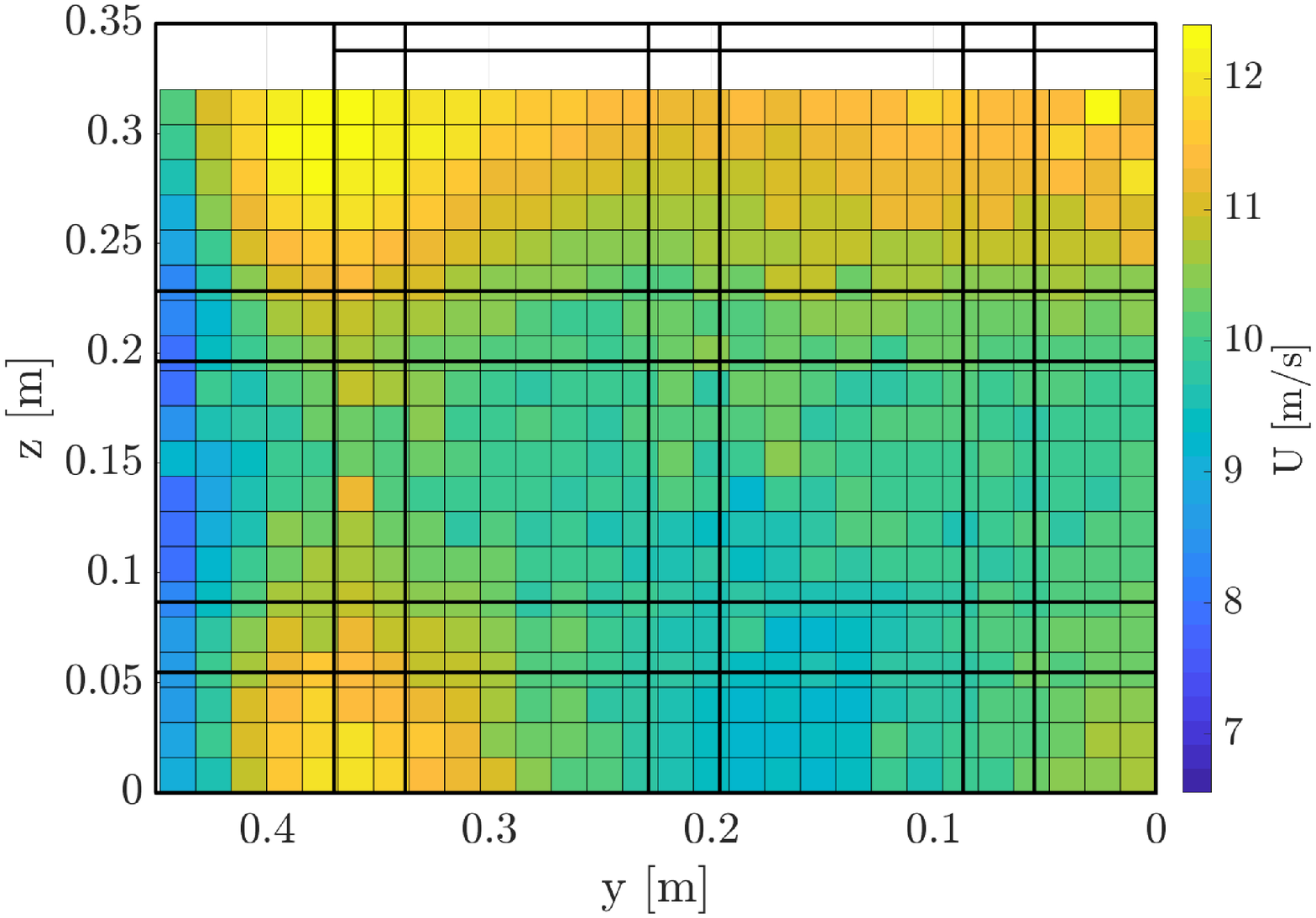}} &
{\footnotesize{}\includegraphics[width=.45\textwidth]{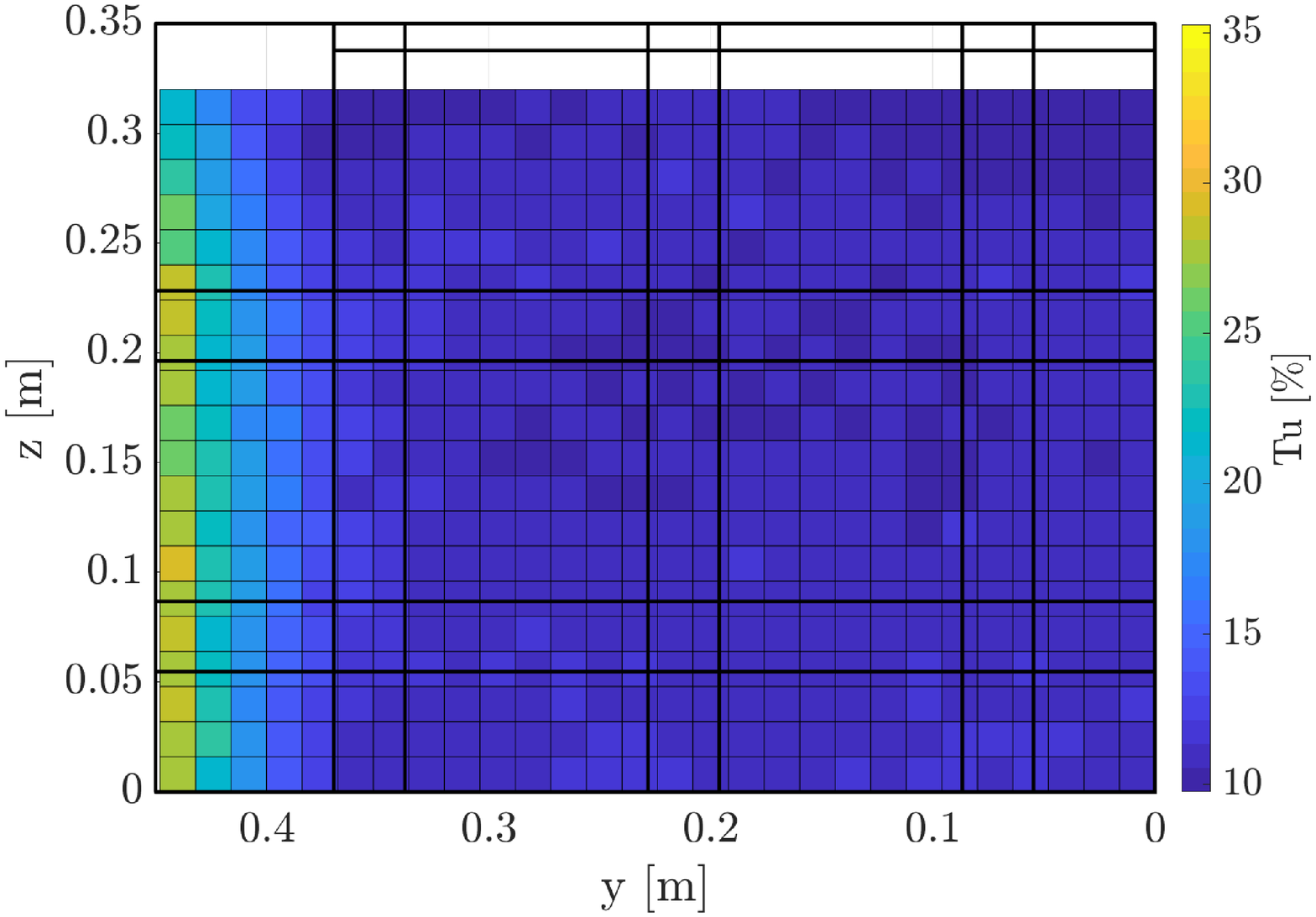}}
\tabularnewline
{\footnotesize{} (a) Grid 2. X--velocity} & {\footnotesize{} (b) Grid 2. Turbulence intensity}
\end{tabular}
\caption{ \label{Fig: MapGrid1} Flow mapping at the airfoil LE plane with turbulence grids installed. x/M = 10. Coordinate (0,0) is the center of the test section.}
\par\end{centering}
\end{figure}

Figure~\ref{Fig: MapGrid1} shows the mapping of the x--velocity and turbulence generated by grid~1 and 2 at the airfoil LE plane, i..e. x/M = 10, where M is the mesh size. Mappings are shown for one--quarter of the entire test section. The coordinate (z = 0, y = 0) is located in the center of the wind tunnel test section. For 10~m/s, a uniform flow and turbulence intensities of20\% and 10\% are observed in the center of the test section for grids~1 and 2, respectively. The larger open area in the left--top corner of grid~1 accelerates the flow and increases the turbulence intensity. Grid~2 significantly attenuates this phenomenon. Grid~2 generates similar turbulence levels and flow uniformity at 20~m/s and 30~m/s inflow velocity. Further details about the turbulence characterization of grid~1 and 2 is found in~\cite{dossantos2021b}. 
%

\begin{figure}[ht]
    \begin{centering}
    \begin{tabular}{cc}
  	    \multicolumn{2}{c}{{\footnotesize{}\includegraphics[width=.45\textwidth] {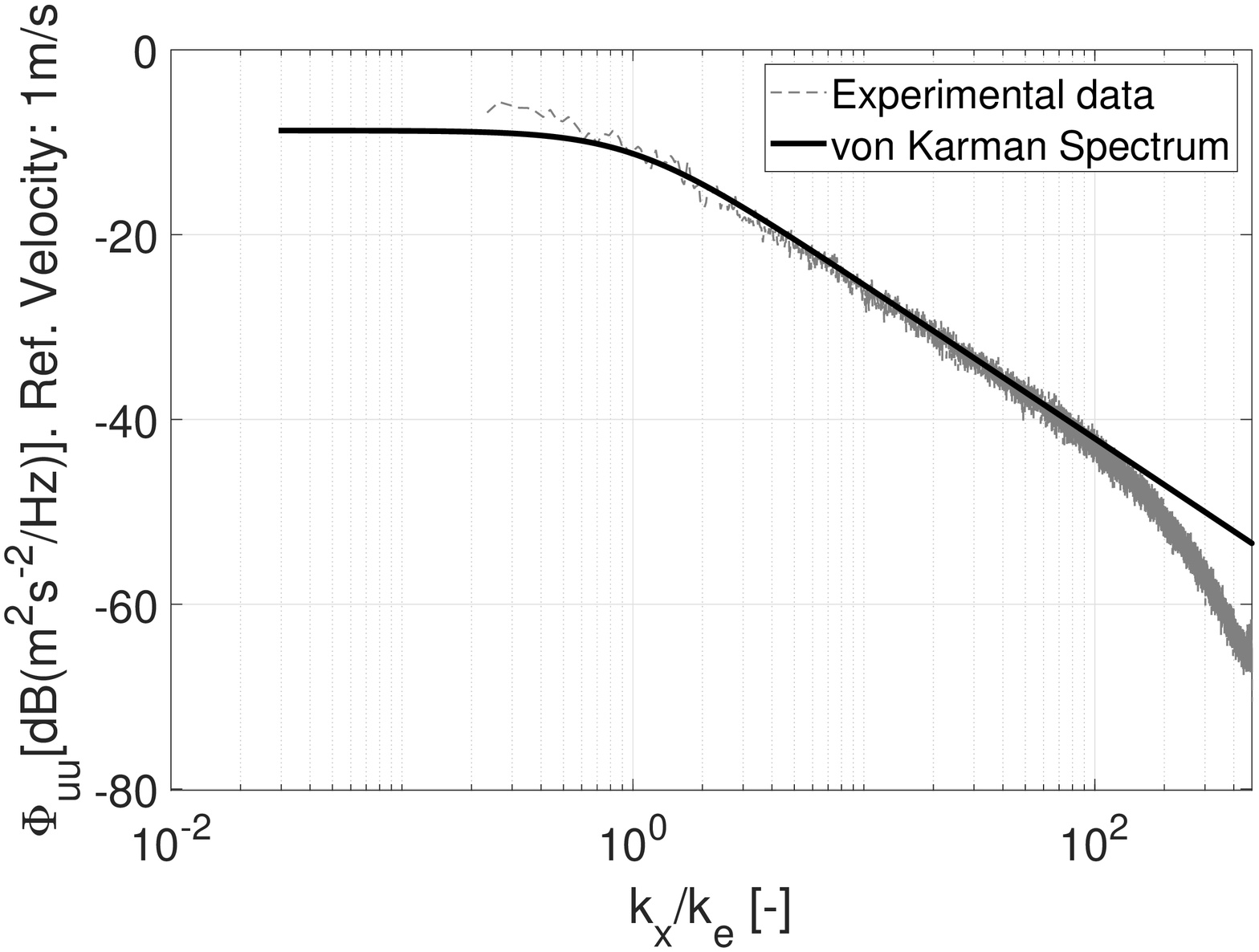}}}
        \tabularnewline
        \multicolumn{2}{c}{{\footnotesize{} (a) Grid 1. U = 10~m/s}}
        \tabularnewline
        {\footnotesize{}\includegraphics[width=.45\textwidth] {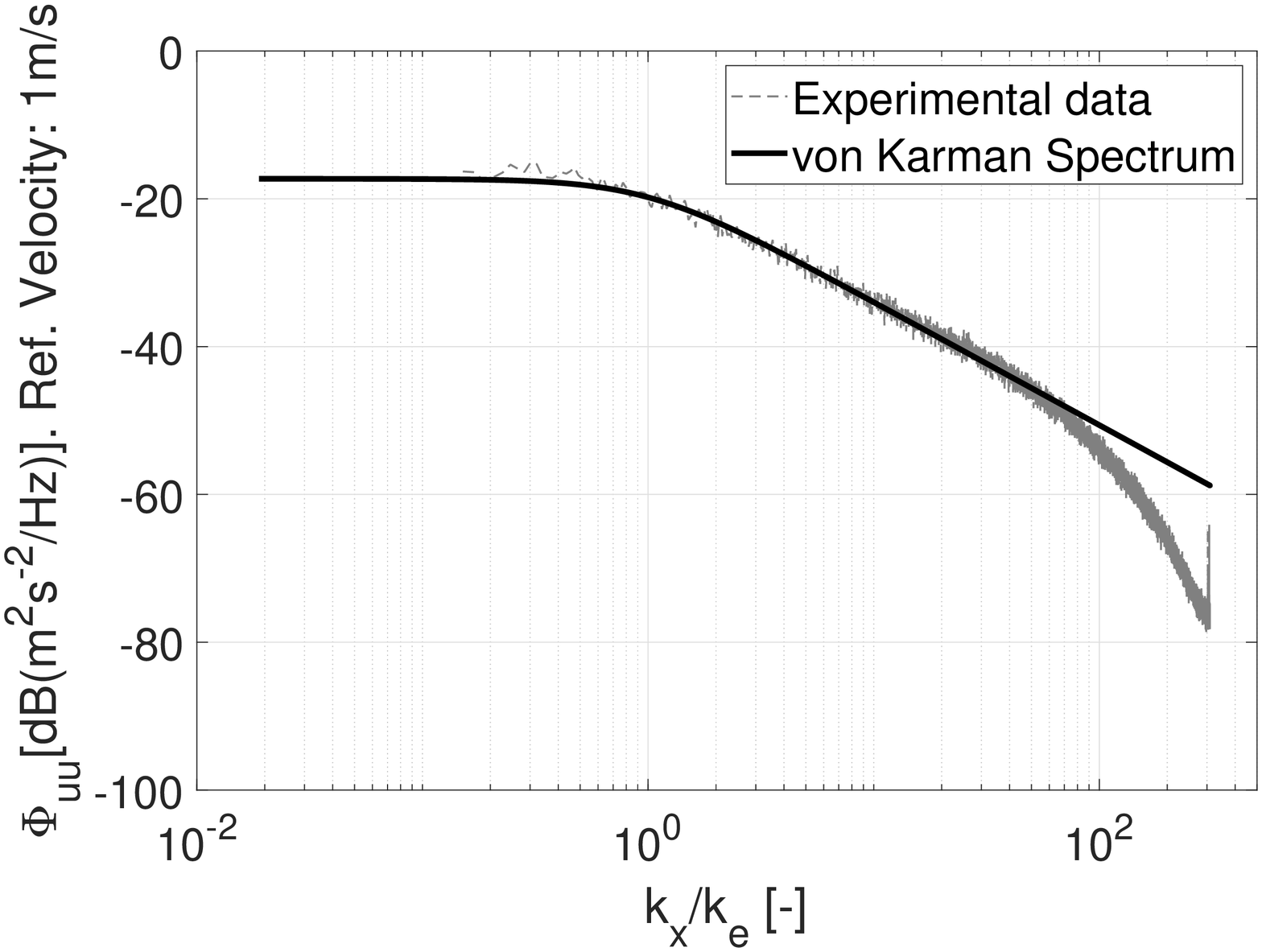}} & 
        {\footnotesize{}\includegraphics[width=.45\textwidth] {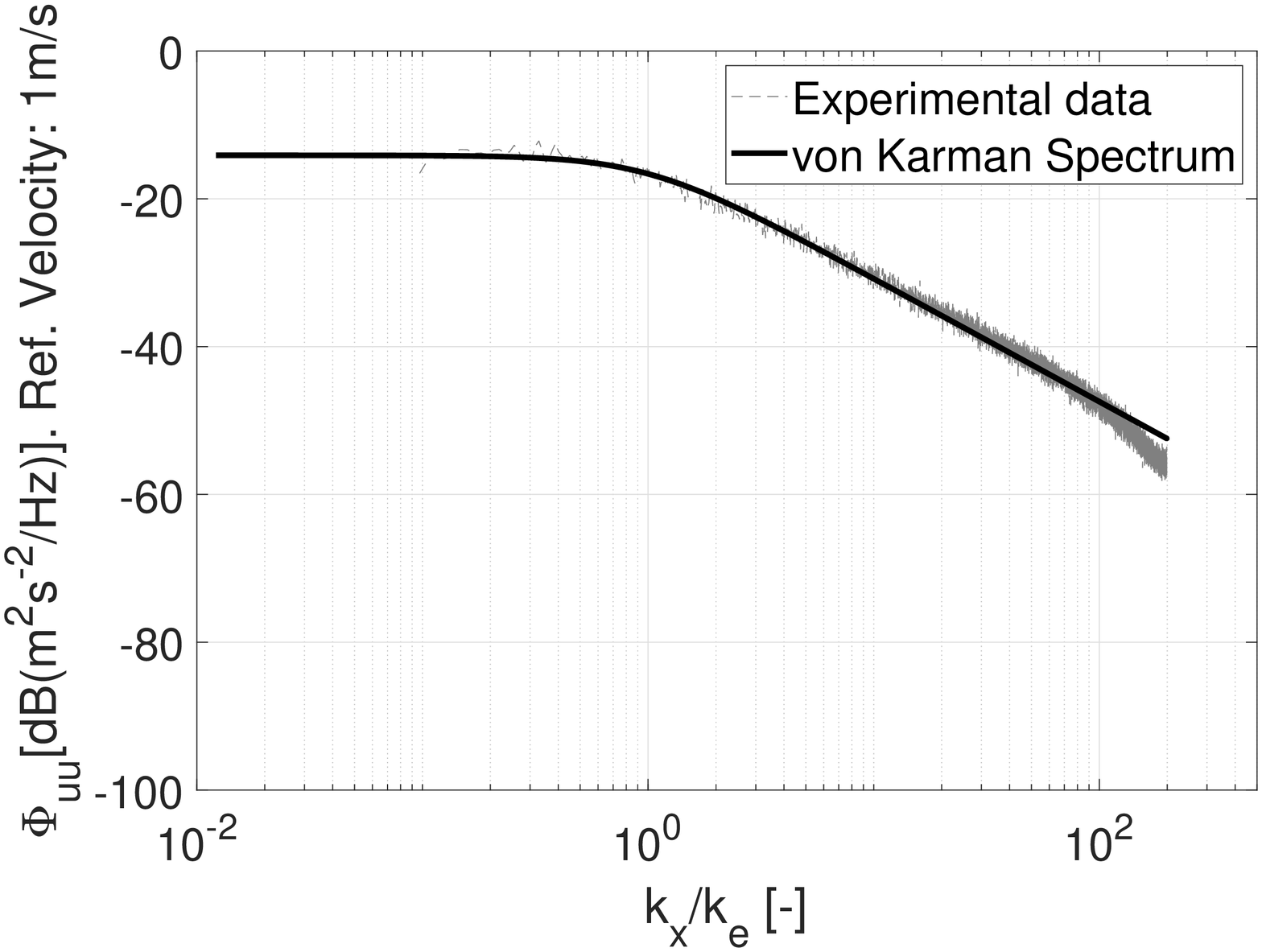}}
        \tabularnewline
        {\footnotesize{} (b) Grid 2. U = 10~m/s} & {\footnotesize{} (c) Grid 2. U = 30~m/s}
            \end{tabular}
        \caption{Experimental velocity spectrum and von Kármán spectrum.}
        \label{Fig: TuSpectrum} 
    \par\end{centering}
\end{figure}

The velocity spectrum is computed using the velocity time history measured at the center of the test section. Figure~\ref{Fig: TuSpectrum} shows the comparison between the experimental velocity spectrum and the von Kármán model, seen in Eq.~\ref{Eq: vonkarman}, for both grids at different velocities. In all cases, the experimental spectrum deviates from the von Kármán spectrum at high frequencies due because the von Kármán spectrum does not model the turbulence dissipation range, as discussed in~\cite{dossantos2021b}~\cite{Pope}. The integral length scales are calculated following the methodology of Hinze~\cite{hinze19720} and are adopted in the calculation of the turbulence spectrum. The integral length scales for the cases presented in Fig.~\ref{Fig: TuSpectrum} are: (a) 63~mm, (b) 39.5~mm, and (c) 48.5~mm.

\subsection{Boundary Layer}
The boundary layer measurements consisted of time--history streamwise velocity at several positions inside the boundary layer and a single measurement in the \FS~ outside the boundary layer. 

\begin{figure}[h!]
 \begin{centering}
    \begin{tabular}{cc}
{\footnotesize{}\includegraphics[width=.42\textwidth]{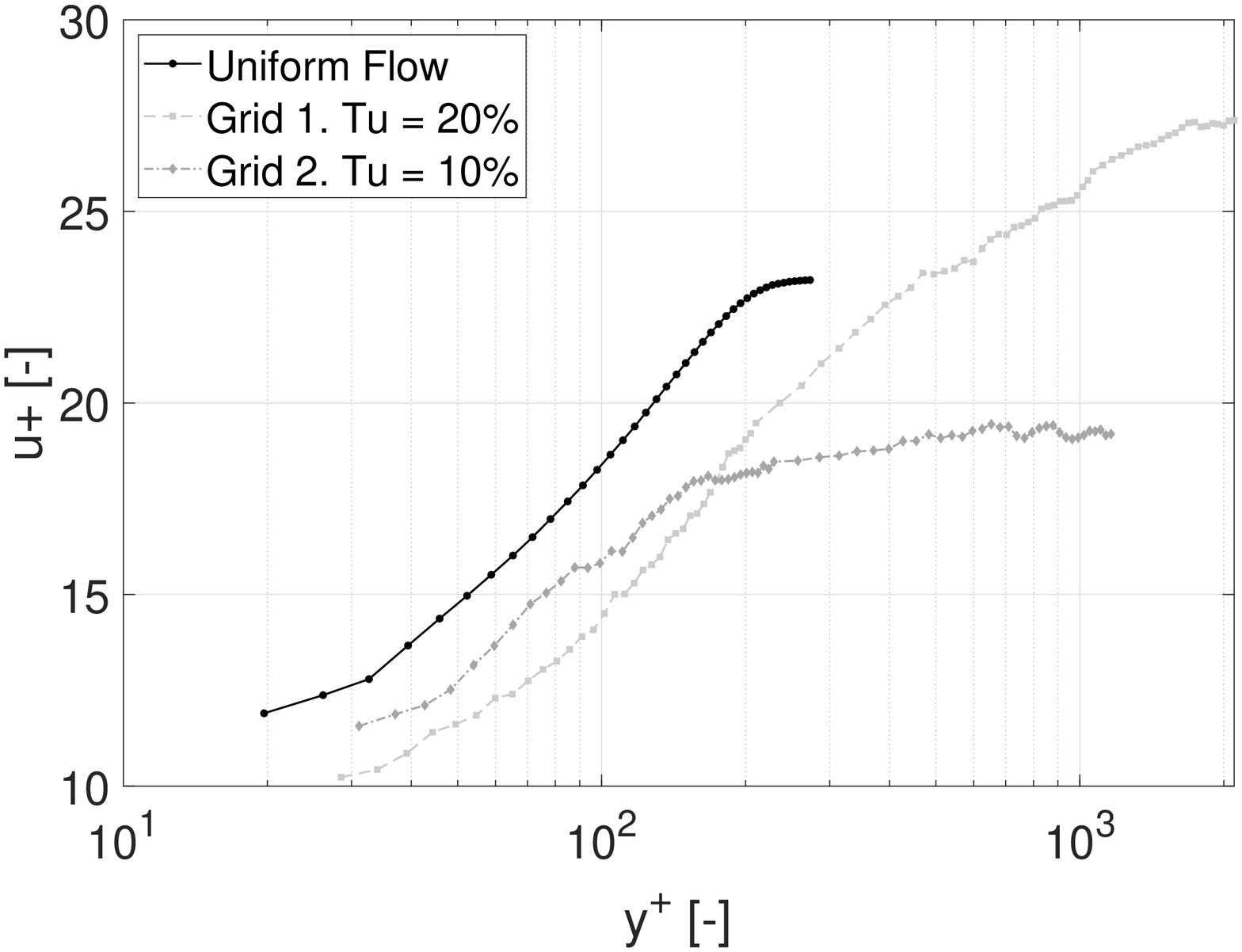}} &
{\footnotesize{}\includegraphics[width=.42\textwidth]{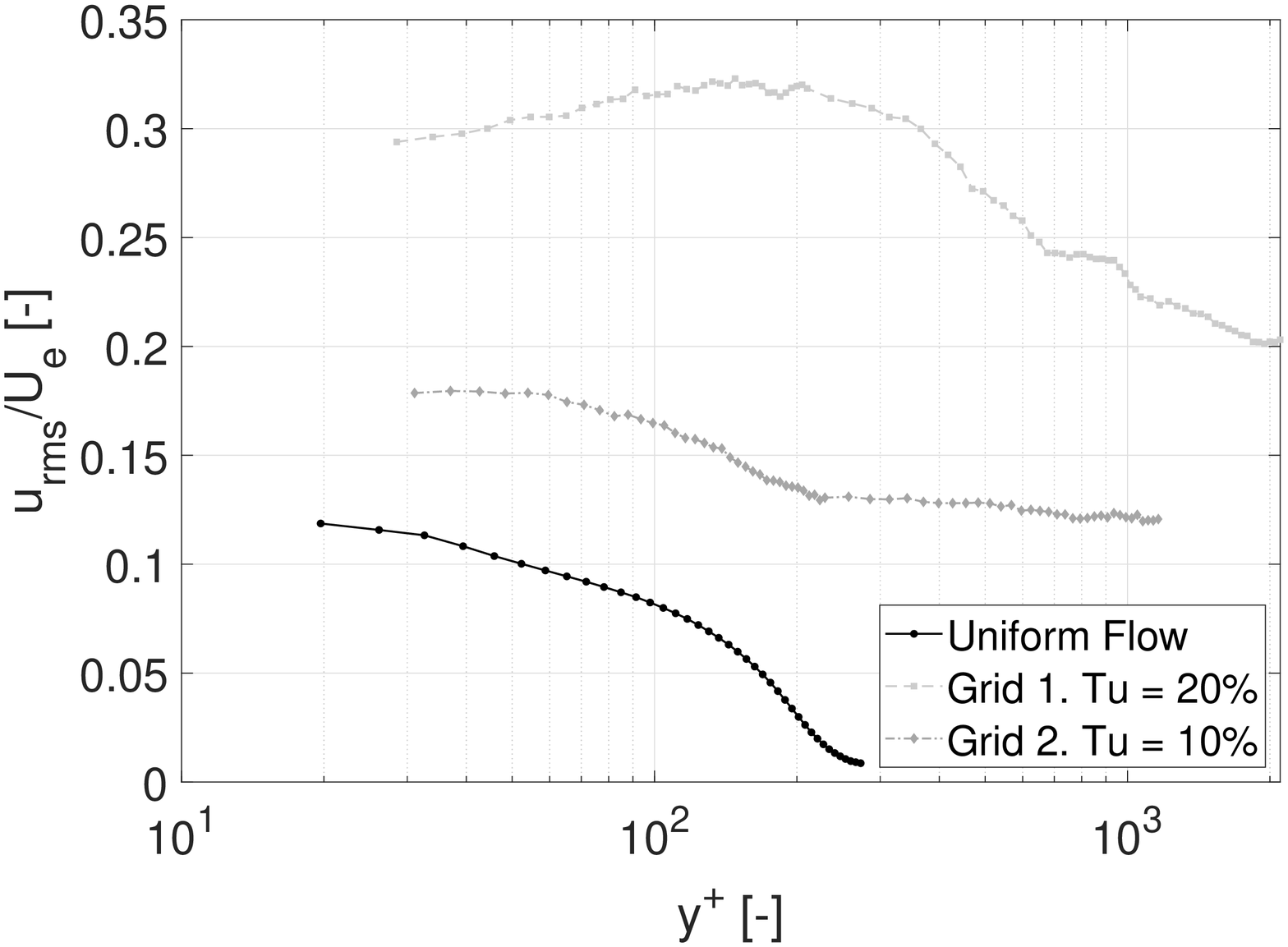}}
\tabularnewline
{\footnotesize{} (a) Mean velocity profile} & {\footnotesize{} (b) Velocity fluctuations profile}
\tabularnewline
{\footnotesize{}\includegraphics[width=.42\textwidth]{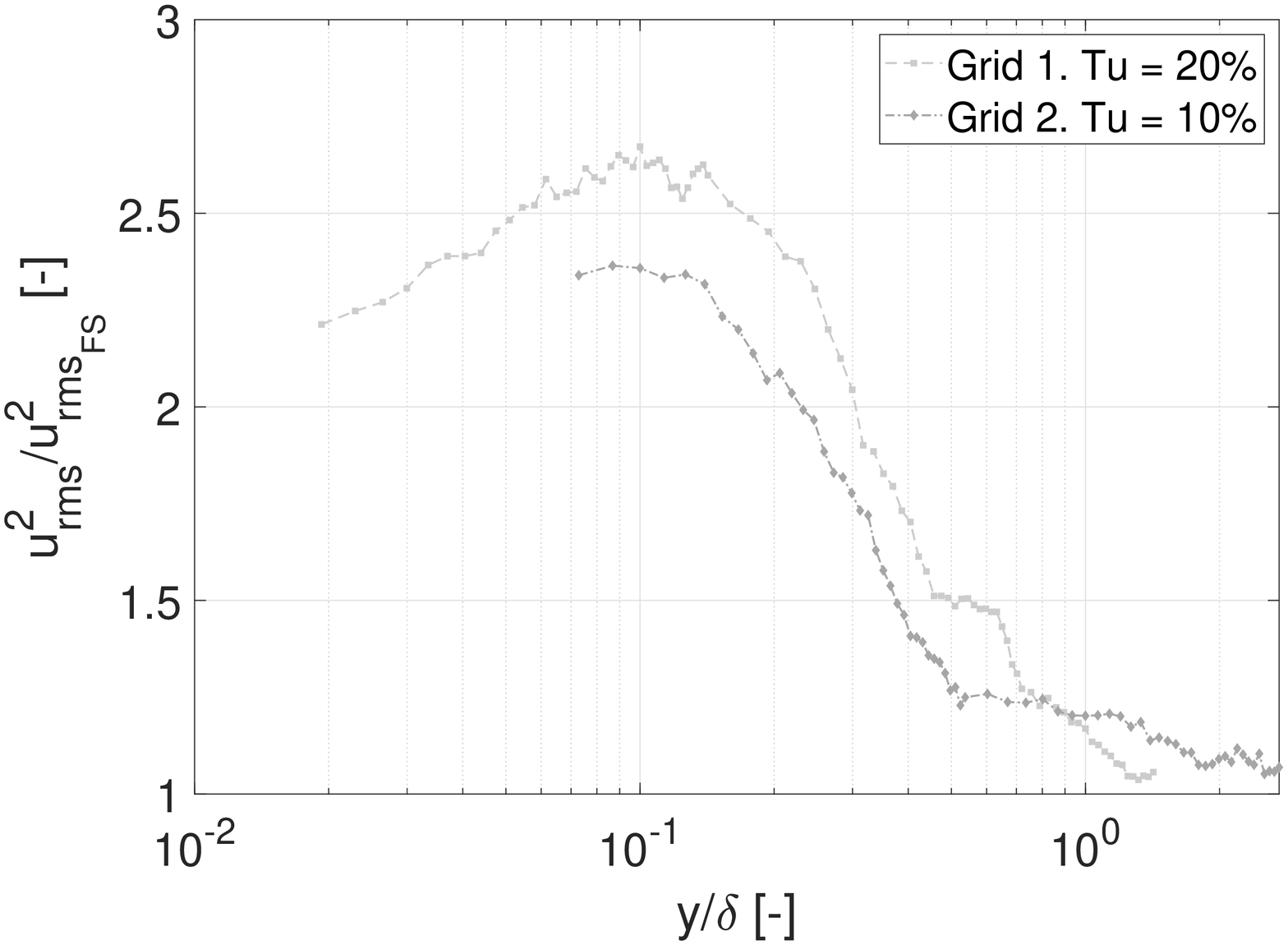}} &
{\footnotesize{}\includegraphics[width=.42\textwidth]{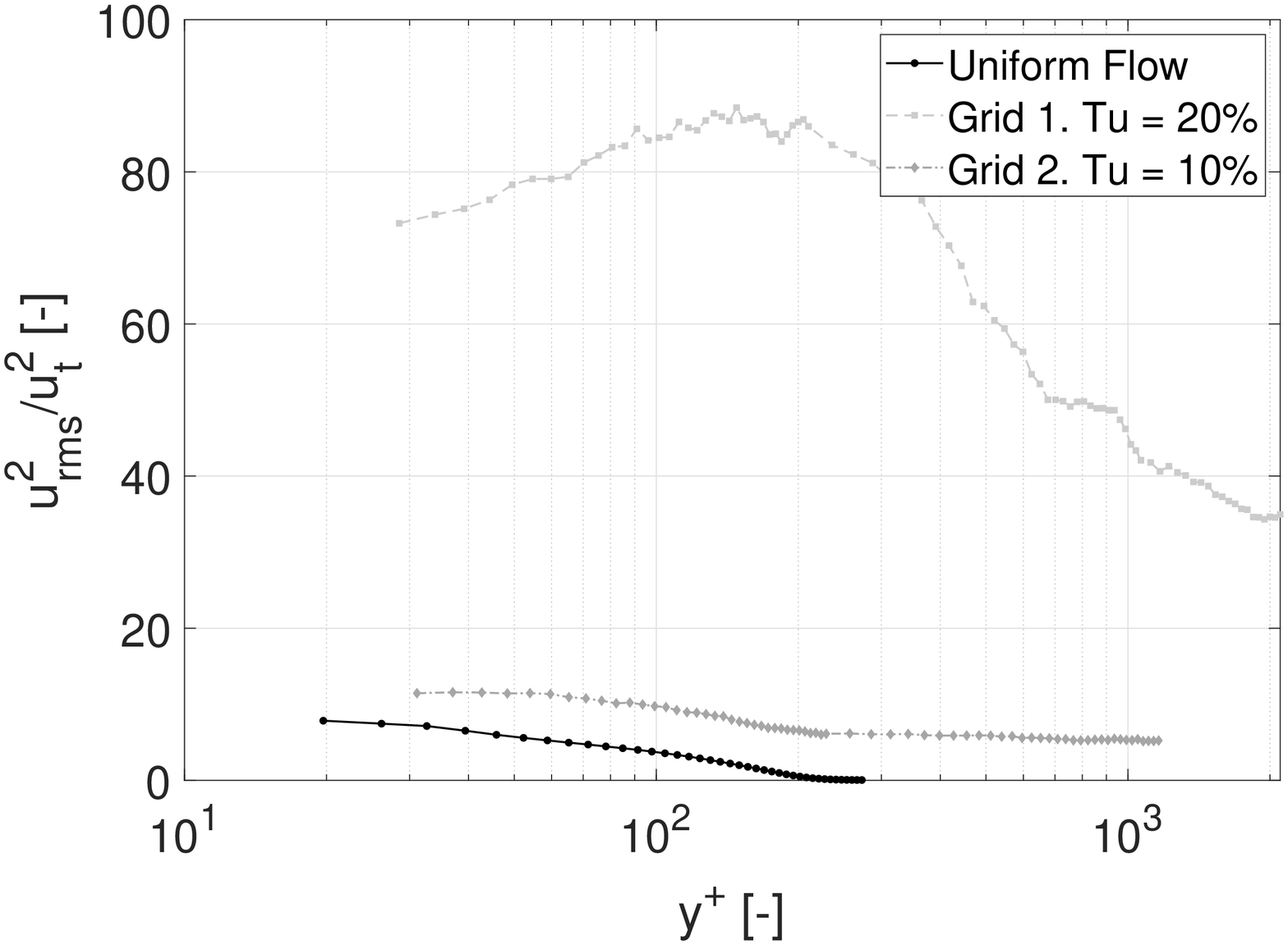}}
\tabularnewline
{\footnotesize{} (c) Normalized velocity fluctuations with outer scales} & {\footnotesize{} (d) Normalized velocity fluctuations with inner scales}
\tabularnewline
{\footnotesize{}\includegraphics[width=.42\textwidth]{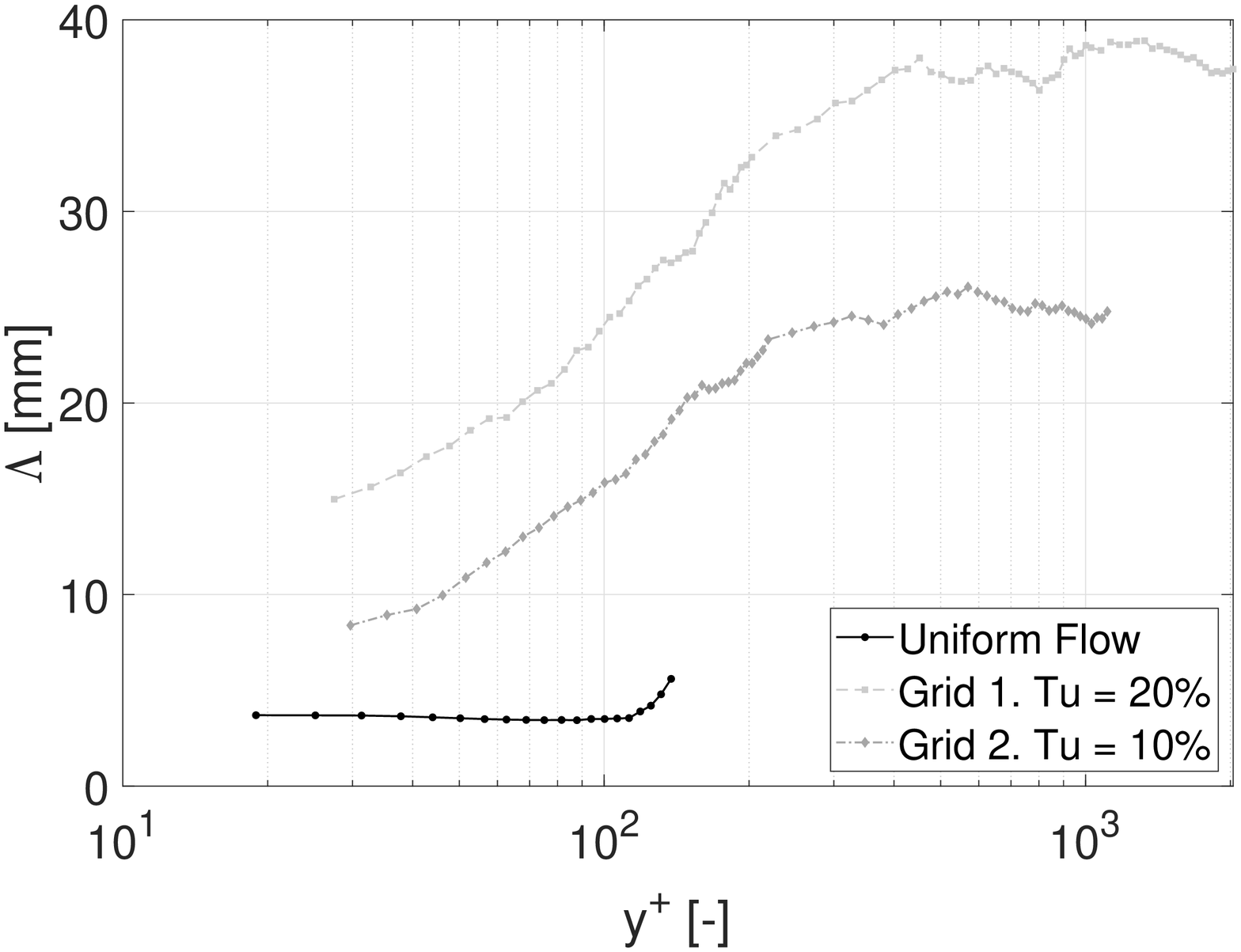}} &
{\footnotesize{}\includegraphics[width=.42\textwidth]{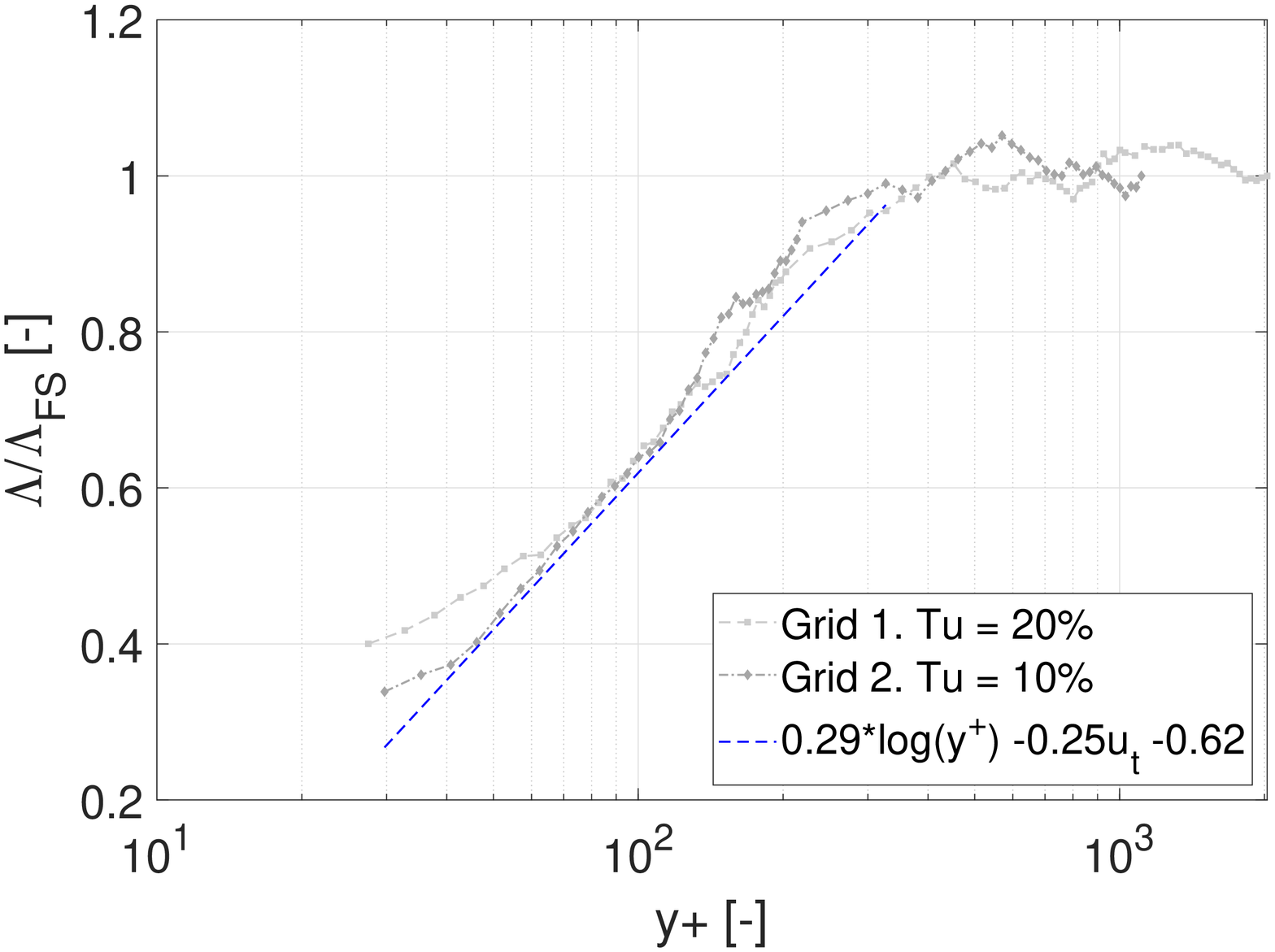}}
\tabularnewline
{\footnotesize{} (e) Integral length scale} & {\footnotesize{} (f) Normalized integral length scale}
\tabularnewline
 \multicolumn{2}{c}{{\footnotesize{}\includegraphics[width=.42\textwidth] {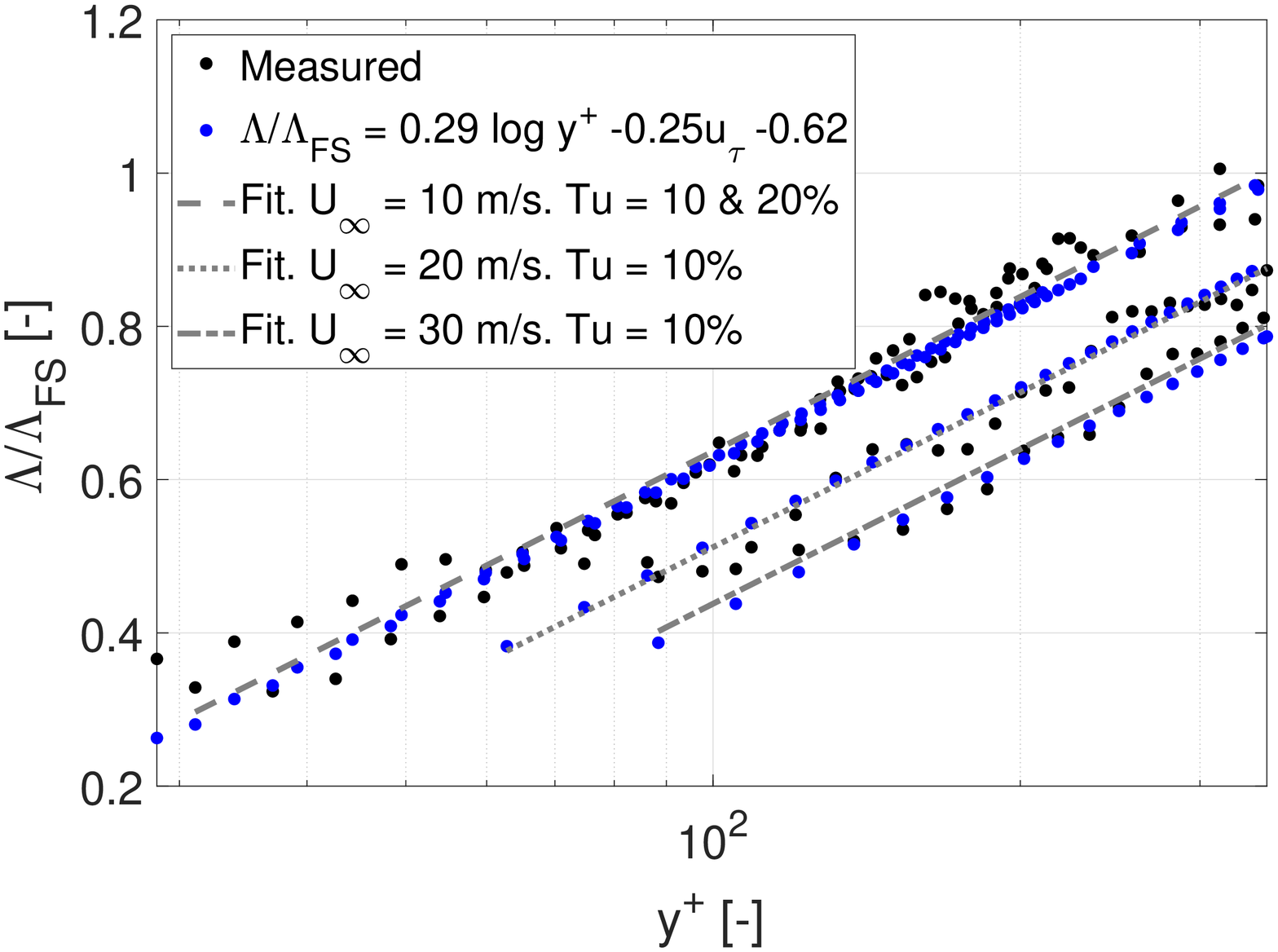}}}
\tabularnewline
\multicolumn{2}{c}{{\footnotesize{} (g) Predicted and measured length scale}}
\tabularnewline
\end{tabular}
\caption{ \label{Fig: BL} Boundary layer measurements. U = 10~m/s.}
\par\end{centering}
\end{figure}

Figure~\ref{Fig: BL} shows the mean velocity, the velocity fluctuations, and the integral length scale along the boundary layer. The results show that even with high inflow turbulence, the mean velocity inside the boundary layer follows the universal Prandtl--von Kármán log--law--of--the--wall, similar to the results obtained by Thole and Bogard~\cite{thole1996high} and Dogan et al.~\cite{dogan2016interactions}.

Table~\ref{tab:BL_results} shows the calculated boundary layer characteristics for inflow speeds of approximately 10~m/s, 20~m/s, and 30~m/s. The \FST~significantly increases the boundary layer thickness. At 10~m/s inflow velocity, it has grown from 7.4~mm for the uniform inflow case to 15~mm and 59~mm for the cases of 10\% and 20\% of inflow turbulence. The increment of the boundary layer thickness is also noticeable at higher velocities for the case of 10\% of inflow turbulence. Oppositely, the inflow turbulence does not significantly influence the friction velocity, as shown in Tab.~\ref{tab:BL_fit}. The friction velocity was obtained from the fitting of the Prandtl--von Kármán log--of--the--wall, see Eq.~\ref{eq: mean_velocity_profile}. The table also presents the calculated values using the information from XFOIL~\cite{drela1989xfoil} for the case of uniform inflow, which agree well with experimental results. 
\begin{table}[htbp!]
\caption{\label{tab:BL_results} Boundary layer thickness ($\mathbf{\delta}$), displacement thikcness ($\mathbf{\delta^*}$), shape factor (H) and external velocity ($\mathbf{U_e}$).}
\centering
\begin{tabular}{lcccccccccccccc}
\hline
        & \multicolumn{4}{c}{Uniform Flow}          && \multicolumn{4}{c}{10$\%$}   \\
        \cline{2-5} \cline{7-10}
        & $\delta$~[mm] & $\delta^*$~[mm] & H~[-] & U\textsubscript{e}~[m/s]  &  & $\delta$~[mm] & $\delta^*$~[mm] & H~[-] & U\textsubscript{e}~[m/s]    \\ \cline{2-5} \cline{7-10}
10 [m/s] & 7.40   & 1.40  & 1.42 & 9.52    
    &    & 15.1   & 0.68  & 1.46 & 7.95          &        \\
20 [m/s] & 4.9   & 0.98  & 1.12 & 16.84         
&       & 7.3   & 1.10  & 1.07 & 17.22       &     \\
30 [m/s] & 4.5   & 0.98  & 1.10 & 25.52        
&       & 11.1  & 0.47  & 1.26 & 25.32      &     \\
\hline
\end{tabular}
\centering
\begin{tabular}{lccccccccc}
        & \multicolumn{4}{c}{$20\%$}  \\ \cline{2-5} 
         & $\delta$~[mm] & $\delta^*$~[mm] & H~[-] & U\textsubscript{e}~[m/s] \\ \cline{2-5} 
10 [m/s] & 57.1   & 12.4  & 1.33 & 10.32   &        \\
\hline
\end{tabular}
\end{table}
\begin{table}[htbp!]
\caption{\label{tab:BL_fit} Friction velocity and Cole's wake factor obtained from the fitting process. The data in parenthesis for the uniform flow case are obtained from XFOIL~\cite{drela1989xfoil}}
\centering
\begin{tabular}{lccccccccccccccccc}
\hline
        & \multicolumn{3}{c}{Uniform Flow}          && \multicolumn{3}{c}{10\%} & & \multicolumn{3}{c}{20\%}   \\
        \cline{2-4} \cline{6-8} \cline{10-12}
        & $u_\tau$ [m/s] & $\Pi_w$ & $R^2$& & $u_\tau$ [m/s] & $\Pi_w$ & $R^2$&& $u_\tau$ [m/s] & $\Pi_w$& $R^2$\\ \cline{2-4} \cline{6-8} \cline{10-12}
10 [m/s] & 0.40 (0.46)   & 1.03 (0.52) & 0.99 &  & 0.40   & 0.060 & 0.87 & & 0.37  & 1.10 & 0.89    \\
20 [m/s]  & 0.77 (0.82)   & 0.54 (0.49)  &  0.99 & & 0.81  & 0.180 &0.81 & & -   & -  & -    \\
30 [m/s]  & 1.13 (1.21)   & 0.51 (0.48)  & 0.99 &  & 1.19   & 0.061& 0.91 && -   & -   & -   \\
\hline
\end{tabular}
\end{table}
%

For the uniform inflow case, the velocity fluctuations in the boundary layer exhibited the same behavior as observed by Dogan et al.~\cite{dogan2016interactions}, i.e., a peak at y\textsuperscript{+} $\approx$ 20. At 10~m/s, the u\textsubscript{rms} peak is 1.35~m/s, which corresponds to 13.5\% Tu, using the \FS~ velocity as reference. The \FST~appreciably increases the velocity fluctuations inside the boundary layer. The turbulence peak increases to 18\%, and 33\% for the cases of 10\% and 20\% inflow turbulence, respectively. For the cases of inflow turbulence, the turbulence inside the boundary layer is higher than in the \FS, likewise for uniform inflow. Nevertheless, the inner peak observed for the uniform flow case is extended into a region between y\textsuperscript{+} = 40 to 400. Furthermore, the case of 20\% inflow turbulence reveals the existence of an outer peak at y\textsuperscript{+} $\approx$ 800, which agrees with Dogan et al.~\cite{dogan2016interactions}.

Figures~\ref{Fig: BL} (c) and (d) present the velocity fluctuations normalized by the outer scales, i.e., the edge velocity fluctuations and boundary layer thickness, and inner scales, i.e., friction velocity and viscosity. The first normalization can be interpreted as a gain factor of the turbulence inside the boundary layer~\cite{dogan2016interactions}, which increases as the \FST~increases. This result is opposite to the findings of Dogan et al.~\cite{dogan2016interactions}, who showed that for low inflow turbulence the gain factor is higher than for high levels of \FST. They suggested that there is a possible saturation limit for the increment of the turbulence inside the boundary layer when subjected to external disturbances. However, they only studied cases of \FST~up to 13\%. For even higher \FST~intensities, e.g., 20\%, this trend should change as shown here and an increase of the gain factor due to the \FST~is still possible. 

 The fact that the \FST~increases the velocity fluctuations even close to the wall, e.g., y\textsuperscript{+} = 30 demonstrates that the \FST~is penetrating along the entire boundary layer~\cite{dogan2016interactions}. Dogan et al.~\cite{dogan2016interactions} stated that the extent of the penetration inside of the boundary layer depends on the level of the \FST. They stated that for lower levels of \FST, the penetration just occurs in the outer part of the boundary layer and not close to the wall. Inside the boundary layer, it seems to be a combination of both phenomena, the \FST~that penetrates the boundary layer and the near--wall turbulence. The increase of the turbulence inside the boundary layer due to 10\%  of inflow turbulence was also noticeable at 20~m/s and 30~m/s. These results are not shown here for sake of simplicity.
\FloatBarrier

Figure~\ref{Fig: BL} (e) shows that the \FST~significantly increases the turbulent integral length scale inside of the boundary layer. At y\textsuperscript{+} = 30, the integral length scale increased from 2.2~mm for the uniform inflow case to 8~mm and 15~mm for the cases of 10\% and 20\%  inflow turbulence. In the \FS, the integral length scales are 26~mm and 38~mm for 10\% and 20\% \FST, respectively. Figure~\ref{Fig: BL} (f) compares the length scale inside of the boundary layer normalized by the length scale in the \FS~($\Lambda_{FS}$) of both cases of inflow turbulence. There is a logarithmic relation, following the expression $a\log y^+ +b$, between $\Lambda/\Lambda_{FS}$ and y\textsuperscript{+} up to y\textsuperscript{+} $\approx$ 350, which is independent of the level of \FST. Figure~\ref{Fig: BL} (g) shows the extension of this analysis to higher velocities for the case of 10\% inflow turbulence. Although the slope of the curve \textit{a} is similar for all the cases, independent of the inflow velocity and turbulence intensity, the level of the length scales \textit{b} varies linearly with the friction velocity. A model to calculate the length scale inside the boundary layer for cases of inflow turbulence is proposed and shown in Eq.~\ref{Eq: LS}. The length scale is modeled as a function of the distance from the wall in wall units, i.e., y\textsuperscript{+}, the friction velocity \textit{u\textsubscript{$\tau$}}, and the length scale in the \FS. A bisquare fitting considering the measurements of y\textsuperscript{+} $<$ 350 for both grids at every inflow velocities defined the constants \textit{a} and \textit{b} and \textit{c}. Note that according to the TNO--Blake model~\cite{Parchen1998, stalnov2016towards} the turbulence and the length scale inside the boundary layer have a significant effect on the surface pressure fluctuations.
\begin{equation}\label{Eq: LS}
\begin{split}
    \frac{\Lambda}{\Lambda_{FS}} &= a\cdot log (y^+) + b\cdot u_\tau + c, \qquad  \; y^+ < 350; \\
    \frac{\Lambda}{\Lambda_{FS}} &= 1, \qquad  \qquad \qquad \qquad \qquad \quad   y^+ > 350; \\
    a &= 0.29,\; b = -0.25 \; c = -0.62,\ \;  R^2 = 0.96;\\
    \end{split}
\end{equation}

\begin{figure}[hbt!]
 \begin{centering}
    \begin{tabular}{cc}
{\footnotesize{}\includegraphics[width=.45\textwidth]{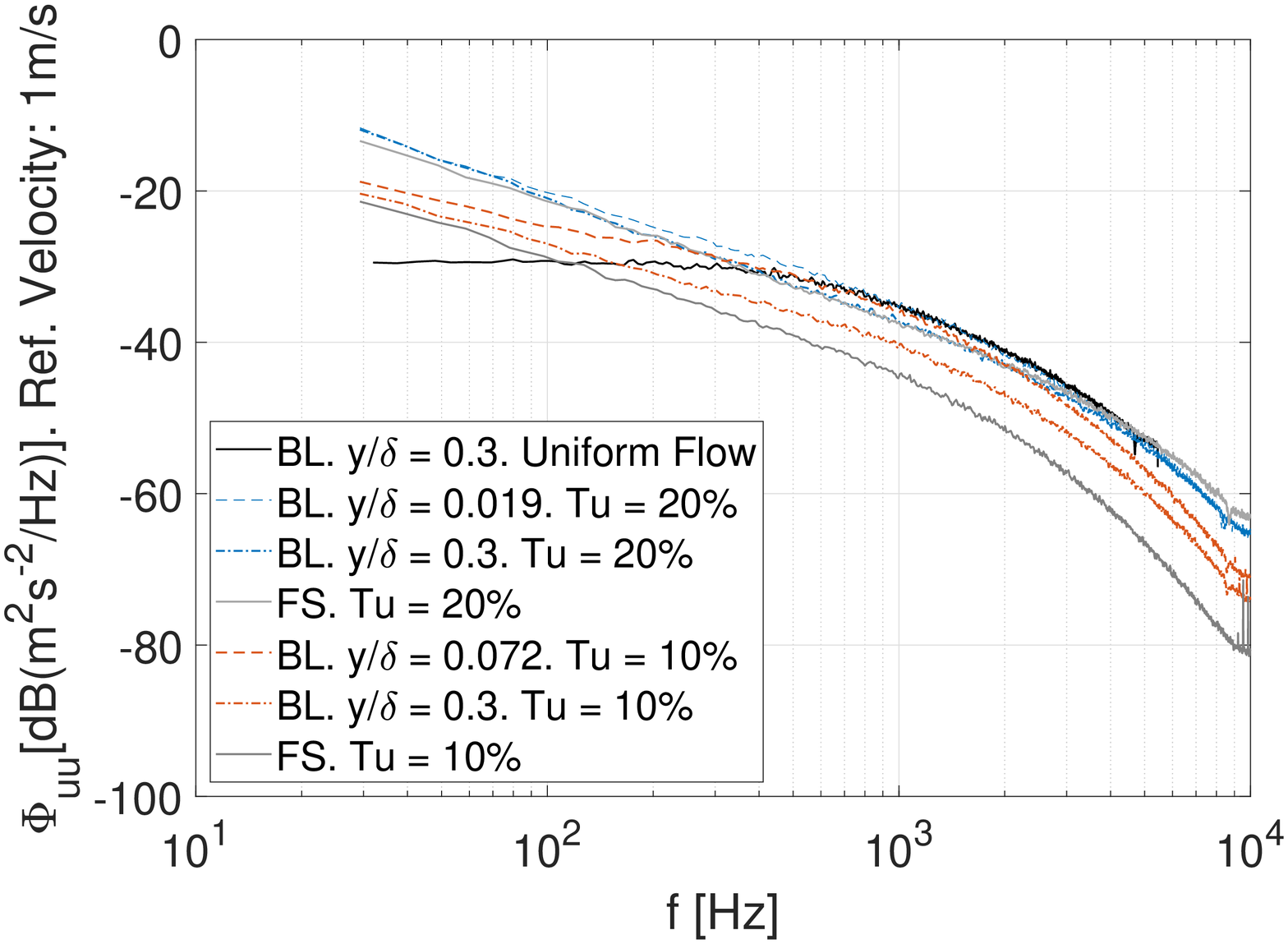}} &
{\footnotesize{}\includegraphics[width=.45\textwidth]{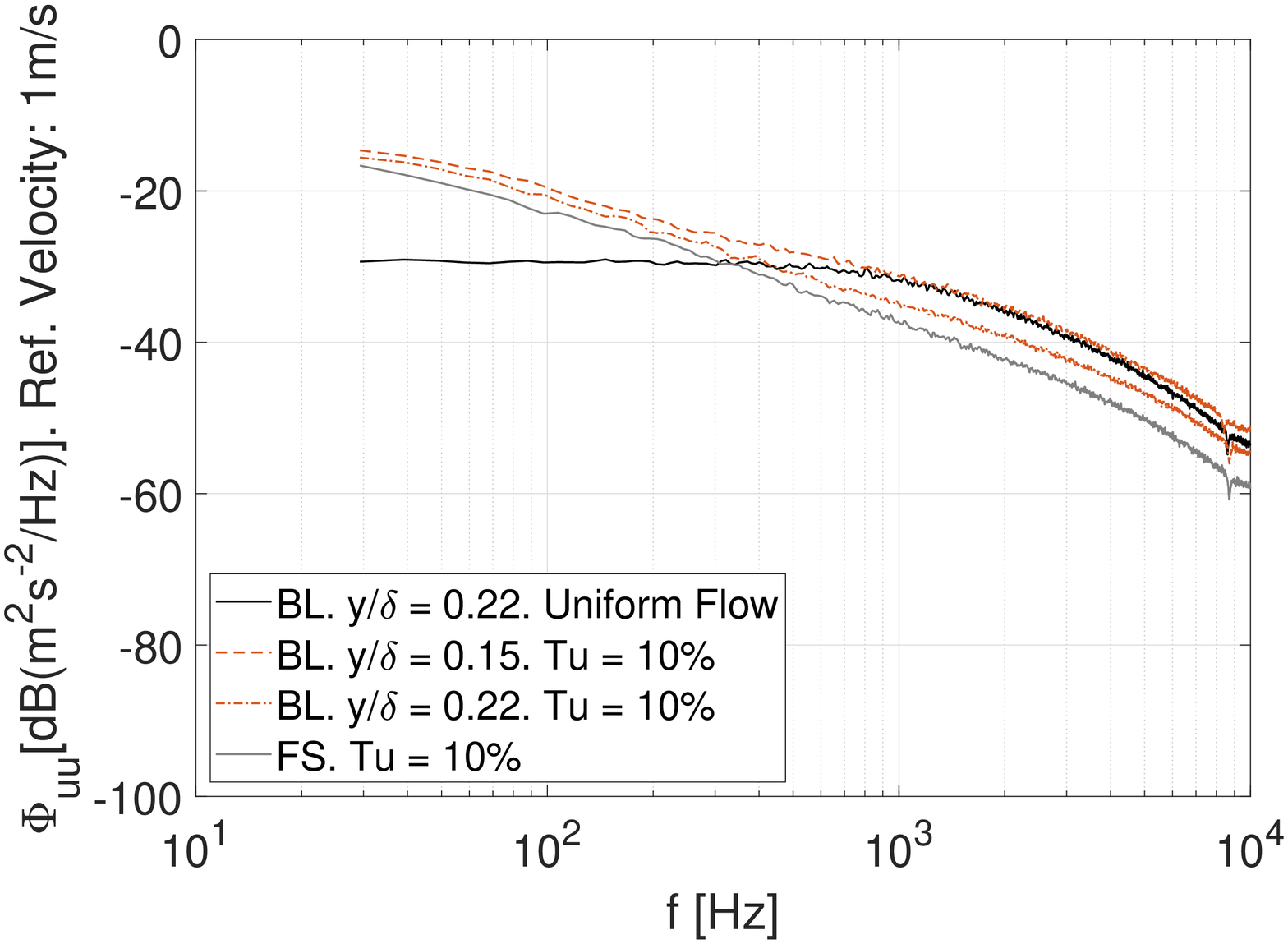}}
\tabularnewline
{\footnotesize{} (a) U = 10~m/s} & {\footnotesize{} (b) U = 20~m/s}
\tabularnewline
\multicolumn{2}{c}{{\footnotesize{}\includegraphics[width=.45\textwidth] {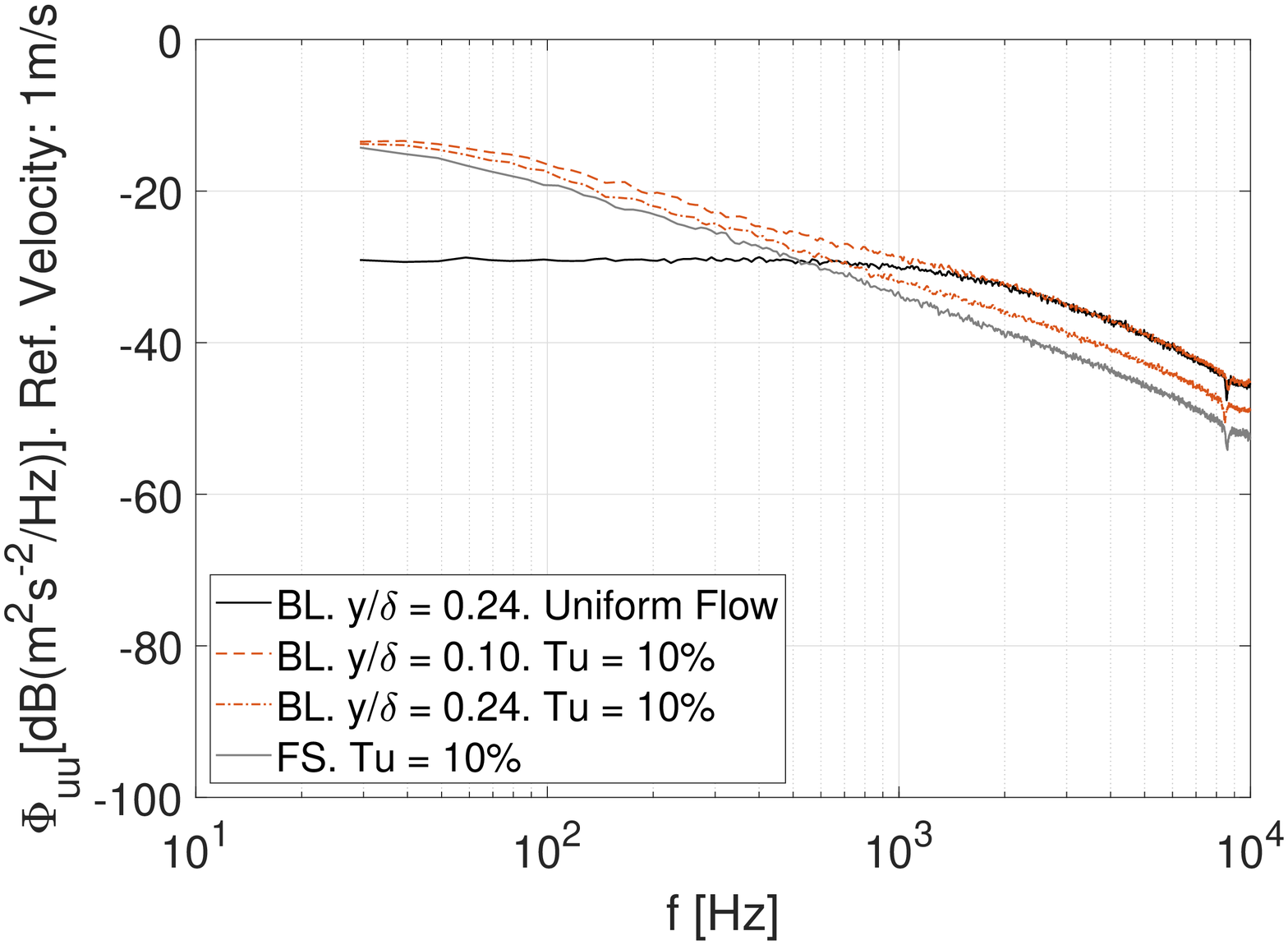}}}
        \tabularnewline
        \multicolumn{2}{c}{{\footnotesize{} (c)  U = 30~m/s}}
        \end{tabular}
\caption{\label{Fig: BL_spectrum} Velocity spectrum in the \FS~(FS) and inside the boundary layer (BL). The two points inside the boundary layer for the cases of inflow turbulence correspond to the point closest to the wall and to match the same $\mathbf{y/\delta}$ of the uniform flow case.}
\par\end{centering}
\end{figure}

Figure~\ref{Fig: BL_spectrum} shows the velocity spectrum inside the boundary layer and in the \FS~ for different inflow conditions at 10~m/s, 20~m/s, and 30~m/s free-stream velocities. For the uniform inflow case, only the measurements closest to the airfoil's wall (y = 1.1~mm) are shown, which correspond to $y/\delta = 0.14$, $0.22$ and $0.24$ and  $y^+ = 20$, $60$ and $87$, for 10~m/s, 20~m/s, and 30~m/s free-stream velocity, respectively. For the cases of \FST, two measurements are presented, namely, the measurement closest to the airfoil wall and to match the same $y/\delta$ of the uniform flow at each velocity. The same y\textsuperscript{+} corresponds to the same y, since there are not significant differences in the friction velocity. The \FS~ corresponds to the measurement outside the boundary layer.

Dogan et al.~\cite{dogan2016interactions} used a cut-off filter to separate the velocity fluctuations caused by the small and large turbulent scales. They stated that for \FST~levels up to 13\%, the velocity fluctuations caused by the smaller or inner turbulent scales remain constant independently of the \FST~level, whereas the \FST~significantly increases the velocity fluctuations caused by the larger scales. This statement is evidenced in this study for the case of 10\% inflow turbulence, at 10~m/s, 20~m/s, and 30~m/s inflow velocity, as shown in Fig.~\ref{Fig: BL_spectrum}. The velocity spectrum inside the boundary layer for the case of 10\% inflow turbulence collapses with the one of the \FS~ up a specific frequency, which are significantly higher than the one for the uniform inflow case at the same distance from the wall. Above such a frequency, the velocity spectra inside the boundary layer for the two incoming flow conditions collapse. Inflow turbulence of 10\% intensity only affects the velocity fluctuations caused by the larger scales of the turbulence. Conversely, for the case of 20\% inflow turbulence, the \FST~seems to influence all the scales of the turbulence inside the boundary layer, since there are no remarkable differences between the velocity spectra at the two positions inside of the boundary layer and the spectrum of the \FS, as shown in Fig.~\ref{Fig: BL_spectrum} (a).

For a turbulence intensity of 10\%, there is a specific frequency in which the velocity fluctuations caused by the near--wall turbulence (inner scales of the turbulence) start to be stronger than these caused by the introduction of \FST. Such frequencies are 380~Hz, 937~Hz, and 1444~Hz, for 10~m/s, 20~m/s, and 30~m/s, respectively, which correspond to a streamwise wavenumber ($\kappa_x = \omega/U_c$) of 630, 472, and 483, respectively. For this calculation U$_c$ is the mean velocity at the same position of the velocity spectrum, i.e., 4.8~m/s, 12.4~m/s, and 18.7~m/s, respectively. Noteworthy is that the velocity spectra inside of the boundary layer for the different inflow conditions match on the same distance from the wall in wall units, i.e., y\textsuperscript{+}, and not on same $y/\delta$.
%

\subsection{Surface pressure fluctuations}
This section analyses the spectra of the unsteady surface pressure for the cases of uniform inflow and 10\% and 20\% inflow turbulence. The surface pressure spectra are normalized by the inner scales, e.g., u\textsubscript{$\tau$} and $\nu$, and outer scales e.g., $\delta$ and U$_{e}$. Moreover, the level of the spectrum is scaled by the boundary layer thickness $\delta$ and the integral length scale \textit{$\Lambda$} calculated with the hot--wire measurements closest to the airfoil's wall for each case. The values of $\delta$ and u\textsubscript{$\tau$} are derived from measurements of the boundary layer, shown in Tabs.~\ref{tab:BL_results} and \ref{tab:BL_fit}, respectively. Furthermore, this section addresses the calculation of the convection velocity and spanwise correlation length. The single point surface pressure spectra are calculated using the pressure time--history measured by the microphone located at the mid--span closest to the trailing edge. 

\begin{figure}[hbt!]
 \begin{centering}
    \begin{tabular}{cc}
{\footnotesize{}\includegraphics[width=.45\textwidth]{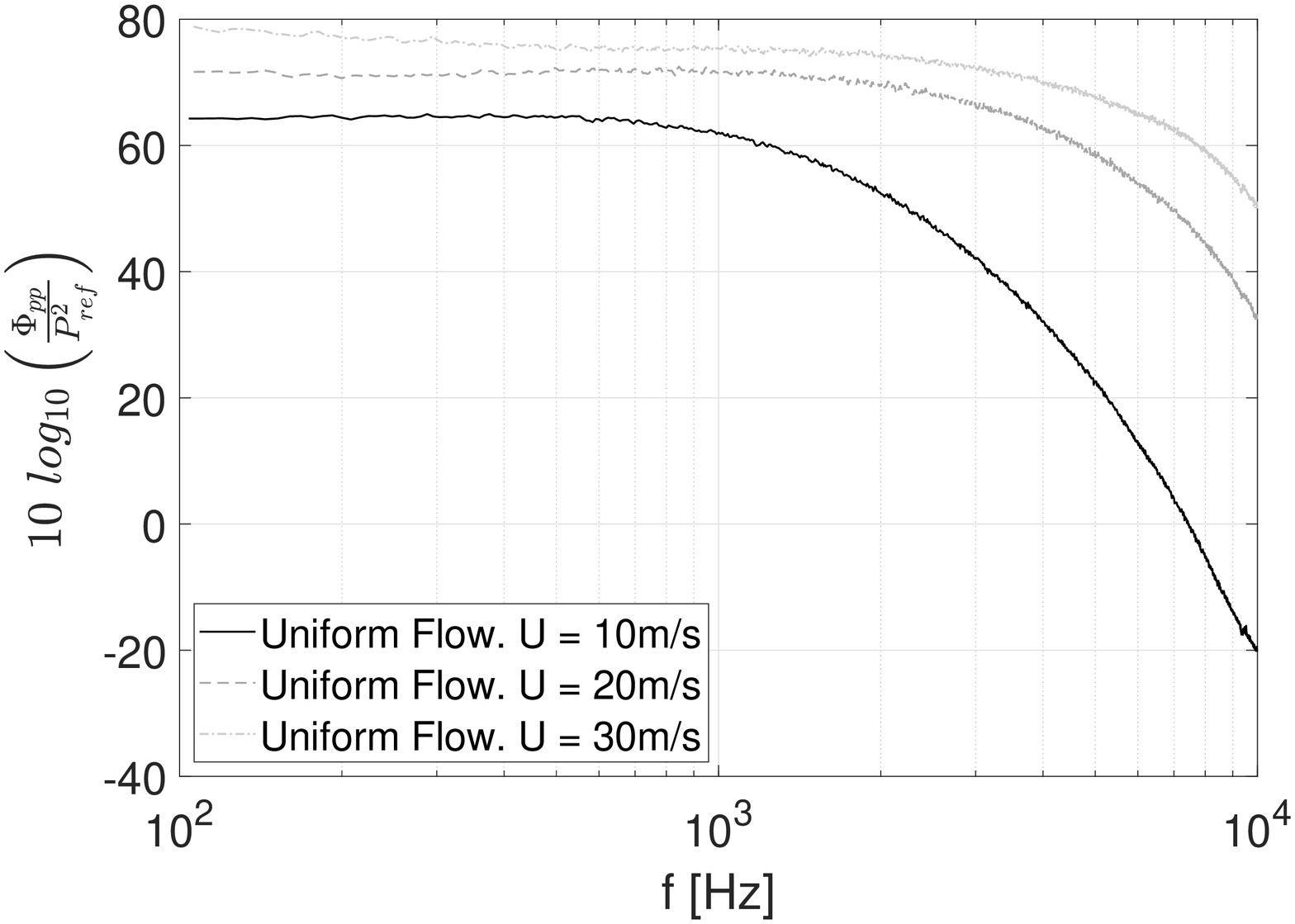}} &
{\footnotesize{}\includegraphics[width=.45\textwidth]{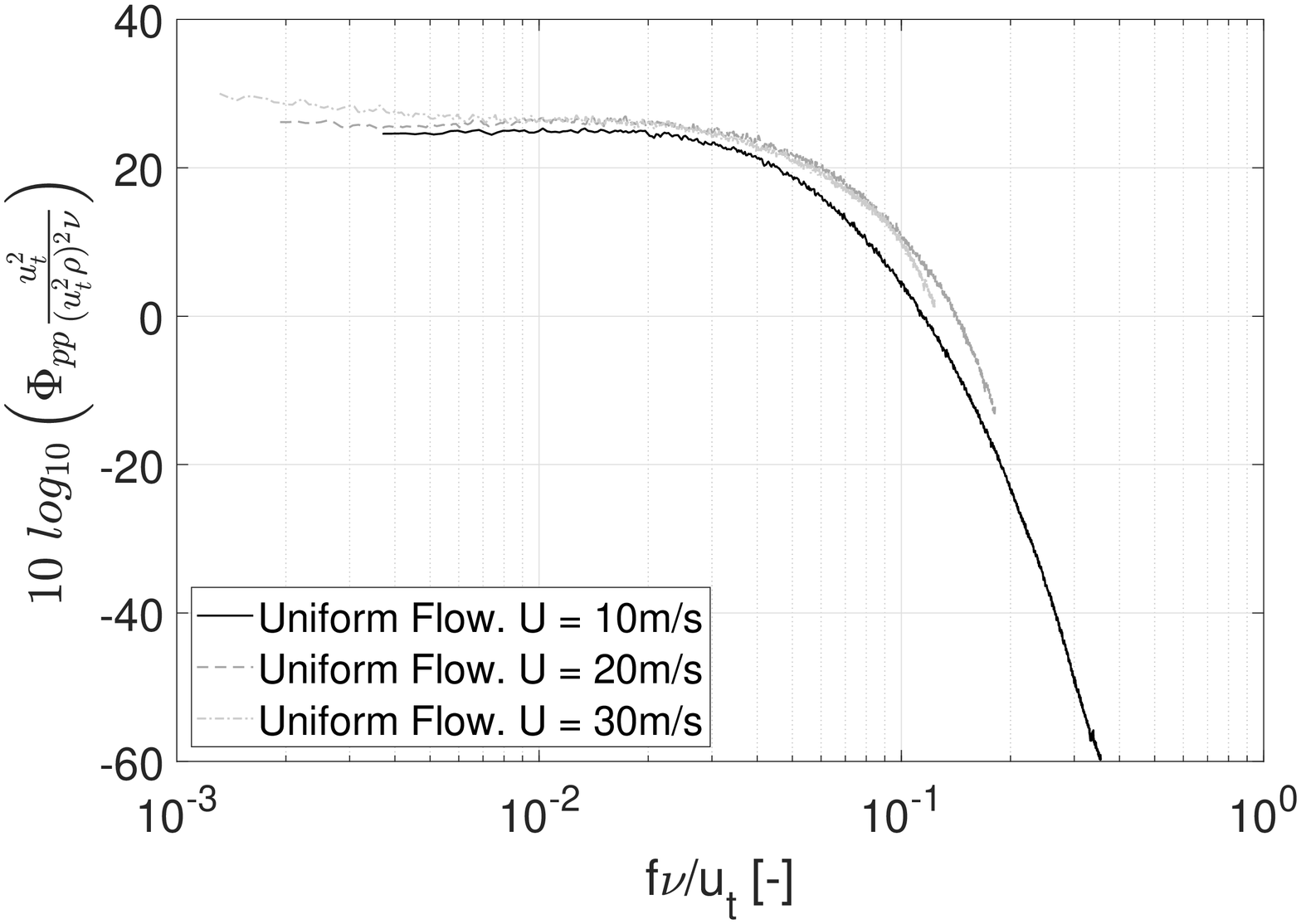}}
\tabularnewline
{\footnotesize{} (a) Not normalized} & {\footnotesize{} (b) Normalized with inner scales}
\tabularnewline
{\footnotesize{}\includegraphics[width=.45\textwidth]{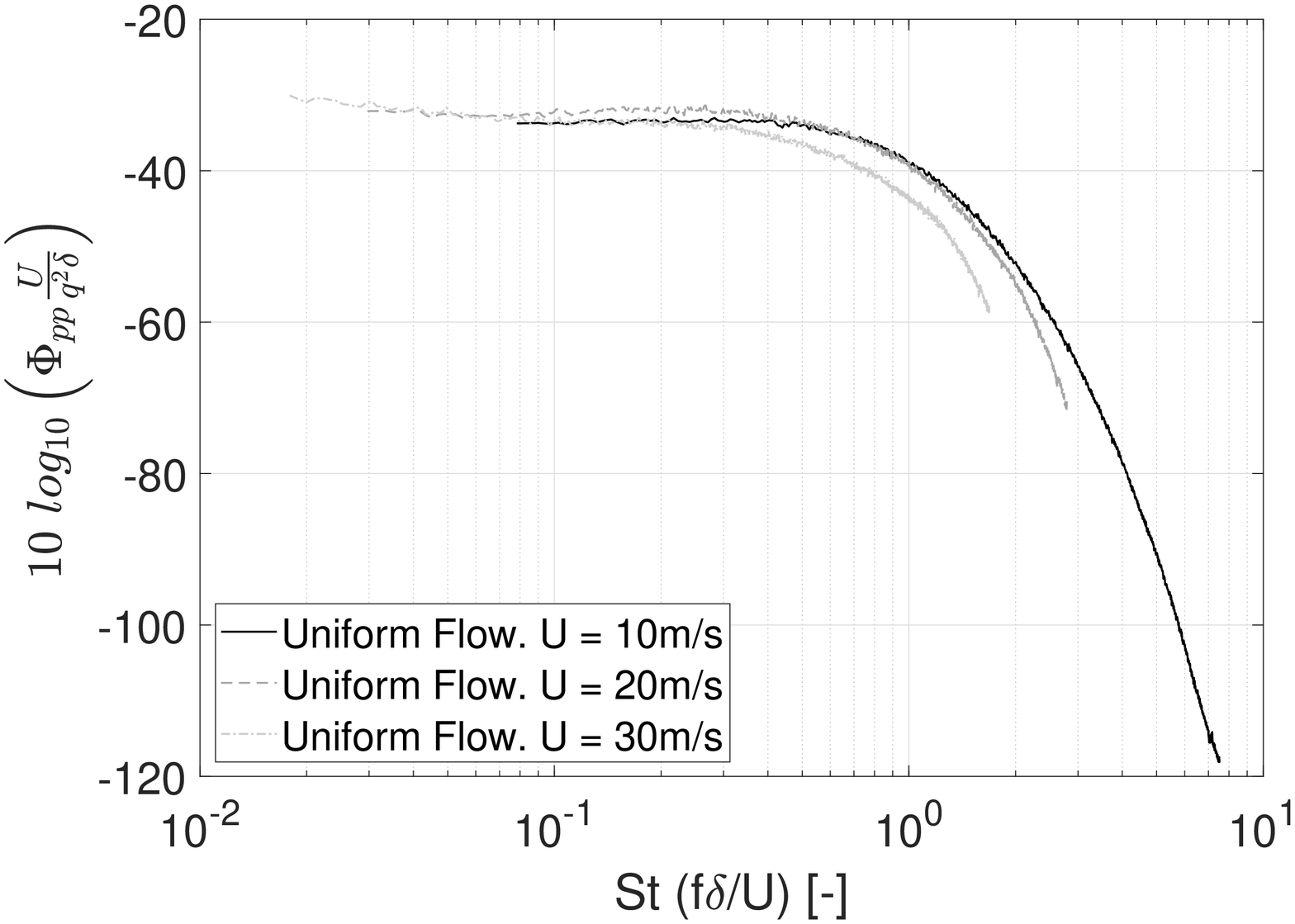}} &
{\footnotesize{}\includegraphics[width=.45\textwidth]{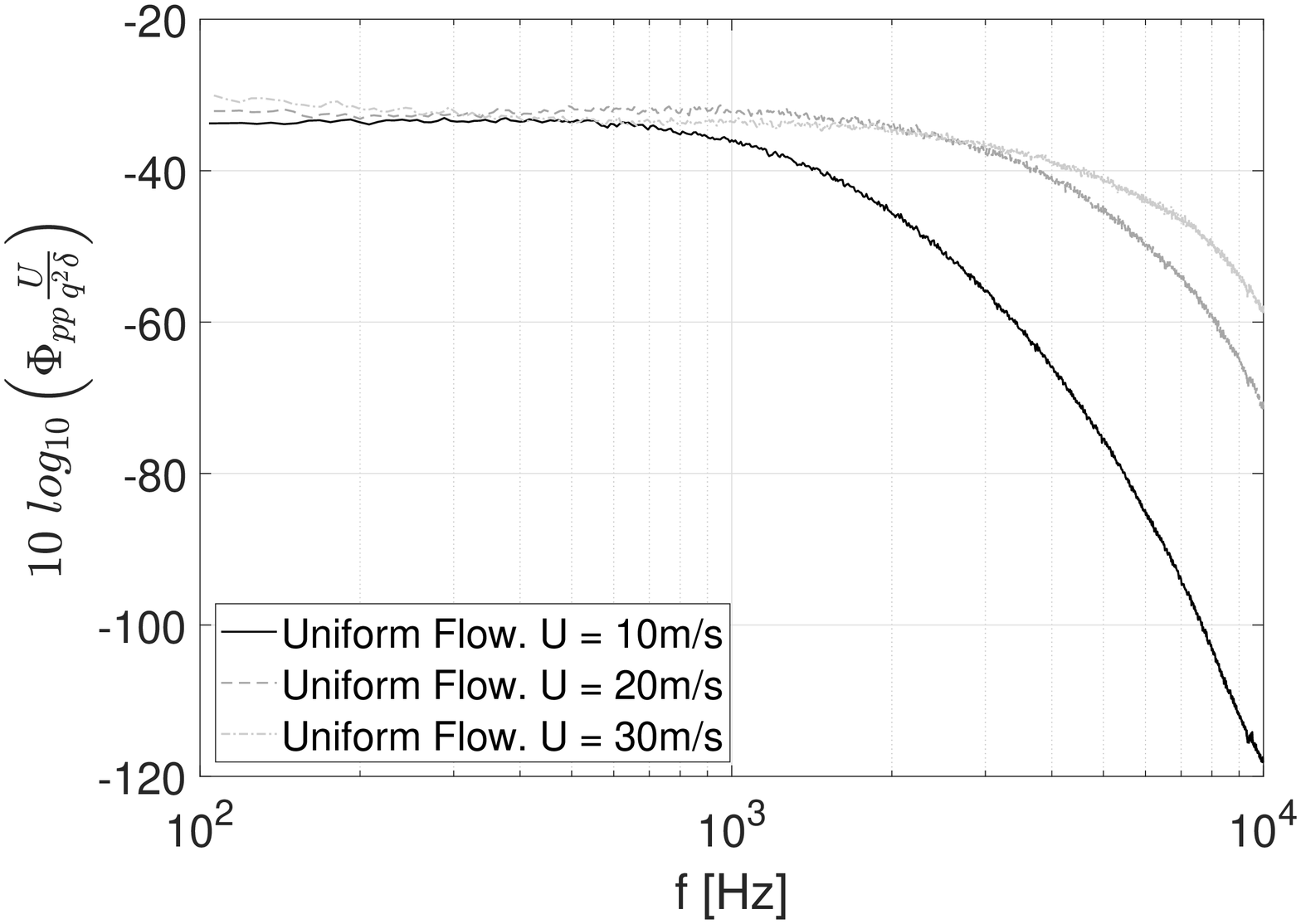}}
\tabularnewline
{\footnotesize{} (c) Normalized with outter scales} & {\footnotesize{} (d) Level scaled by $\delta$}
\tabularnewline
\multicolumn{2}{c}{{\footnotesize{}\includegraphics[width=.45\textwidth] {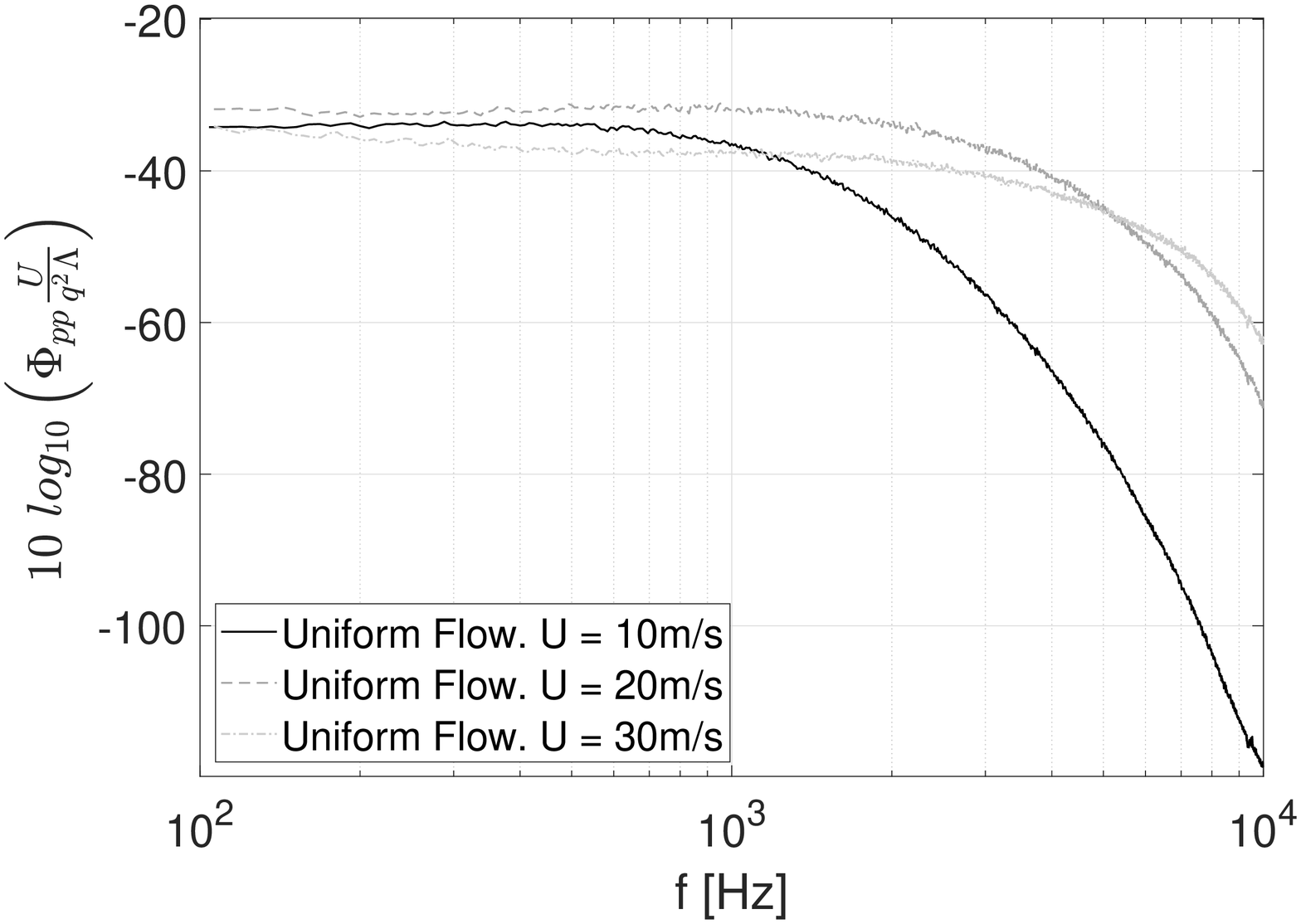}}}
\tabularnewline
\multicolumn{2}{c}{{\footnotesize{} (e)  Level scaled by $\Lambda$}}
\end{tabular}
\caption{ \label{Fig: P_spectrum_UF} Normalized and not normalized surface pressure spectra for uniform flow and FST. $\mathbf{P_{\mathrm{ref}}}$ = 20 \textmu Pa, $\mathbf{q = 1/2 \rho U^2}$. U is the free stream velocity.}
\par\end{centering}
\end{figure}

Figure~\ref{Fig: P_spectrum_UF} shows the surface pressure spectrum normalized and not normalized for the uniform inflow case at 10~m/s, 20~m/s, and~30~m/s free-stream velocity. Figure~\ref{Fig: P_spectrum_UF} (b) and (c) show that the pressure spectrum scales with the inner scales at high frequency and with the outer scales at low frequency, mainly for 20~m/s and 30~m/s inflow velocity. Additionally, Fig.~\ref{Fig: P_spectrum_UF} (d) and (e) show that the level of the surface pressure spectrum scales within 2dB with the boundary layer thickness and integral length scale up to 1~kHz frequency.

\begin{figure}[hbt!]
 \begin{centering}
    \begin{tabular}{cc}
{\footnotesize{}\includegraphics[width=.45\textwidth]{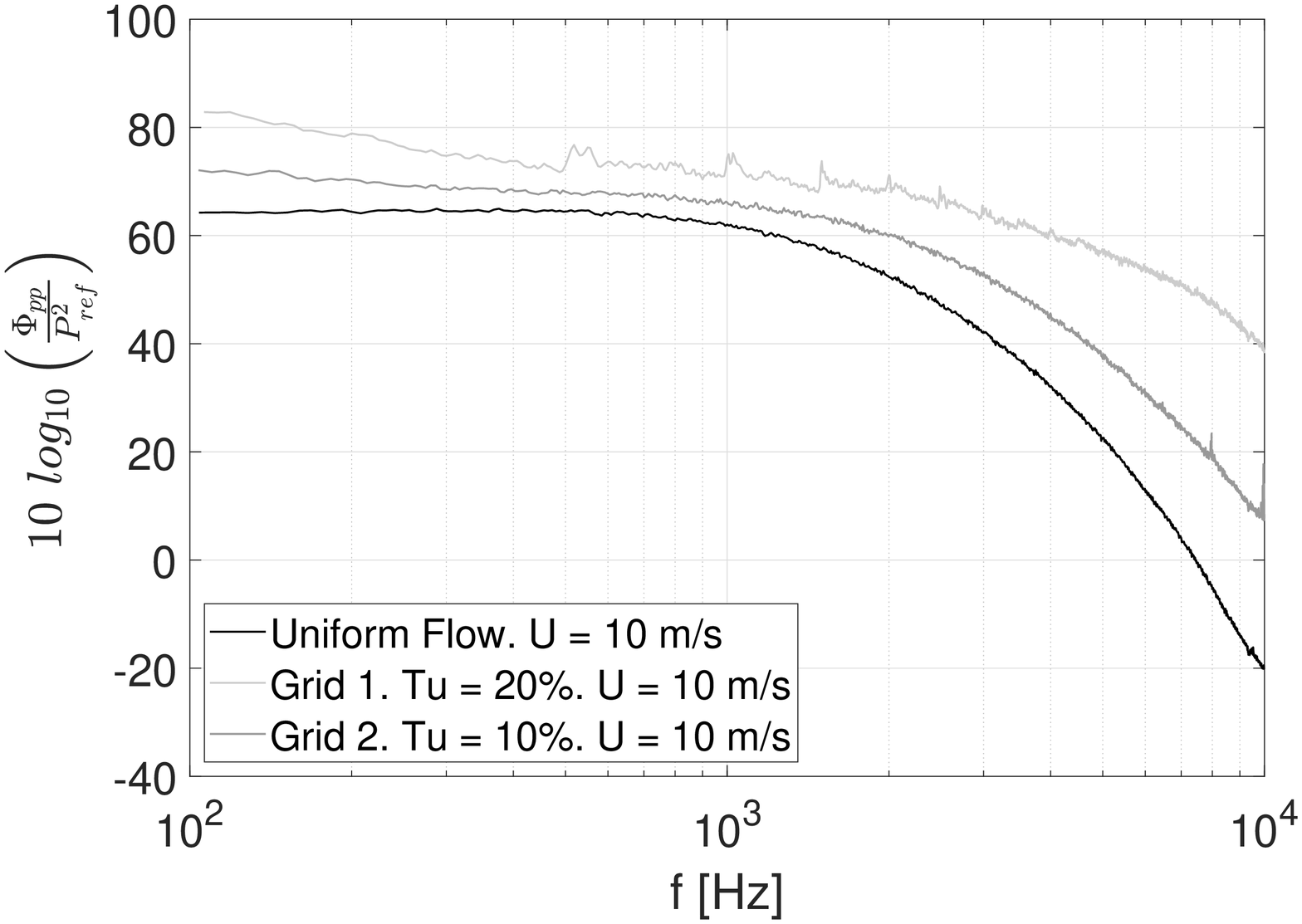}} &
{\footnotesize{}\includegraphics[width=.45\textwidth]{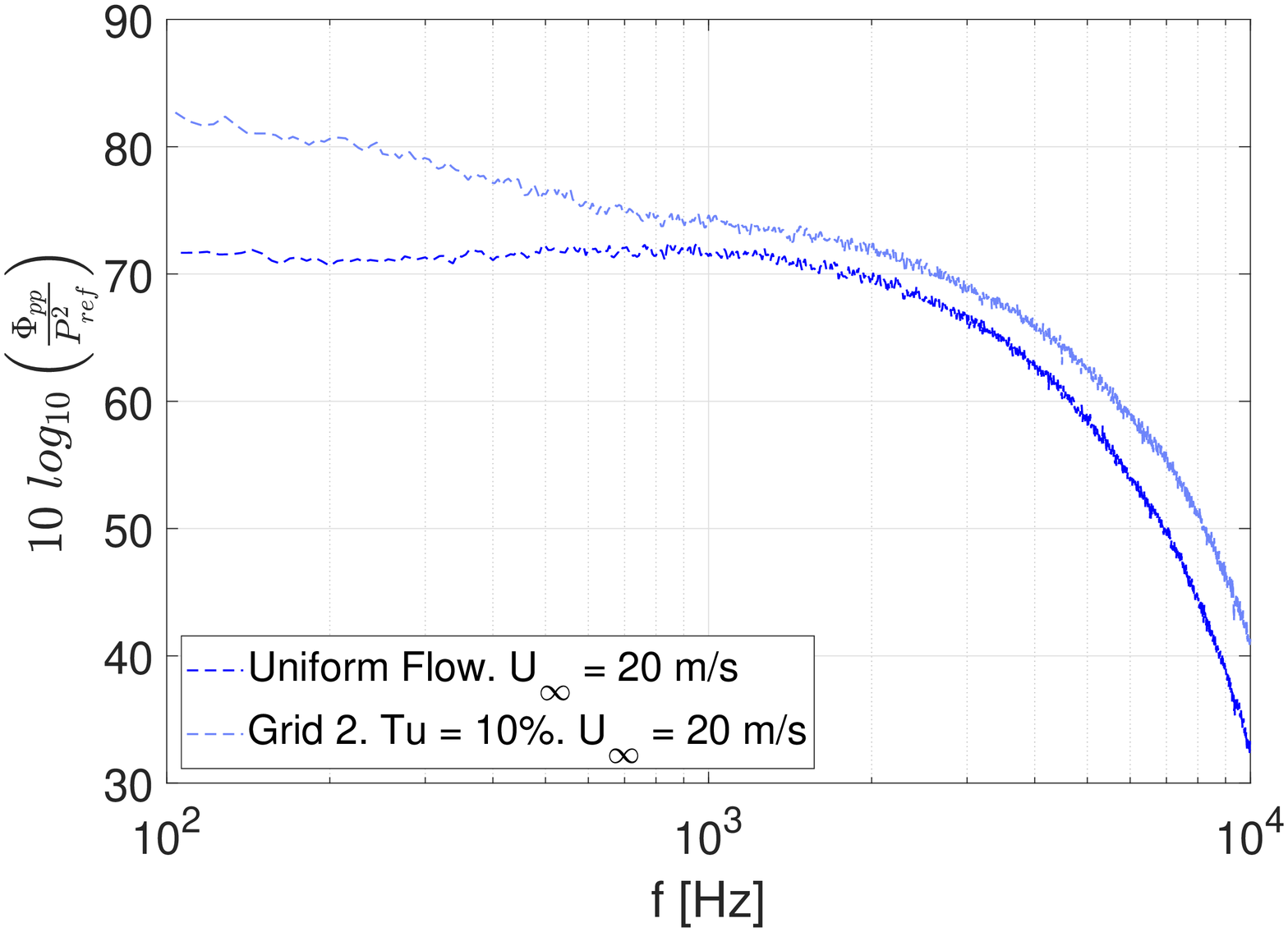}}
\tabularnewline
{\footnotesize{} (a) Not normalized. U = 10~m/s} & {\footnotesize{} (b) Not normalized. U  = 20~m/s}
\tabularnewline
{\footnotesize{}\includegraphics[width=.45\textwidth]{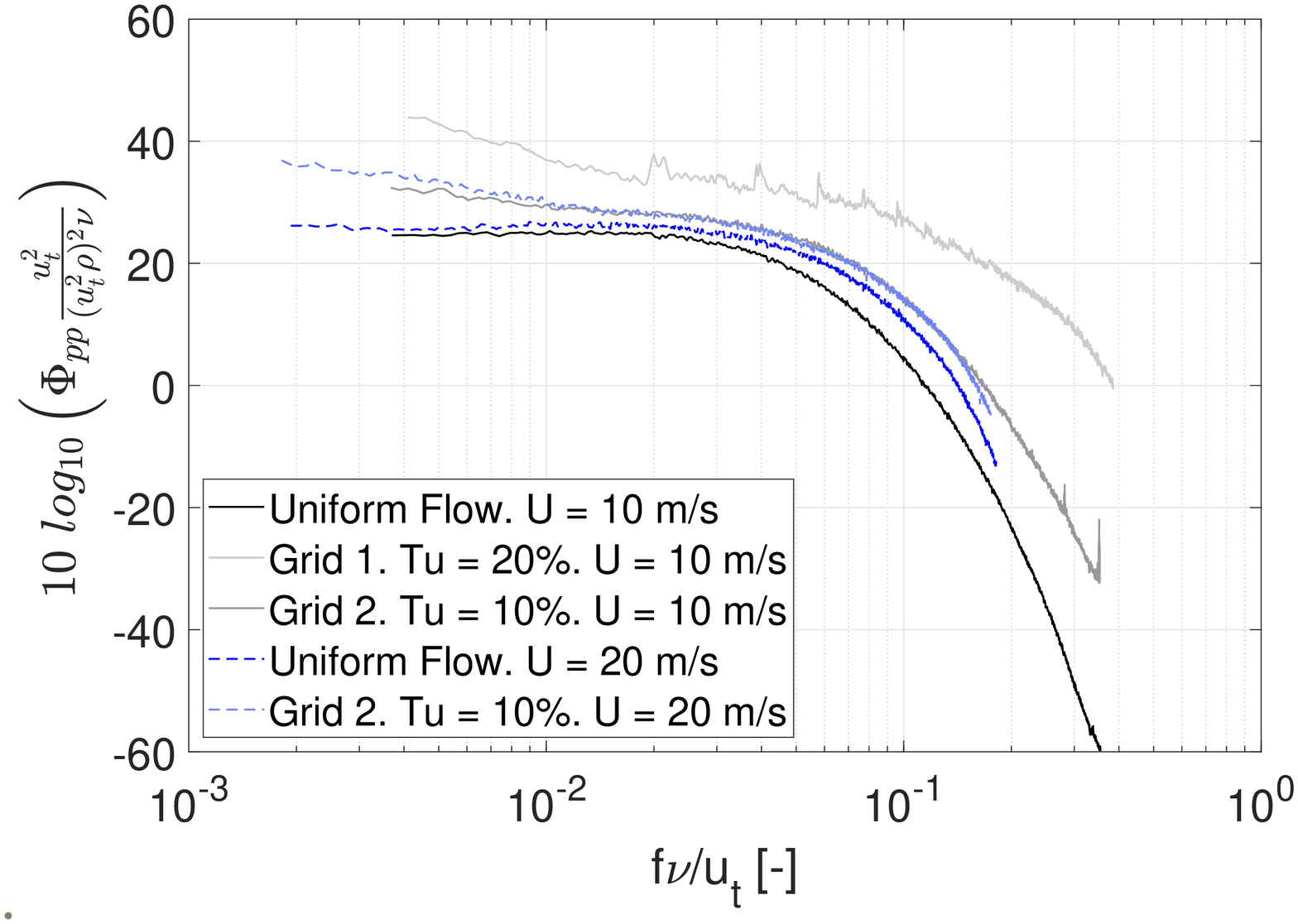}} &
{\footnotesize{}\includegraphics[width=.45\textwidth]{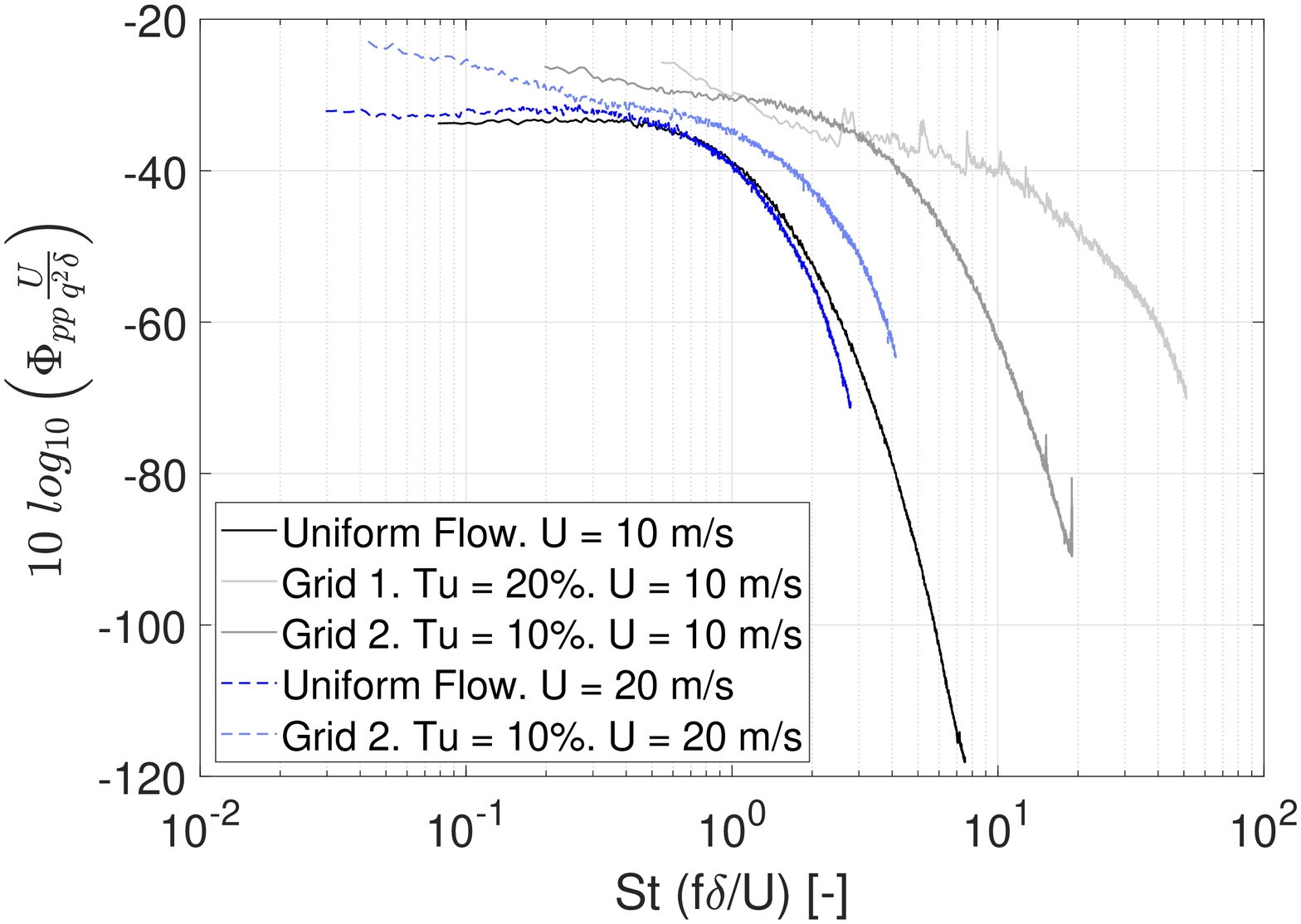}}
\tabularnewline
{\footnotesize{} (c) Normalized with inner scales} & {\footnotesize{} (d) Normalized with outter scales }
\tabularnewline
\tabularnewline
{\footnotesize{}\includegraphics[width=.45\textwidth]{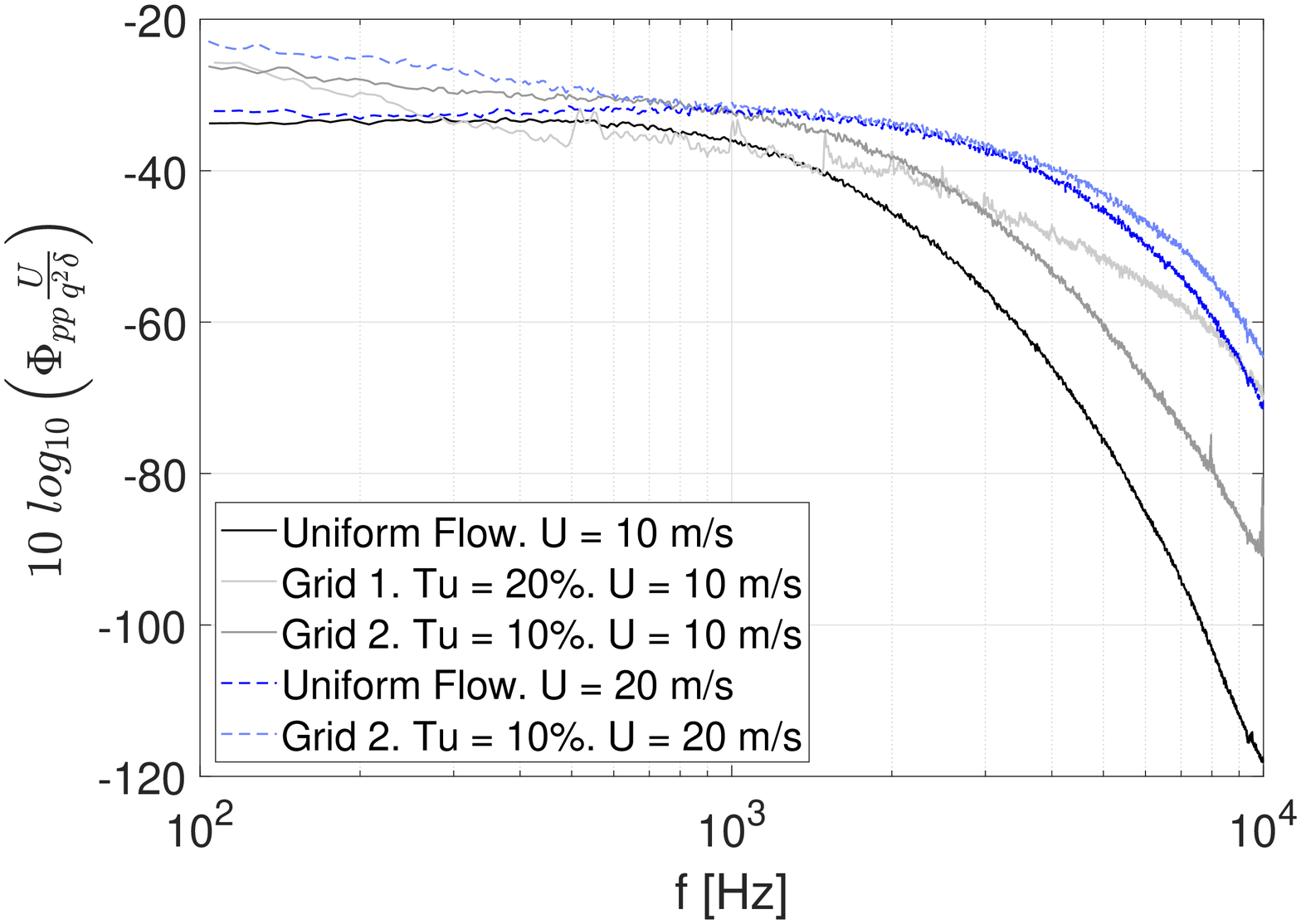}} &
{\footnotesize{}\includegraphics[width=.45\textwidth]{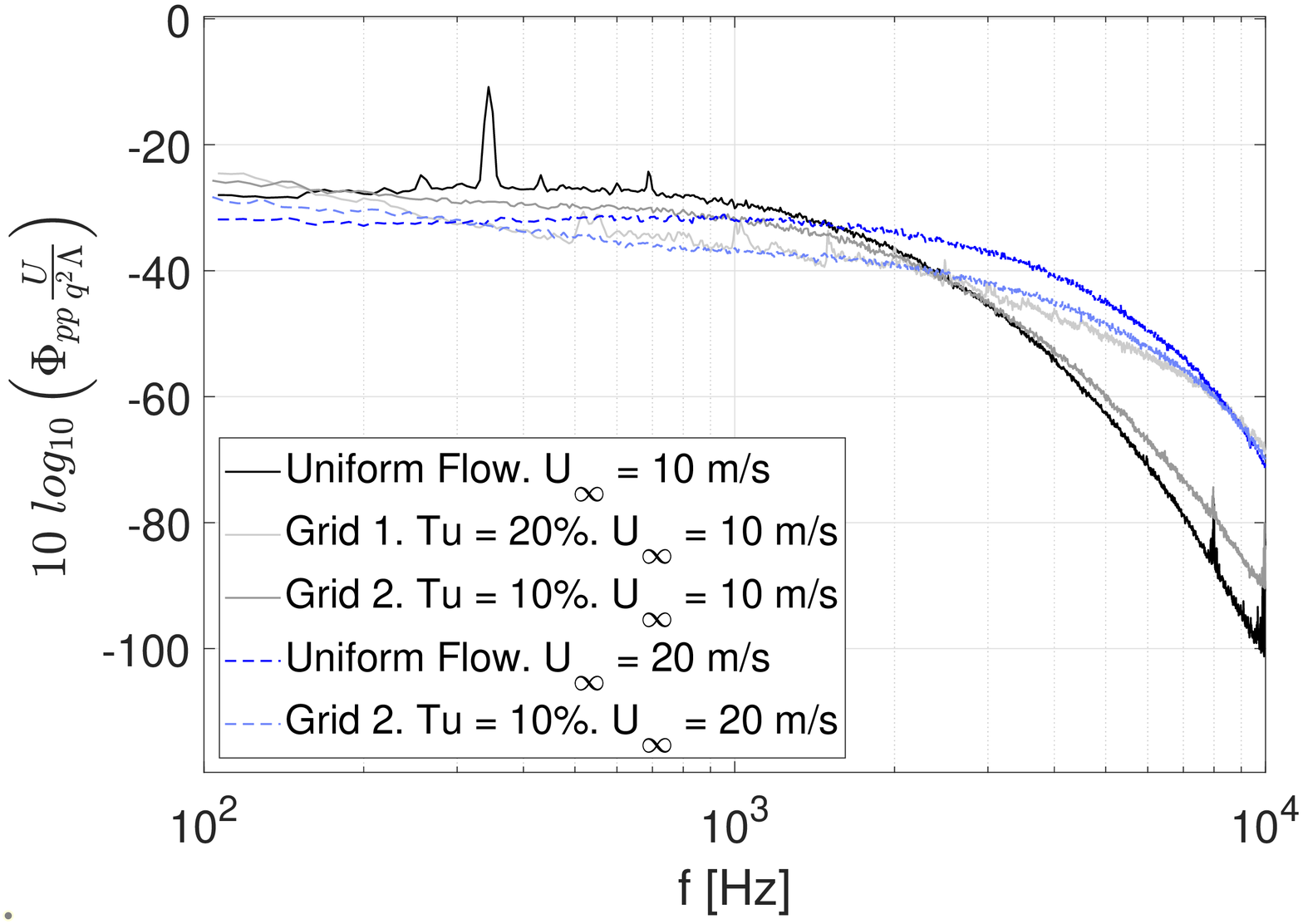}}
\tabularnewline
{\footnotesize{} (e) Level scaled by $\delta$} & {\footnotesize{} (f) Level scaled by $\Lambda$}
\tabularnewline
\end{tabular}
\caption{ \label{Fig: P_spectrum_grids} Normalized and not normalized surface pressure spectra for uniform flow and FST. $\mathbf{P_{\mathrm{ref}}}$ = 20 \textmu Pa, $\mathbf{q = 1/2 \rho U^2}$. U is the free stream velocity.}
\par\end{centering}
\end{figure}

The increment of the surface pressure fluctuations due to the \FST~is evidenced in Figs.~\ref{Fig: P_spectrum_grids} (a) and (b) for 10~m/s and 20~m/s, respectively. The 20\% \FST~increases the surface pressure spectrum by 10~dB in the 500~Hz to 3000~Hz frequency range and even more at other frequencies. The 10\% \FST~also increases the surface pressure fluctuations in the entire frequency range. For frequencies above 500~Hz, the \FST~increases on average 6~dB for both velocities. The increase of the surface pressure spectrum in the low--frequency range at 20~m/s is related to the increase of the velocity spectrum in the boundary layer, shown in Fig.~\ref{Fig: BL_spectrum}, in the same frequency range, in which the larger turbulent structures introduced by the \FS~ stronger influence the turbulence inside the boundary layer. As stated before, for cases of inflow turbulence, the turbulence inside the boundary layer is composed of the penetration of the \FST~and the near--wall turbulence. 

Figure~\ref{Fig: P_spectrum_grids} (c) shows that the surface pressure spectrum for the case of 10\% inflow turbulence scales with the inner scales in the entire frequency range when the inflow velocity is increased.  
Furthermore, the surface pressure spectrum level for frequencies up to 1~kHz scales with the boundary layer thickness and the integral length scale of the boundary layer for all inflow conditions, as shown in Figs.~\ref{Fig: P_spectrum_grids} (e) and (f), respectively. This level--scaling could be a start to model the surface pressure spectrum when the airfoil is subjected to inflow turbulence.  
 
\subsubsection{Convection velocity and spanwise correlation length}
\begin{figure}[hbt!]
 \begin{centering}
    \begin{tabular}{cc}
{\footnotesize{}\includegraphics[width=.45\textwidth]{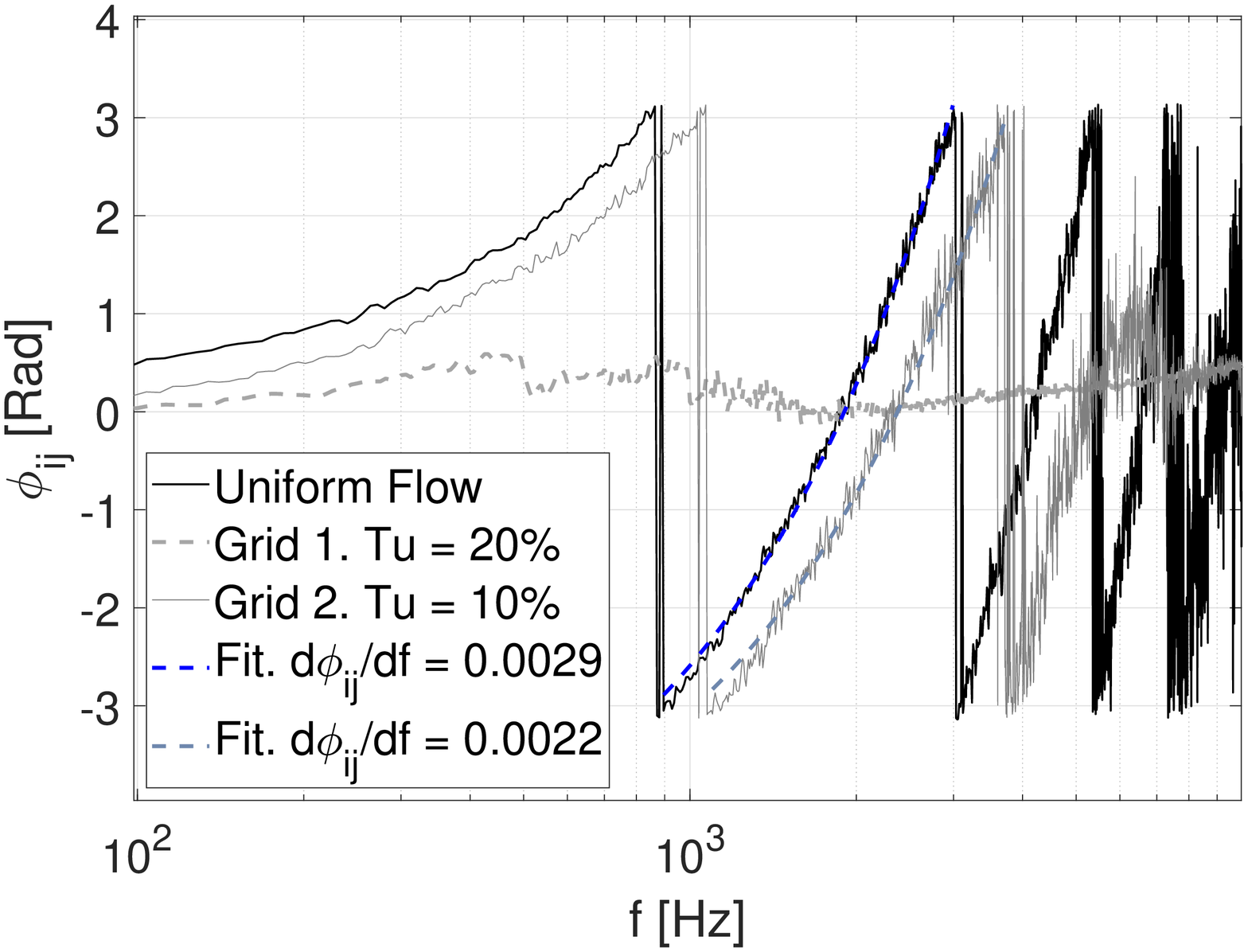}} &
{\footnotesize{}\includegraphics[width=.45\textwidth]{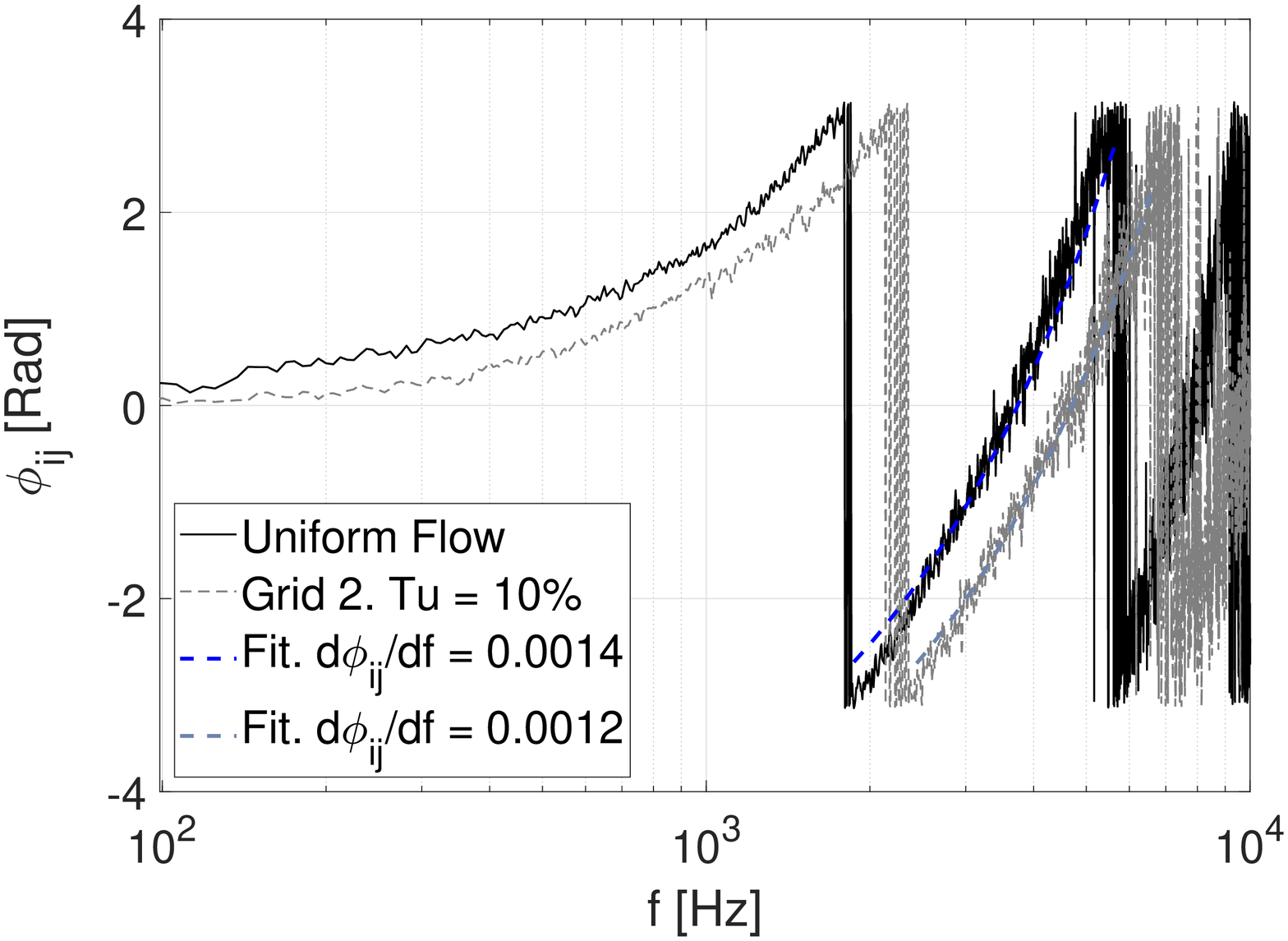}}
\tabularnewline
{\footnotesize{} (a) U = 10~m/s} & {\footnotesize{} (b) U = 20~m/s}
\end{tabular}
\caption{ \label{Fig: convel} Phase between two chordwise microphones located at midspan. n\textsubscript{x} = 3~mm}
\par\end{centering}
\end{figure}

The determination of the convection velocity and spanwise corrrelation length are crucial for the prediction of the trailing edge noise. Figure~\ref{Fig: convel} (a) and (b) show the phase ($\phi_{ij}$) between the two central microphones located streamwise for the cases of a \FS~ velocity of 10~m/s and 20~m/s, respectively. For the cases of uniform inflow and 10\% inflow turbulence, a clear linear region at each the $\phi_{ij}$ versus f curve is identified for both velocities. A linear function is used for fitting the curve in the linear region, which results in the values of the slope shown in the graphs, yielding the values of $U_c/U$ presented in Tab~\ref{tab:con_vel}, where U is the free-stream velocity, i.e., the velocity outside the boundary layer. The 10\% \FST~increases the convection velocity, reaching values of $U_c/U$ equal to unity at 10~m/s. For uniform inflow, $U_c/U$ does not change with the \FS~velocity. However, for the case of 10\% \FST, $U_c/U$ is reduced when the inflow velocity increases. The velocity at which the vortices convect plays an important role in the scattering of the hydrodynamic waves at the trailing edge. An increment in the convection velocity should be reflected in an increment in the trailing edge noise.
\begin{table}[htbp!]
\caption{\label{tab:con_vel} Normalized convection velocity ($\mathbf{U_c/U}$).}
\centering
\begin{tabular}{lcccccccccccccc}
\hline
        & \multicolumn{1}{c}{Uniform flow}          && \multicolumn{1}{c}{10\% } & &  \\ \cline{2-4} 
       
10 [m/s] & 0.73 & & 1.07 \\
20 [m/s] & 0.76 & & 0.89 \\   
\hline
\end{tabular}
\end{table}

\begin{figure}[hbt!]
 \begin{centering}
    \begin{tabular}{cc}
{\footnotesize{}\includegraphics[width=.45\textwidth]{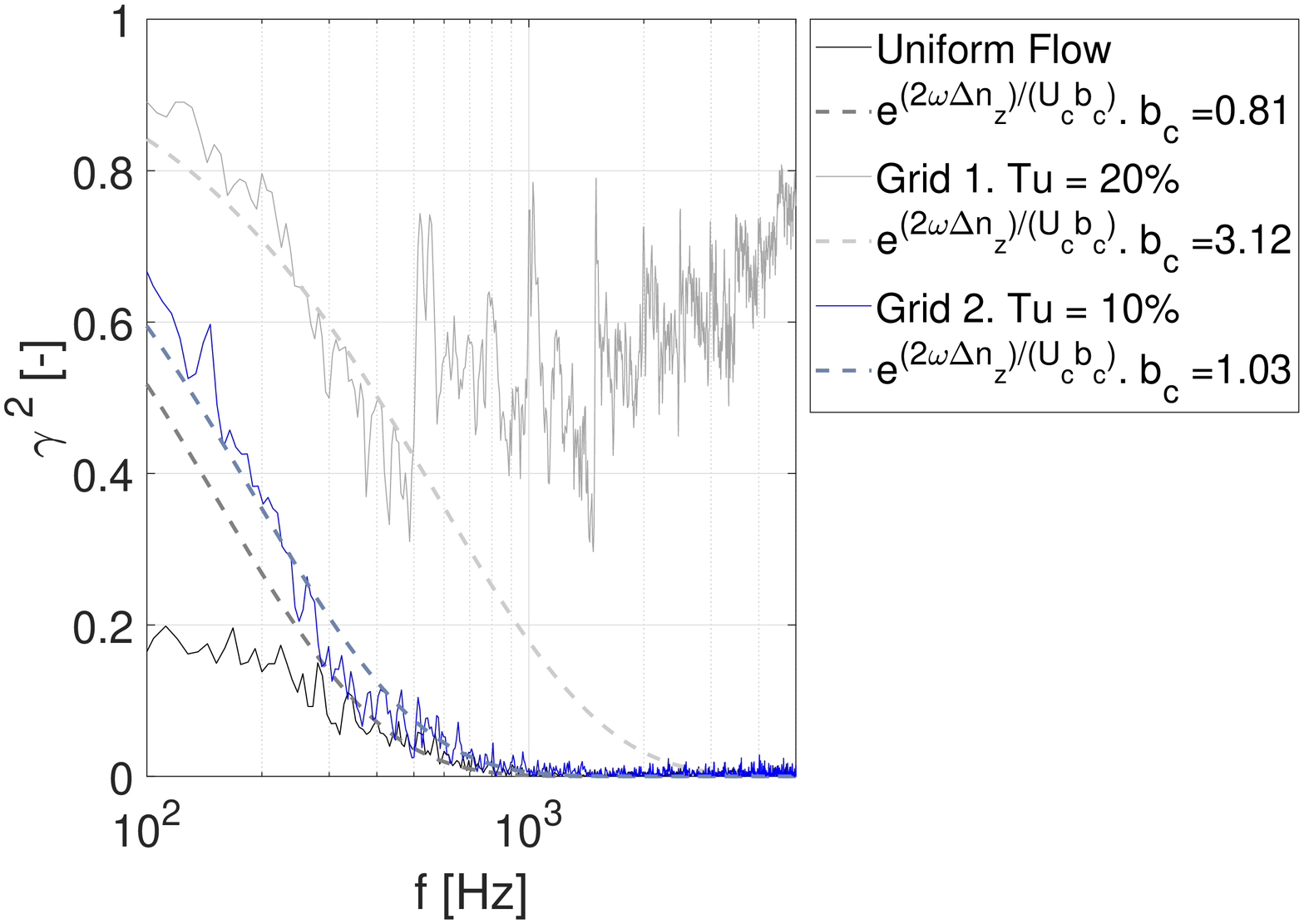}} &
{\footnotesize{}\includegraphics[width=.45\textwidth]{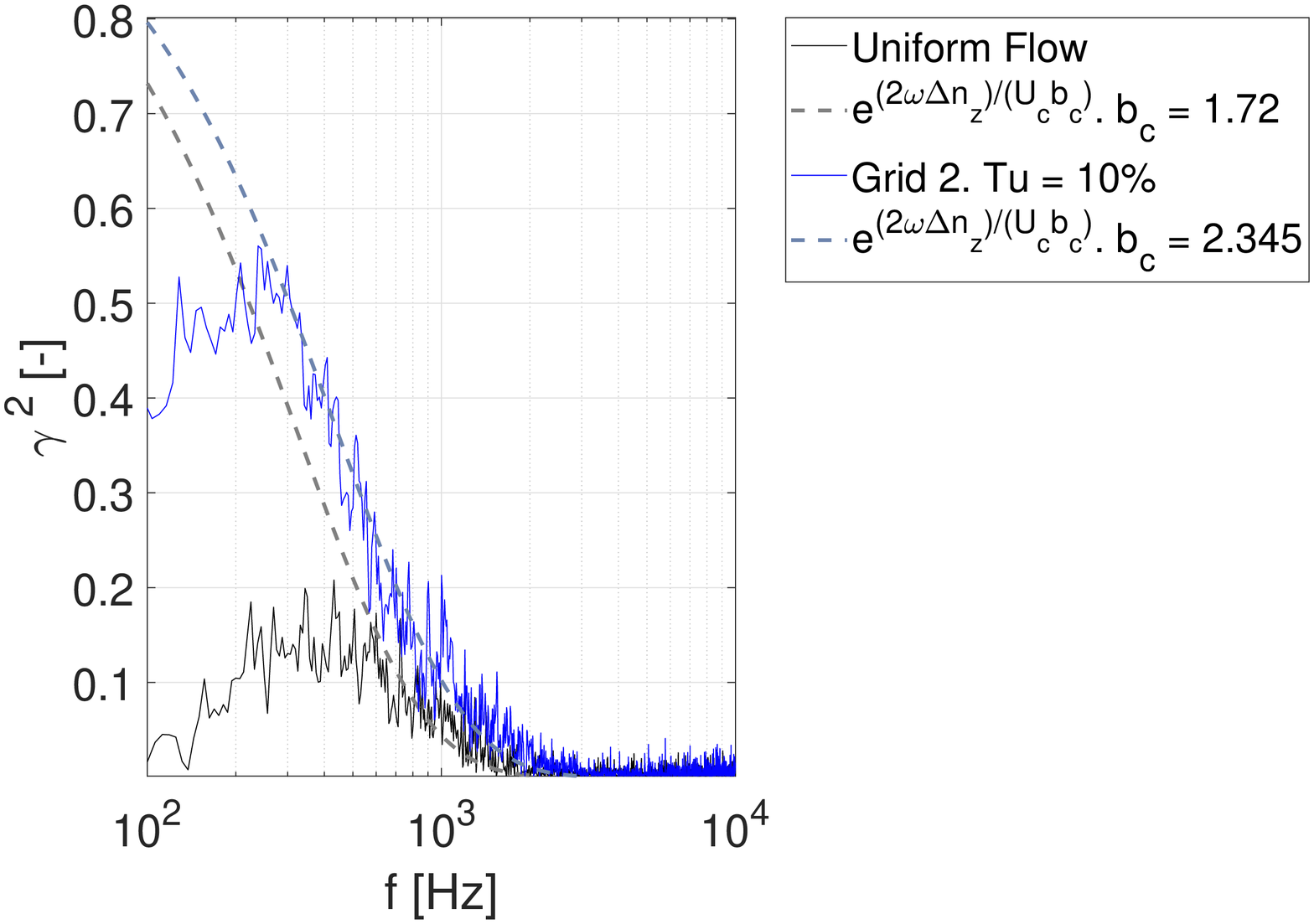}}
\tabularnewline
{\footnotesize{} (a) U = 10~m/s} & {\footnotesize{} (b) U = 20~m/s}
\end{tabular}
\caption{ \label{Fig: coherence} Coherence between two microphones in the spanwise direction located at x/c = -0.0675.  $\mathbf{\Delta n_z = 3}$ [mm].}
\par\end{centering}
\end{figure}

Different from the observation for the cases of uniform inflow and 10\% inflow turbulence, Fig.~\ref{Fig: convel} (a) shows that for the case of 20\% \FST, the phase between the two chordwise microphones remains almost zero in the entire frequency range, suggesting that both microphones are measuring the unsteady pressure caused by the same turbulent structure. The smallest turbulent structures that have sufficient energy to cause instabilities in the surface pressure are larger than the distance between the microphones, i.e., 3~mm. The same discussion applies when analyzing the coherence between two microphones in the spanwise direction for the case of 20\% \FST, shown in Fig.~\ref{Fig: coherence} (a). 

With 20\% inflow turbulence, the coherence does not follow the model proposed by Corcos, see Fig.~\ref{Fig: coherence} (a). The coherence does not decay as a function of the frequency and remains at values higher than 0.5, suggesting that the two spanwise microphones are measuring the same turbulent structure. Similar coherence values are obtained when doubling the distance between the microphones (these results are not shown here for sake of simplicity). The boundary layer measurements support this analysis, where it seems that the 20\% \FST~influences the velocity fluctuations inside the boundary layer in the entire frequency range, including the smallest turbulent structures. According to Sharp et al.~\cite{sharp2009effects}, the \FST~changes the interaction between the turbulent structures inside of the boundary layer. Hence, the smallest scales could have suffered a significant increase due to the insertion of the large scales of the 20\% \FST. 

\begin{figure}[hbt!]
 \begin{centering}
    \begin{tabular}{cc}
{\footnotesize{}\includegraphics[width=.45\textwidth]{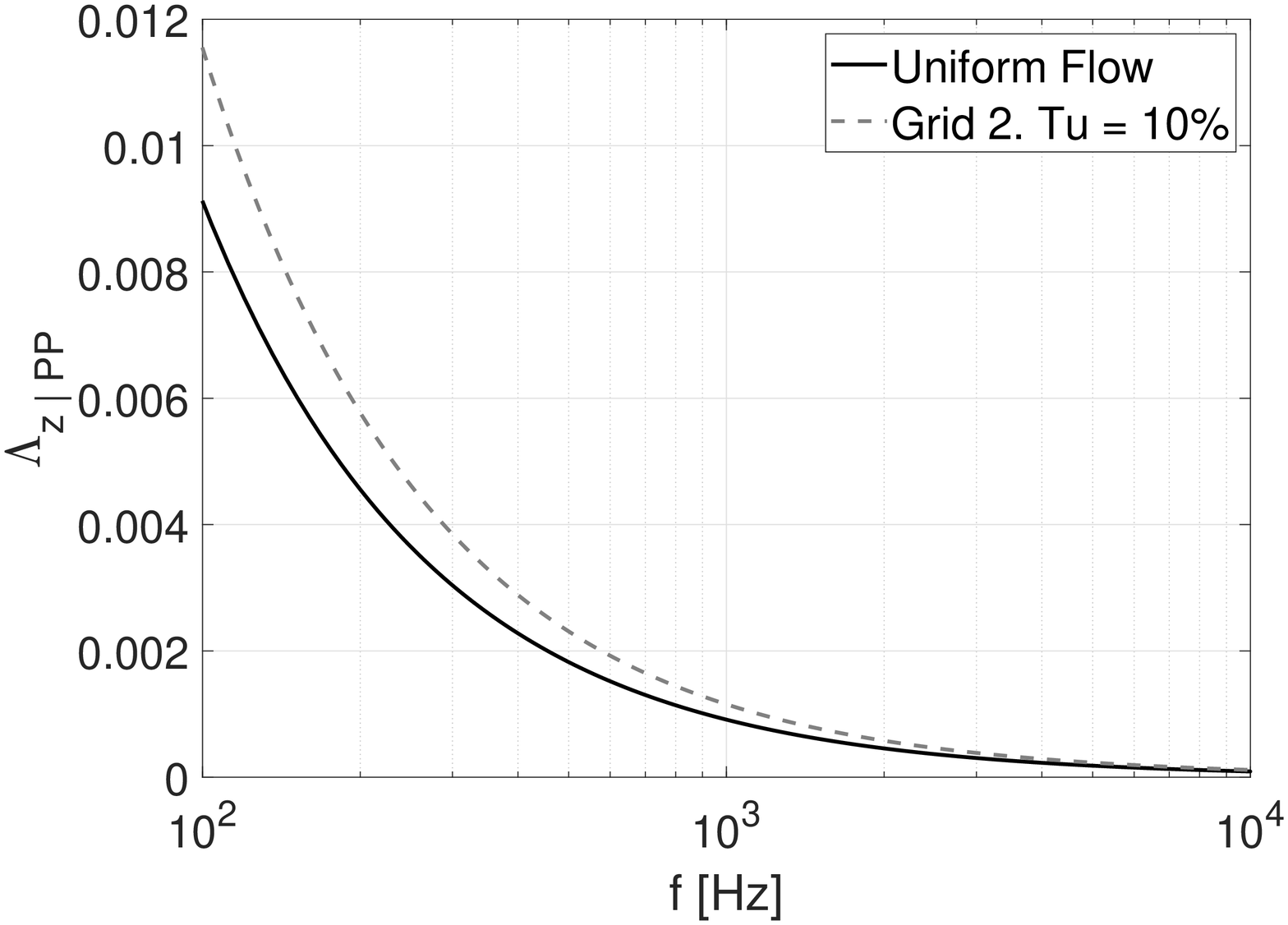}} &
{\footnotesize{}\includegraphics[width=.45\textwidth]{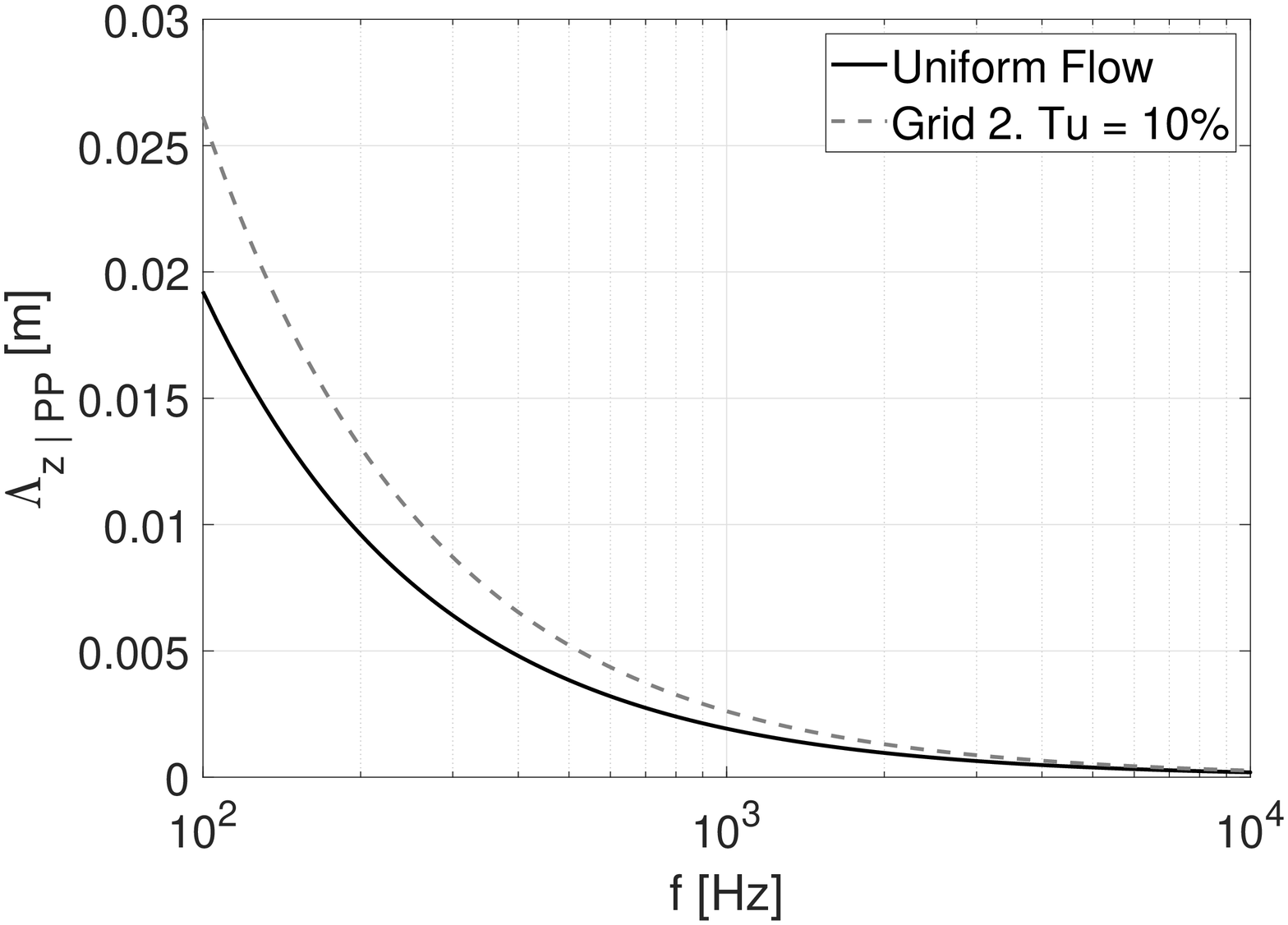}}
\tabularnewline
{\footnotesize{} (a) U = 10~m/s} & {\footnotesize{} (b) U = 20~m/s}
\end{tabular}
\caption{ \label{Fig: cor_length}Spanwise correlation length.}
\par\end{centering}
\end{figure}

For the case of 10\% inflow turbulence, the coherence between two spanwise microphones follows the Corcos model with values close to zero for high frequencies, as for the case of uniform inflow. The 10\% inflow turbulence increases the coherence considerably for frequencies up to 1~kHz in comparison with the uniform incoming flow case, for both 10~m/s and 20~m/s inflow velocity, as shown in Fig.~\ref{Fig: coherence}. Figure~\ref{Fig: cor_length} shows the spanwise correlation length calculated with the fitted Corcos' constant at 10~m/s and 20~m/s. The increase of the spanwise correlation length reflects the increase of the coherence due to the \FST. 

\subsubsection{Prediction of surface pressure fluctuations and far--field noise}
\begin{figure}[hbt!]
 \begin{centering}
    \begin{tabular}{cc}
{\footnotesize{}\includegraphics[width=.45\textwidth]{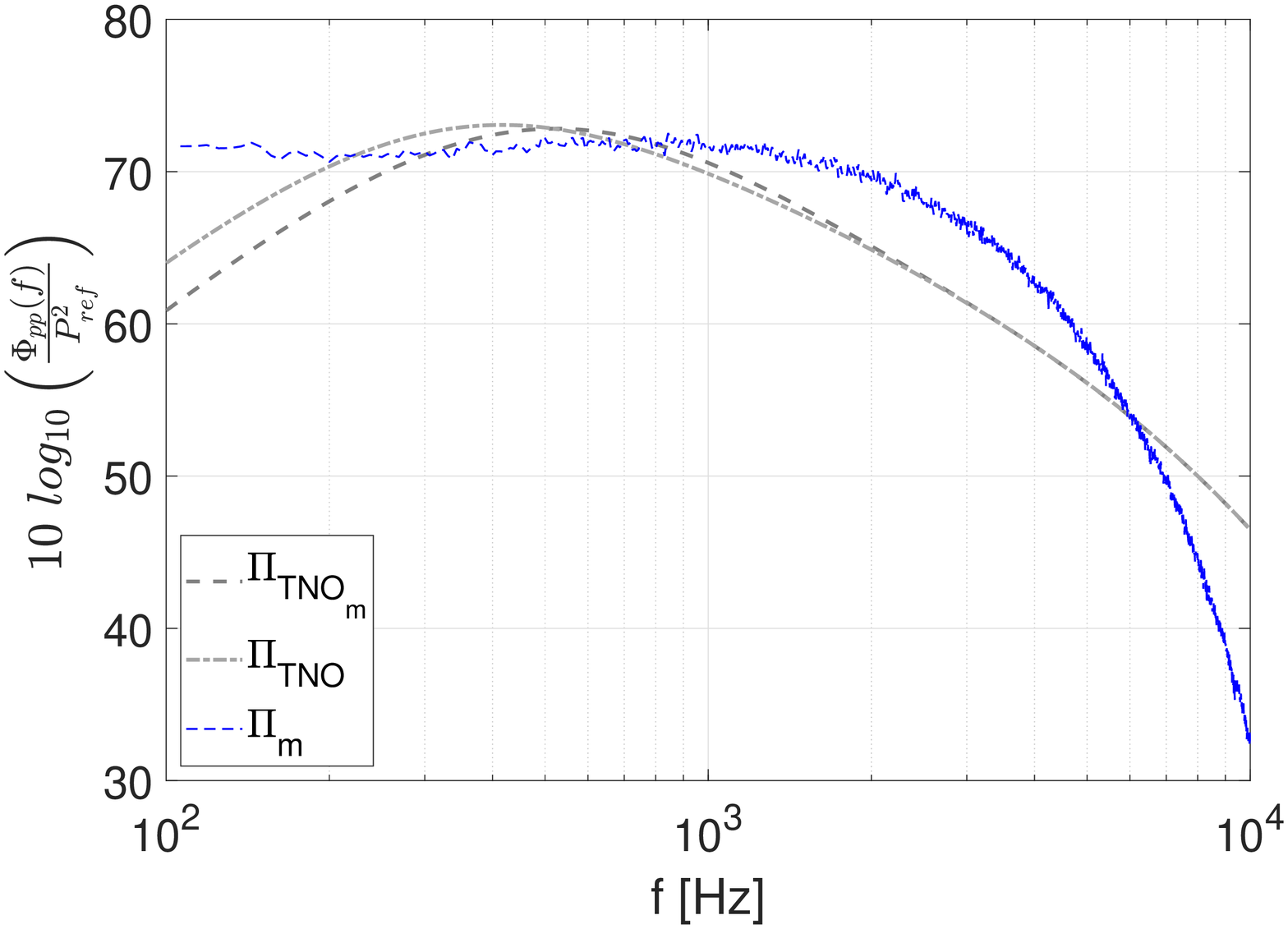}} &
{\footnotesize{}\includegraphics[width=.45\textwidth]{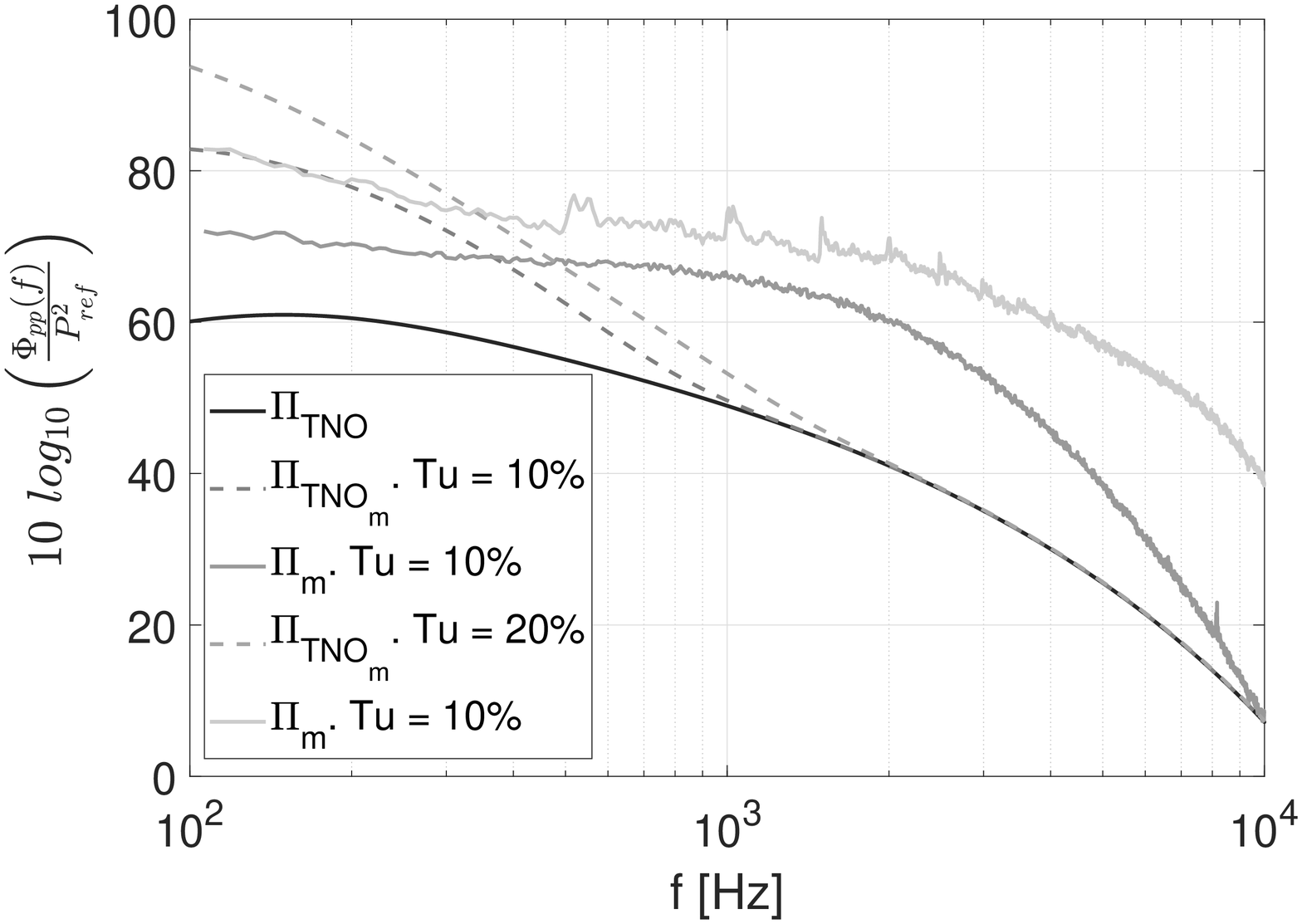}}
\tabularnewline
{\footnotesize{} (a) Uniform incoming flow. U = 20~m/s} & {\footnotesize{} (b) Turbulent incoming flow. U = 10~m/s}
\end{tabular}
\caption{ \label{Fig: SPM_prediction} Predicted and measured surface pressure: predicted by TNO--Blake model using as input XFOIL parameters ($\mathbf{\Pi_{TNO}}$), predicted by TNO--Blake model using as input measured mean--velocity profile of the boundary layer ($\mathbf{\Pi_{TNO_m}}$) and measured by surface sensors ($\mathbf{\Pi_{m}}$). For $\mathbf{\Pi_{TNO}}$ with FST, the $\mathbf{N_{crit}}$ in XFOIL was set to 1 ($\mathbf{FST = 1.9\%}$).} 
\par\end{centering}
\end{figure}

The TNO--Blake model is a semi--empirical model that predicts the surface pressure spectral density having as input parameters of the boundary layer that are easily determined by fast turn--around codes like XFOIL~\cite{drela1989xfoil}. The input parameters are $\delta, \delta^*, \theta,$ and c\textsubscript{f}. Figure~\ref{Fig: SPM_prediction} (a) shows the measured and the predicted surface pressure fluctuations using as input the measurement data and XFOIL data. The TNO--Blake model does not consider the influence of the \FST~on the boundary layer statistics and consequently on the surface pressure fluctuations. For low inflow turbulence levels, i.e., up to 1.9\%, the critical number \textit{N\textsubscript{crit}} in XFOIL can be set to model the effect of the inflow turbulence on the boundary layer parameters~\cite{drela1989xfoil}. However, for cases of high inflow turbulence, where penetration occurs, the model becomes inadequate.

Figure~\ref{Fig: SPM_prediction} (b) shows the prediction of surface pressure fluctuations for 10\% and 20\% of inflow turbulence by the TNO--Blake model. The prediction is calculated using the experimental measurements of the boundary layer, as well as, the XFOIL data for the boundary layer, considering N\textsubscript{crit} = 1, i.e.,Tu = 1.9\% as input. Despite using the experimental measurements as input, the model does not properly predict the \SPF~at high levels of inflow turbulence. There is no agreement between the prediction and the experimental measurements. The TNO--Blake model only predicts an increase of the pressure spectrum for frequencies up to 1~kHz, which is caused mainly by the increase of the boundary layer thickness. A deeper analysis of how the incrase of the turbulence inside the boundary layer caused by the \FST~is reflected in the \SPF~is required to propose a novel prediction model that accounts for this phenomenon.  

\begin{figure}[hbt!]
 \begin{centering}
    \begin{tabular}{cc}
{\footnotesize{}\includegraphics[width=.45\textwidth]{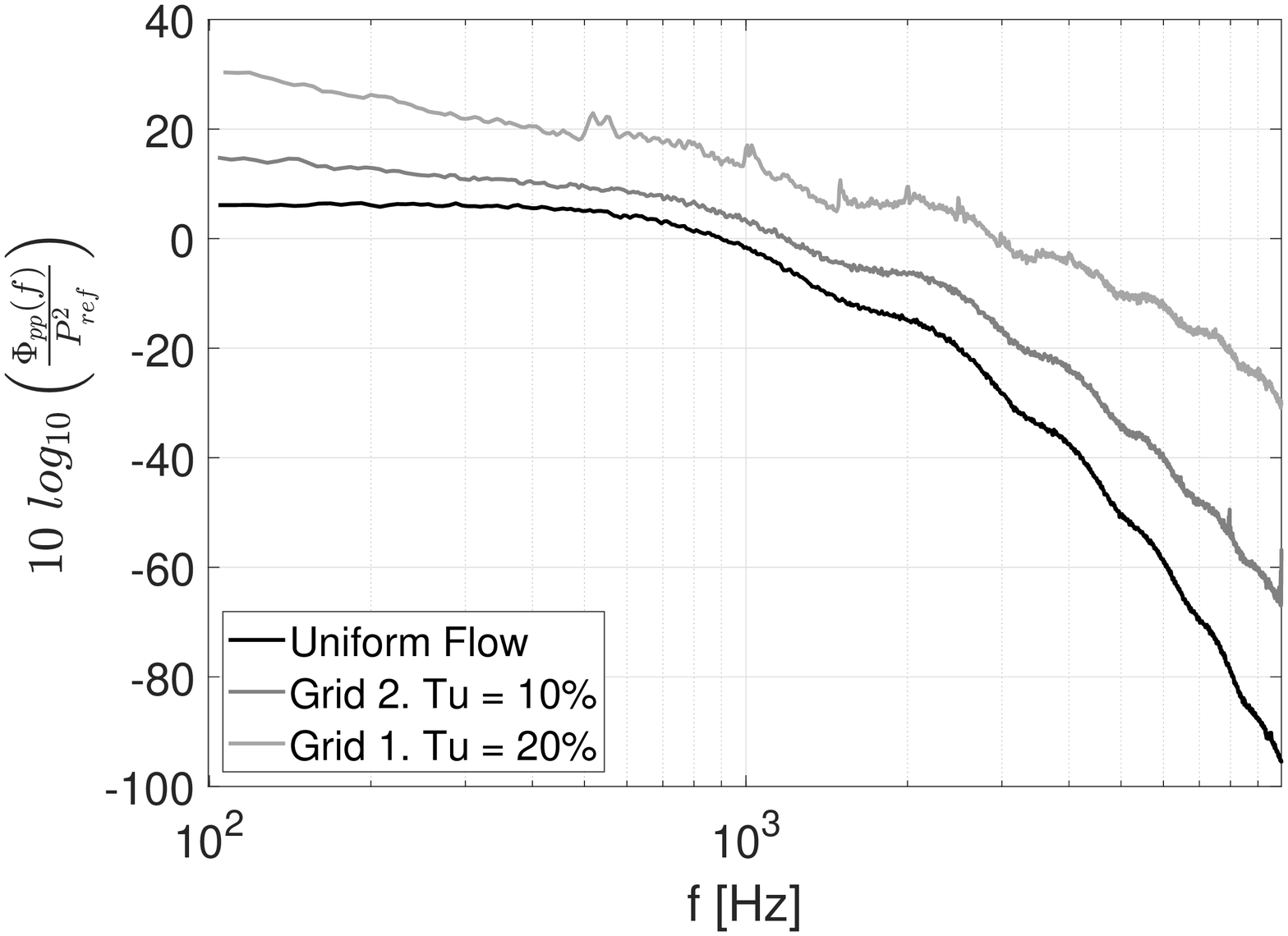}} &
{\footnotesize{}\includegraphics[width=.45\textwidth]{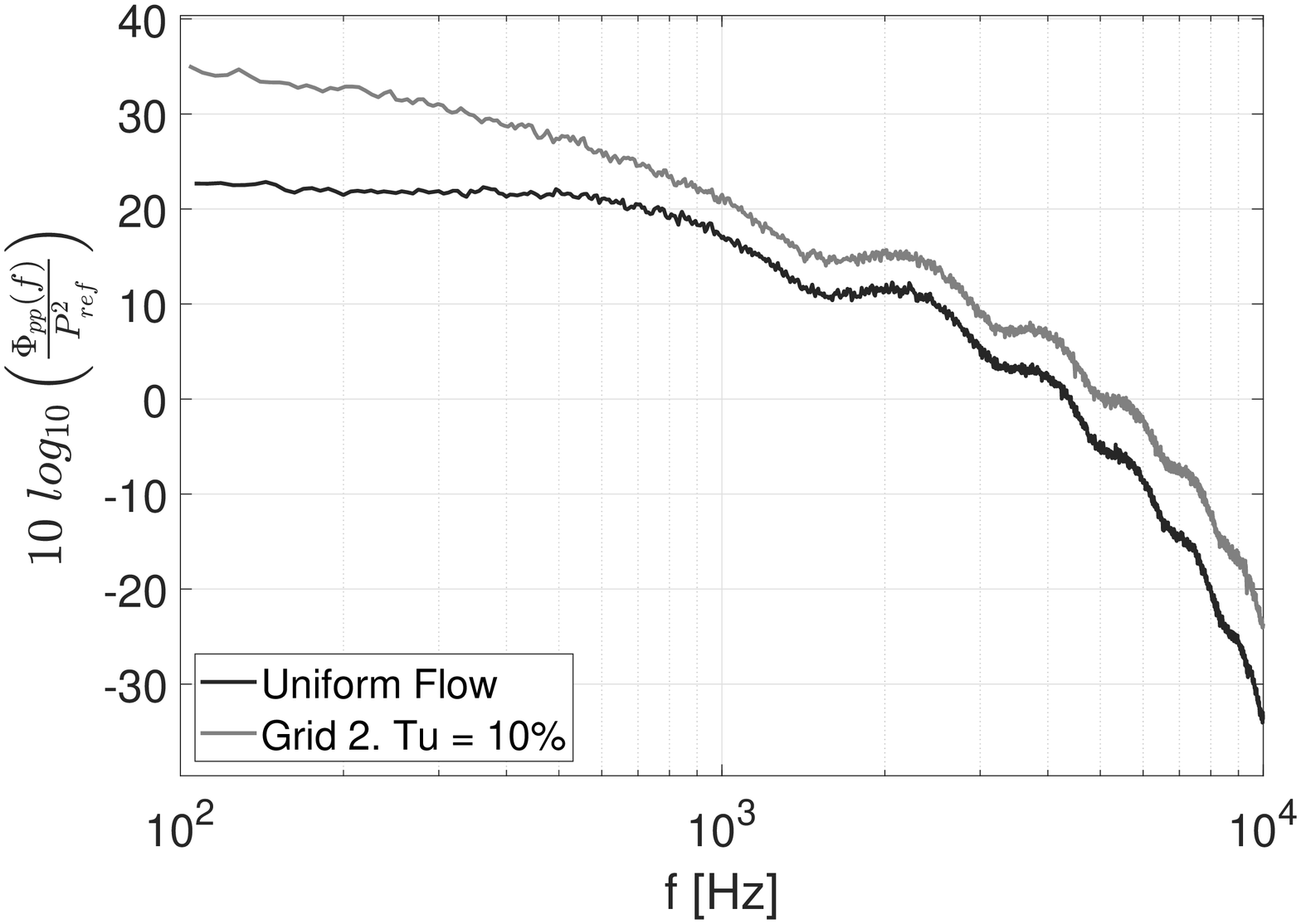}}
\tabularnewline
{\footnotesize{} (a) U  = 10~m/s} & {\footnotesize{} (b) U  = 20~m/s}
\end{tabular}
\caption{ \label{Fig: FFN} Predicted far--field trailing edge noise spectrum using as input the measurements with the surface pressure sensors.}
\par\end{centering}
\end{figure}

Figure~\ref{Fig: FFN} shows the predicted the far--field noise due to the \FST~using Amiet theory, ref. Eq.~\ref{Eq:Amiet}. The experimental measurements of the surface pressure fluctuations and fitted Corcos' constant served as input for each case. The theory predicts that high levels of \FST~can significantly increase the far--field trailing edge noise by more than 10~dB and 6~dB for the cases of 20\% and 10\% inflow turbulence, respectively. However, further experiments have to be conducted to verify that the propagation of the surface pressure fluctuations and the scattering effect that occurs at the trailing edge discontinuity are still valid for the cases of high \FST. Furthermore, the increment of the trailing edge noise for low frequencies has to be compared to the leading edge noise in order to verify which noise source would be the dominant for this frequency range. However, it is also important to highlight that the free-stream turbulence is also affecting the high-frequency range where the trailing edge noise is the dominant noise source even for applications with inflow turbulence.

\section{Conclusions and future works}
This research aimed to determine the influence of the inflow turbulence on the surface pressure fluctuations near the trailing edge and the predicted far--field noise. Experimental tests were done with on a NACA~0012 airfoil in the open jet of the Aeroacoustic Wind Tunnel of the University of Twente at the Reynolds number range of 170 $\times$ 10\textsuperscript{3} to 500 $\times$ 10\textsuperscript{3}. Two passive grids were used to generate uniform turbulence of 10\% and 20\% at the plane of the airfoil leading edge. The experiments discussed in this paper included the velocity fluctuations inside of the boundary layer and surface pressure fluctuations close to the trailing edge. 

Even for high \FST, the velocity profile of the boundary layer follows the universal Prandtl -- von Kármán log--of--the--wall, demonstrating the formation of a boundary layer. The \FST~slightly influences the friction velocity. However, it significantly increases the boundary layer thickness at the trailing edge and the velocity fluctuations and the integral length scale along the boundary layer. For the cases of inflow turbulence, the integral length scale inside the boundary layer scales with the length scale of the \FST, the friction velocity, and the distance from the wall in wall units y\textsuperscript{+}. We propose a logarithmic model to calculate the integral length scale along the boundary layer, which is independent of the inflow turbulence intensity. 

The presence of \FST~changes the interaction of the different scales of turbulence. Results showed that the 20\% \FST~penetrates the boundary layer and increases the velocity fluctuations along the entire boundary layer in the entire frequency range, influencing even the smallest turbulent structures. For 10\% of inflow turbulence, the \FST~penetrates the boundary layer only in the low--frequency range, increasing the velocity fluctuations caused by the larger scales of the turbulence. 

This work demonstrated that the \FST~significantly increases the turbulence intensity in the boundary layer, which is reflected in a considerable increase of the surface pressure spectrum in the entire frequency range, i.e., more than 10~dB and 6~dB for 20\% and 10\% inflow turbulence, respectively, at 10~m/s. A similar trend is observed for a higher inflow velocity. The level of the surface pressure spectrum for the uniform and turbulent inflow conditions scales with the boundary layer thickness and the integral length scale of the boundary layer for frequencies up to 1~kHz. The Amiet theory was used to predict the far--field noise, and a similar increase of the \SPF~was found.

We adopted the TNO--Blake model to predict the surface pressure fluctuations. The model agrees with the experimental results for the uniform inflow case. However, for cases of inflow turbulence, the model cannot predict the \SPF~correctly, even when using the boundary layer measurements as input. This shows the need to extend the model to account for the effect of the penetration of \FST on the boundary layer.

Future efforts will be focused on modeling the effect of the \FST~on the turbulence inside of the boundary layer and surface pressure fluctuations near the trailing edge and determining the relation between the inflow turbulence and length scale and the frequency range of the penetration. 

\section*{Acknowledgments}
The authors would like to Acknowledge Ing. W. Lette, ir. E. Leusink, S. Wanrooij, and the technicians from the metal workshop of the University of Twente for the technical support. The authors also acknowledge TNO, and the Maritime Research Institute Netherlands (MARIN) for the insightful discussions. Furthermore, this research has received funding from the European Union’s Horizon 2020 research and innovation program under the Marie Sklodowska--Curie grant agreement No. 860101.
\bibliography{sample}

\end{document}